\documentclass[11pt]{article}
\usepackage{graphics}
\usepackage{psfrag}
\usepackage{epsfig}
\usepackage{psfig}
\usepackage{epsf}
\usepackage{float}
\usepackage{amssymb,stmaryrd,latexsym}
\newcommand{\fet}[1]{\mbox{\boldmath $#1$}}

\newcommand{\beq}{\begin{equation}}
\newcommand{\eeq}{\end{equation}}
\newcommand{\beqa}{\begin{eqnarray}}
\newcommand{\eeqa}{\end{eqnarray}}
\newcommand{\nn}{\nonumber \\ }

\newcommand{\vs}{\vspace{-0.2cm}}

\newcommand{\no}{\nonumber}
\newcommand{\Mp}{M_\pi}

\newcommand{\Mpn}{M_{\pi^0}}
\newcommand{\Mpp}{M_{\pi^+}}

\newcommand{\Mppz}{M_{\pi^+}^2}

\newcommand{\LOI}{{\rm LO}\!\!\!\!/}

\begin{document}

%%{\bf DRAFT} 

\hfill {\tiny JLAB-THY-04-228}

\hfill {\tiny HISKP(TH)-04/07}

\hfill {\tiny FZJ-IKP-TH-2004-04}
%%{\bf \today} 

%\hfill nucl-th/0405048

\vspace{1cm}

\begin{center}

{{\Large\bf 
The two--nucleon system at next-to-next-to-next-to-leading order}}\footnote{Work 
supported in part by U.S.~Department of Energy under contract number DE-AC05-84ER40150.}

\end{center}

\vspace{.3in}

\begin{center}

{\large 
E. Epelbaum,$^\ast$\footnote{email: epelbaum@jlab.org}
W. Gl\"ockle,$^\dagger$\footnote{email:
                           walter.gloeckle@tp2.ruhr-uni-bochum.de}
Ulf-G. Mei{\ss}ner$^\star$$^\ddagger$\footnote{email: 
                           meissner@itkp.uni-bonn.de}}

\bigskip

$^\ast${\it Jefferson Laboratory, Theory Division, Newport News, VA 23606, USA}

\bigskip 

$^\dagger${\it Ruhr-Universit\"at Bochum, Institut f{\"u}r
  Theoretische Physik II\\ D-44870 Bochum, Germany}\\

\bigskip

$^\star${\it Universit\"at Bonn, Helmholtz-Institut f{\"u}r
  Strahlen- und Kernphysik (Theorie)\\ D-53115 Bonn, Germany}\\

\bigskip

$^\ddagger${\it Forschungszentrum J\"ulich, Institut f\"ur Kernphysik 
(Theorie)\\ D-52425 J\"ulich, Germany}

\end{center}

\vspace{.6in}

\thispagestyle{empty} 

\begin{abstract}
\noindent 
We consider the two--nucleon system at next-to-next-to-next-to-leading order (N$^3$LO) 
in chiral effective field theory. The two--nucleon potential at N$^3$LO consists 
of one-, two- and three-pion exchanges and a set of contact interactions with zero, 
two and four derivatives. In addition, one has to take into account various 
isospin--breaking and relativistic corrections. We employ spectral function 
regularization for the multi--pion exchanges. Within this framework,
it is shown that the three-pion exchange contribution is negligibly small.
The low--energy constants (LECs) related to pion-nucleon vertices  
are taken consistently from studies of pion-nucleon scattering in
chiral perturbation theory. The total of 26 four--nucleon LECs has been determined by
a combined fit  to some $np$ and $pp$ phase shifts from the Nijmegen analysis
together with the $nn$ scattering length. The description of nucleon--nucleon 
scattering and the deuteron observables at N$^3$LO is improved 
compared to the one at NLO and NNLO. The theoretical uncertainties in observables are 
estimated based on the variation of the cut--offs in the spectral function representation 
of the potential and in the regulator utilized in the Lippmann--Schwinger equation.  
\end{abstract}

\vfill

\pagebreak

%%%%%%%%%%%%%%%%%%%%%%%%%%%%%%%%%%%%%%%%%%%%%%%%%%%%%%%%%%%%%%%%%%%%%%%%%%%%%%%%%
\section{Introduction}
\def\theequation{\arabic{section}.\arabic{equation}}
\label{sec:intro}

Since the seminal work of Weinberg \cite{wein} to derive the forces between
two, three, $\ldots$ nucleons from chiral effective field theory,
there has been a flurry of activities to work out the consequences of such an
approach, to improve that scheme or to construct alternatives, for reviews see
\cite{border,BvK}. Here, we will be dealing with a modified Weinberg scheme,
in which pions are treated nonperturbatively and the power counting is 
applied to the nucleon-nucleon potential. The potential consists of
one-, two-, $\ldots$ pion--exchanges (1PE, 2PE, $\ldots$) and a string
of contact interactions with an increasing number of derivatives (zero, two,
four, $\ldots$) that parameterize the shorter ranged components of the
nuclear force  (the precise framework is specified
in more detail below). Such an approach has a variety of advantages over 
more conventional schemes or phenomenological models. First, it offers a systematic
method to improve calculations by going to ever increasing orders in the
power counting and it allows to give theoretical uncertainties. 
Second, one can consistently derive two- and three-nucleon forces (see e.g.
\cite{E3NF}), which has never been achieved before and paves e.g. the way for
a new look at the problem of  nuclear matter. Third, nucleon and nuclear
properties can be calculated from one effective Lagrangian, which is of
particular importance if one intends to extract neutron properties from 
(electromagnetically induced) measurements on light nuclei in a controlled theoretical way. 
In this paper, we present the nucleon-nucleon potential
at next-to-next-to-next-to-leading order (N$^3$LO) in the chiral expansion, 
extending our earlier work, and we apply this potential to observables in 
two--nucleon systems.
Our work differs from the one of Entem and Machleidt (EM), who first presented an
N$^3$LO potential in Ref.~\cite{EM3}, in various ways. First, they use the
two--pion exchange contributions based on dimensional regularization,
which have a very singular short-range behavior. We employ spectral function
regularization, which allows for a better separation between the long- and
short-distance contributions. Second, EM present results only for one 
choice of the cut-off necessary to regulate the high-momentum components
in the Lippman-Schwinger equation to generate the scattering and the bound
states.\footnote{The results for two different choices of the cut--off in 
the Lippmann--Schwinger equation are shown in \cite{Machl_INT}.} 
We perform systematic variations of this cut-off and other parameters
which allows us to give not only central values but also theoretical uncertainties.
Third, our treatment of the isospin breaking effects differs from the one
of EM (which is based on our earlier work \cite{WME}). Fourth, we employ
a relativistic version of the Schr\"odinger equation, which allows to 
calculate consistently relativistic corrections also in three and four nucleon
systems. Other less significant differences will be discussed in due course.
We believe that with the material presented here an important step has been
made to put precision calculations in nuclear physics on a firm theoretical basis,
which not only allows to readdress many issues that have already been investigated
in quite a detail but will also open  new areas of testing chiral dynamics in
few-nucleon systems or shed more light on the issue of the nuclear forces in the
limit of vanishing quark masses (see \cite{Betal,BeSa1,CHlim,BeSa2} for earlier work 
on that topic).

\medskip
\noindent Our manuscript is organized as follows. In sec.~\ref{sec:pot}, we
explicitly give the potential at N$^3$LO. The contributions up to NNLO have 
already been extensively discussed in \cite{ubi,fr94,norb,EGM1,EGM2,EGMs1}. 
The N$^3$LO corrections due to two-- and three--pion 
exchange have been derived recently in \cite{NK21,NK31,NK32} using dimensional regularization
to regularize divergent loop integrals. Throughout this work, we use a 
different regularization scheme for the potential, namely the 
spectral function regularization (SFR). This approach has been recently 
proposed and successfully applied at NLO and NNLO \cite{EGMs1,EGMs2}. As demonstrated in these
references, the SFR scheme allows to significantly improve 
the convergence of chiral Effective Field Theory (EFT) for the two--nucleon system.
We also give an overview of various isospin--breaking interactions including 
electromagnetic forces and discuss the regularization procedure necessary to
render the (iterated) potential finite. 
In section \ref{sec:LS} we deal with the scattering equation. In order to account for the 
relativistic corrections to the nucleon kinetic energy, we have decided to use the Lippmann--Schwinger 
equation with the relativistic expression for the kinetic energy. Such an approach can naturally be extended 
to few--nucleon systems and to processes with external probes.
We also discuss how to cast the relativistic Lippmann--Schwinger equation into a nonrelativistic form, which 
might be useful in certain applications. 
Various deuteron properties are considered in section \ref{sec:bs} using both relativistic 
and nonrelativistic Schr\"odinger equations. 
The fitting procedure to determine the low--energy constants (LECs) 
and the accuracy of the fits are detailed in sec.~\ref{sec:fit}. 
Results for phase shifts
and the deuteron (bound state) properties are displayed and discussed in 
sec.~\ref{sec:res}. Our findings are summarized in sec.~\ref{sec:summ}. The appendices
contain details on the kinematics, the partial wave decomposition, 
the momentum space treatment of the Coulomb interaction
and the effective range expansion.
\bigskip
\noindent

%%%%%%%%%%%%%%%%%%%%%%%%%%%%%%%%%%%%%%%%%%%%%%%%%%%%%%%%%%%%%%%%%%%%%%%%%%%%%%%%%
\section{The two--nucleon potential at N$^3$LO}
\def\theequation{\arabic{section}.\arabic{equation}}
\setcounter{equation}{0}
\label{sec:pot}

\subsection{General remarks}
\label{sec:GenRem}

Before going into details of calculations, we would like to make certain general remarks.   
As already pointed out in the introduction, we strictly follow the scheme suggested by
Weinberg \cite{wein}. In this approach one uses the EFT technique to derive nuclear forces 
from the most general (approximately) chiral invariant effective Lagrangian. 
The NN S--matrix is obtained via (non--perturbative) solution of the Lippmann--Schwinger 
(LS) equation. In most practical calculations (including the present one), the later step 
can only be performed numerically. 

\medskip\noindent
Starting from the most general chiral invariant effective Hamiltonian
density for pions and nucleons one can derive energy-independent and hermitean nuclear forces by
a variety of methods including the method of unitary transformation, see e.g. \cite{CHlim}.
The resulting nucleonic forces are ordered by the power of the generic low--momentum scale $Q$
related to the  three--momenta of nucleons, the pion mass and typical four--momenta 
of virtual pions:
\beq
V \sim  \mathcal{O} \left[ \frac{1}{F_\pi^2} \left( \frac{Q}{\Lambda_\chi} \right)^\nu \right]
\eeq
where $F_\pi$ is the pion decay constant and $\Lambda_\chi$ is the chiral symmetry breaking scale
or, more generally, the smaller of the chiral symmetry breaking scale and 
the scale $\Lambda_{\rm LEC}$ 
associated with the LECs accompanying four--nucleon contact interactions (as discussed below).  
The power $\nu$ for a given diagram can be calculated using the rules of dimensional analysis
\cite{wein}
\beq
\label{powc}
\nu = -2 + 2 E_n + 2 (L -C) + \sum_i V_i \Delta_i~,
\eeq
where $E_n$, $L$, $C$ and $V_i$ are the numbers of nucleons, loops, separately connected pieces and vertices of
type $i$, respectively. Further, the quantity $\Delta_i$, which defines the dimension of a vertex of
type $i$,  is given by
\beq
\label{deltai}
\Delta_i = d_i + \frac{1}{2} n_i - 2~,
\eeq
with $d_i$ the number of derivatives or $M_\pi$ insertions and $n_i$ the number of 
nucleon lines at the vertex $i$.  One has $\Delta_i \geq 0$   
as a consequence of chiral invariance.
This leads to $\nu \geq 0$ for connected diagrams  with two and more nucleons. 
One also recognizes that the graphs with loops are suppressed and that $(n+1)$--nucleon 
forces appear at higher orders than the $n$--nucleon ones. We note, however, that 
the formula (\ref{powc}) does not apply to a specific sort of diagrams, sometimes referred to
as reducible, whose contributions are
enhanced due to the 
presence of anomalously small energy denominators resulting from purely nucleonic intermediate states.
Such reducible diagrams are responsible for the non--perturbative aspect in the few--nucleon problem and 
must be summed up to infinite order. They, however, do not contribute to the nuclear potential and result 
from iteration of the potential in the Lippmann--Schwinger equation. 

\medskip\noindent
It remains to specify our way of counting the nucleon mass. In the single--nucleon sector it appears to be 
natural to treat the nucleon mass $m$ in the same way as the chiral symmetry breaking scale $\Lambda_\chi \sim 1\,$GeV.
As argued in \cite{wein}, in the few--nucleon sector consistency requires that the nucleon mass is 
considered as a much larger scale compared to the chiral symmetry breaking scale. If one adopts the 
counting rule $m \sim \Lambda_\chi$, no nonperturbative resummation of the amplitude is required from the 
point of view of the chiral power counting. In this work we adopt the counting rule $m/Q \sim 
(\Lambda_\chi/Q )^2$, which has also been used in \cite{ubi}. 
%(XXX really needed??)

\medskip\noindent
In the following sections we will discuss various contributions to the NN potential up to N$^3$LO 
including isospin--breaking corrections. 

\subsection{Contact terms}
We  consider first the contact terms of the two--nucleon potential. To the accuracy
we are working, the potential in the center--of--mass
system (cms) for initial and final nucleon momenta $\vec{p}$ and $\vec{p}~'$, 
respectively, takes the form:\footnote{We use the notation of ref.~\cite{EGM2} 
(ref.~\cite{Machl_INT}) for $V^{(0)}$ and $V^{(2)}$ ($V^{(4)}$).} 
\beqa\label{Vcon}
V_{\rm cont} &=& V^{(0)}_{\rm cont} + V^{(2)}_{\rm cont} + V^{(4)}_{\rm cont}~, \no\\
V^{(0)}_{\rm cont} &=& C_S  + C_T \, \vec{\sigma}_1 \cdot  \vec{\sigma}_2~,\no\\
V^{(2)}_{\rm cont} &=& C_1 \, \vec{q}\,^2 + C_2 \, \vec{k}^2 +
( C_3 \, \vec{q}\,^2 + C_4 \, \vec{k}^2 ) ( \vec{\sigma}_1 \cdot \vec{\sigma}_2)
+ iC_5\, \frac{1}{2} \, ( \vec{\sigma}_1 + \vec{\sigma}_2) \cdot ( \vec{q} \times
\vec{k})\no\\
&& {} + C_6 \, (\vec{q}\cdot \vec{\sigma}_1 )(\vec{q}\cdot \vec{\sigma}_2 ) 
+ C_7 \, (\vec{k}\cdot \vec{\sigma}_1 )(\vec{k}\cdot \vec{\sigma}_2 )~,\nn
V^{(4)}_{\rm cont} &=& D_1 \, \vec q \,^4 + D_2 \, \vec k \,^4 + D_3 \, \vec q\,^2  \vec k \,^2
+ D_4 \, (\vec q \times \vec k )^2 \no\\
&& {} +  \Big( D_5 \, \vec q \,^4 + D_6 \, \vec k \,^4 + D_7 \, \vec q\,^2 \vec k \,^2 
+ D_8 \, (\vec q \times \vec k )^2 \Big)  (\vec \sigma_1 \cdot \vec \sigma_2 ) \no\\
&& {} +  i \,\Big(  D_{9} \,  \vec q \,^2 + D_{10} \, \vec k \,^2 \Big) \frac{\vec \sigma_1
+ \vec \sigma_2 }{2} \cdot (\vec q \times \vec k ) \nn 
&& {} + \Big( D_{11} \,  \vec q \,^2 + D_{12} \, \vec k \,^2 \Big) (\vec \sigma_1 \cdot \vec q ) 
(\vec \sigma_2 \cdot \vec q \Big) 
+ \Big( D_{13} \,  \vec q \,^2 + D_{14} \, \vec k \,^2 \Big) (\vec \sigma_1 \cdot \vec k ) 
(\vec \sigma_2 \cdot \vec k ) \no\\
&& {} + D_{15} \, \Big(\vec \sigma_1 \cdot ( \vec q \times \vec k ) \, \vec 
\sigma_2 \cdot (\vec q \times \vec k) \Big) 
\eeqa
with $\vec{q} = \vec{p}~'-\vec{p}$ and $\vec{k} = (\vec{p}+\vec{p}~')/2$. 
The superscripts denote the corresponding chiral order as defined in eq.~(\ref{powc}). 
Notice that 
the contact operator basis in eq.~(\ref{Vcon}) represents just one particular choice among many others.
One could equally well use another set of 24 independent contact operators
including for instance terms which contain the product of isospin matrices 
$\vec \tau_1 \cdot \vec \tau_2$. A one--to--one correspondence between different sets of 
contact operators can be established upon performing antisymmetrization of the potential,
see ref.~\cite{Ephd} for more details. 
Notice that we have only shown isospin--invariant terms in eq.~(\ref{Vcon}).
Isospin--breaking short--range corrections will be specified below.     

\medskip\noindent
The terms in eq.~(\ref{Vcon}) feed into the matrix--elements of the two S--waves
($^1S_0$, $^3S_1$), the four P--waves 
($^1P_1$, $^3P_1$, $^3P_2$, $^3P_0$), the four D--waves 
($^1D_2$, $^3D_2$, $^3D_3$, $^3D_1$) 
and the two lowest transition potentials
($^3D_1 - ^3S_1$, $^3F_2 - ^3P_2$) in the following way:
\beqa\label{VC}
\langle ^1S_0 | V_{\rm cont}| ^1S_0 \rangle&=& \tilde C_{1S0} + C_{1S0} ( p^2 + p '^2) +
D_{1S0}^1 \, p^2 \, {p'}^2 + D_{1S0}^2 \, ({p}^4+{p}'^4)~,\nn
\langle ^3S_1 | V_{\rm cont}| ^3S_1 \rangle&=& \tilde C_{3S1} + C_{3S1} ( p^2 + p '^2) +
D_{3S1}^1 \, p^2 \, {p'}^2 + D_{3S1}^2 \, ({p}^4+{p}'^4)~,\nn
\langle ^1P_1 | V_{\rm cont}| ^1P_1 \rangle&=& C_{1P1} \, p \, p' +
D_{1P1} \,p \, p' \, ({p}^2 +  p' \, ^2)~,\nn
\langle ^3P_1 | V_{\rm cont}| ^3P_1 \rangle&=& C_{3P1} \, p  \, p' + D_{3P1} \,p \, p' \, ({p}^2 \, 
 +p ' \, ^2 )~,\nn
\langle ^3P_0 | V_{\rm cont}| ^3P_0 \rangle&=& C_{3P0} \, p  \, p' + D_{3P0} \,p \, p' \, ({p}^2  
 +p ' \, ^2 )~,\nn
\langle ^3P_2 | V_{\rm cont}| ^3P_2 \rangle&=& C_{3P2} \, p  \, p' + D_{3P2} \,p \, p' \, ({p}^2  
 +p ' \,^2 )~,\nn
\langle ^1D_2 | V_{\rm cont}| ^1D_2 \rangle&=& D_{1D2} \, {p}^2\,{p}'^2~,\nn
\langle ^3D_2 | V_{\rm cont}| ^3D_2 \rangle&=& D_{3D2} \, {p}^2\,{p}'^2~,\nn
\langle ^3D_1 | V_{\rm cont}| ^3D_1 \rangle&=& D_{3D1} \, {p}^2\,{p}'^2~,\nn
\langle ^3D_3 | V_{\rm cont}| ^3D_3 \rangle&=& D_{3D3} \, {p}^2\,{p}'^2~,\nn
\langle ^3S_1 | V_{\rm cont}| ^3D_1 \rangle&=& C_{3D1 - 3S1} \, p^2 
+ D_{3D1 - 3S1}^1 \, {p}^2\,{p}'^2 + D_{3D1 - 3S1}^2 \, p^4~,\nn
\langle ^3D_1 | V_{\rm cont}| ^3S_1 \rangle&=& C_{3D1 - 3S1} \, {p '}^2 
+ D_{3D1 - 3S1}^1 \, {p}^2\,{p}'^2 + D_{3D1 - 3S1}^2 \, {p '}^4~,\nn
\langle ^3P_2 | V_{\rm cont}| ^3F_2 \rangle&=& D_{3F2 - 3P2}\, {p}^3\,{p}'~, \nn
\langle ^3F_2 | V_{\rm cont}| ^3P_2 \rangle&=& D_{3F2 - 3P2}\, {p}\,{p '}^3~, 
\label{V4ct}
\eeqa
with $p = |\vec{p}\,|$ and ${p}' = |\vec{p}\,'|$.
The spectroscopic LECs are related to the ones in eq.~(\ref{Vcon}) according 
to the following relations:
\beqa\label{VC_Vcon}
\tilde C_{1S0} &=& 4 \pi (C_S - 3 C_T )~, \nn
\tilde C_{3S1} &=& 4 \pi (C_S + C_T )~, \nn
C_{1S0} &=& \pi (4 C_1 + C_2 -12C_3
-3C_4 -4C_6 -C_7)~, \nn
C_{3S1} &=& \frac{\pi}{3} \, ( 12C_1 + 3C_2 +12C_3
+3C_4 +4C_6 +C_7) ~,\no\\
C_{1P1} &=& \frac{2\pi}{3} \, ( -4C_1 + C_2 +12C_3
-3C_4 +4C_6 -C_7) ~, \no \\
C_{3P1} &=& \frac{2\pi}{3} \, ( -4C_1 + C_2 - 4C_3
+C_4 + 2C_5 -8C_6 + 2C_7)~\no\\
C_{3P2} &=& \frac{2\pi}{3} \, ( -4C_1 + C_2 - 4C_3
+C_4 - 2C_5 )~, \no \\
C_{3P0} &=& \frac{2\pi}{3} \, ( -4C_1 + C_2 - 4C_3
+C_4 + 4C_5 +12C_6 - 3C_7)~, \no \\
C_{3D1 - 3S1} &=& C_{\epsilon 1} = \frac{2\sqrt{2}\pi}{3} \, ( 4C_6 + C_7)~,\nn
D_{1S0}^1 &=& \frac{\pi}{6}  ( 80 D_1 +5 D_2 + 4 D_3 + 16 D_4 - 240 D_5 - 15 D_6 -12 D_7 
-48 D_8 -80 D_{11} -4 D_{12} \nn
&& \mbox {\hskip 0.5 true cm}  -4 D_{13}  -5 D_{14}  -16 D_{15} ) \nn
D_{1S0}^2 &=& \frac{\pi}{4}  ( 16 D_1 + D_2 + 4 D_3  - 48 D_5 - 3 D_6 - 12 D_7 - 16 D_{11} - 4 D_{12} 
 - 4 D_{13}  - D_{14} )~, \nn
D_{3S1}^1 &=& \frac{\pi}{18}  ( 240 D_1 +15 D_2 + 12 D_3 + 48 D_4 + 240 D_5 + 15 D_6 + 12 D_7 
+ 48 D_8 +80 D_{11} + 4 D_{12} \nn
&& \mbox {\hskip 0.5 true cm}  + 4 D_{13}  + 5 D_{14}  + 16 D_{15} ) \nn
D_{3S1}^2 &=& \frac{\pi}{12}  ( 48 D_1 + 3 D_2 + 12 D_3  + 48 D_5 + 3 D_6 + 12 D_7 + 16 D_{11} + 4 D_{12} 
 + 4 D_{13}  + D_{14} )~, \nn
D_{1P1} &=& - \frac{\pi}{3} ( 16 D_1 - D_2 - 48 D_5 + 3 D_6 - 16 D_{11} + D_{14} )~, \nn
D_{3P1} &=& - \frac{\pi}{6} ( 32 D_1 - 2 D_2 + 32 D_5 - 2 D_6 - 8 D_9 - 2  D_{10} + 48 D_{11} 
+4 D_{12} - 4 D_{13} - 3 D_{14} )~, \nn
D_{3P2} &=& - \frac{\pi}{30} ( 160 D_1 - 10 D_2 + 160 D_5 - 10 D_6 + 40 D_9 + 10  D_{10} + 16 D_{11} 
- 4 D_{12} + 4 D_{13} -  D_{14} )~, \nn
D_{3P0} &=& - \frac{\pi}{3} ( 16 D_1 -  D_2 + 16 D_5 -  D_6 - 8 D_9 - 2  D_{10} - 32 D_{11} 
- 4 D_{12} + 4 D_{13} + 2 D_{14} )~, \nn
D_{1D2} &=&  \frac{2 \pi}{15} ( 16 D_1 + D_2 - 4 D_3 - 4 D_4 - 48 D_5 - 3 D_6 + 12 D_7 + 12 D_8 
- 16  D_{11} + 4 D_{12}+ 4 D_{13}
\nn
&& \mbox {\hskip 0.5 true cm}  -  D_{14} + 4 D_{15})~, \nn
D_{3D2} &=&  \frac{\pi}{15} ( 32 D_1 + 2 D_2 - 8 D_3 - 8 D_4 + 32 D_5 + 2 D_6 - 8 D_7 - 8 D_8 
- 8 D_9 + 2 D_{10} + 48  D_{11} \nn 
&& \mbox {\hskip 0.5 true cm} - 12 D_{12} - 12 D_{13} + 3 D_{14} + 16 D_{15})~, \nn
D_{3D1} &=&  \frac{\pi}{45} ( 96 D_1 + 6 D_2 - 24 D_3 - 24 D_4 + 96 D_5 + 6 D_6 - 24 D_7 - 24 D_8 
- 72  D_9  +18 D_{10} \nn
&& \mbox {\hskip 0.5 true cm} - 80 D_{11} + 20 D_{12} + 20 D_{13} - 5 D_{14} - 
64 D_{15})~, \nn
D_{3D3} &=&  \frac{2 \pi}{15} ( 16 D_1 + D_2 - 4 D_3 - 4 D_4 + 16 D_5 + D_6 - 4 D_7 - 4 D_8 + 8 D_9  
- 2  D_{10} - 4 D_{15})~, \nn
D_{3D1 - 3S1}^1 &=& D_{\epsilon 1}^1 = \frac{\sqrt{2} \pi}{18} (112  D_{11} -  4 D_{12} - 4 D_{13} + 7 D_{14} - 16 D_{15} )~, \nn
D_{3D1 - 3S1}^2 &=& D_{\epsilon 1}^2 = \frac{\sqrt{2} \pi}{6} (16 D_{11} + 4 D_{12} + 4 D_{13} + D_{14} )~, \nn
D_{3F2 - 3P2} &=& D_{\epsilon 2} = - \frac{\sqrt{6} \pi}{15} (  16 D_{11} 
- 4 D_{12} + 4 D_{13} -  D_{14} )~, 
\eeqa
These 24 constants are not fixed by chiral symmetry and have to be determined
by a fit to data or phase shifts and mixing parameters in the corresponding channels.  
{}From each of the two S--waves, we can determine four parameters, whereas each of the four
P--waves and the mixing parameter $\epsilon_1$ contain two free
parameters. Further, one free parameter contributes to each of the four D--waves and 
to the mixing parameter $\epsilon_2$.
Of course, we have to account for the channel coupling
in the mixed spin--triplet partial waves.
Once the spectroscopic LECs  have been
determined, the original $C_S, C_T, C_1,\ldots,C_7$ and $D_1, \ldots , D_{15}$  
are fixed uniquely. 
%We remark that all LECs related to contact operators and 
%discussed in the present work have to be understood as renormalized quantities.

\subsection{One--, two-- and three--pion exchange}
\label{sec:pion_pot}
Consider now one--, two--  and three--pion exchange (3PE) contributions
$V_{1\pi}$,  $V_{2\pi}$ and  $V_{3\pi}$, respectively. At N$^3$LO ($Q^4$) in the low--momentum 
expansion $Q$ they can be written as:
\beqa
\label{Vschem}
V_{1\pi} &=&   V_{1\pi}^{(0)} +  V_{1\pi}^{(2)} +  V_{1\pi}^{(3)} + V_{1\pi}^{(4)} + \ldots \,, \nn
V_{2\pi} &=&   V_{2\pi}^{(2)} +  V_{2\pi}^{(3)} + V_{2\pi}^{(4)} + \ldots \,, \nn
V_{3\pi} &=&   V_{3\pi}^{(4)}  + \ldots \,. 
\eeqa
Here the superscripts denote the corresponding chiral order and the ellipses refer to 
$Q^5$-- and higher order terms which are not considered in the present work. 
Contributions due to exchange of four-- and more pions are further suppressed:
$n$--pion exchange diagrams start to contribute at the order $Q^{2n-2}$,
see e.g.~\cite{wein}. 

\medskip\noindent
In the following we will give explicit expressions for the individual contributions 
in eq.~(\ref{Vschem}). The pion--exchange NN potential in the two--nucleon center--of--mass system (c.m.s) 
takes the  form:
\beqa
V &=& V_C + \fet \tau_1 \cdot \fet \tau_2 \, W_C + \left[   
V_S + \fet \tau_1 \cdot \fet \tau_2 \, W_S \right] \, \vec \sigma_1 \cdot \vec \sigma_2 
+ \left[ V_T + \fet \tau_1 \cdot \fet \tau_2 \, W_T \right] 
\, \vec \sigma_1 \cdot \vec q \, \vec \sigma_2 \cdot \vec q \\
&& {}+   \left[   
V_{LS} + \fet \tau_1 \cdot \fet \tau_2 \, W_{LS} \right] \, i ( \vec \sigma_1 + \vec \sigma_2 )
\cdot ( \vec q \times \vec k  ) 
+   \left[   
V_{\sigma L} + \fet \tau_1 \cdot \fet \tau_2 \, W_{\sigma L} \right] \,   \vec \sigma_1 
\cdot (\vec  q \times \vec k  ) \vec \sigma_2 \cdot (\vec  q \times \vec k  ) \,,
\nonumber
\eeqa
where the superscripts $C$, $S$, $T$, $LS$ and $\sigma L$ of the scalar functions 
$V_C$, $\ldots$, $W_{\sigma L}$ refer to central, spin--spin, tensor, spin--orbit and 
quadratic spin--orbit components, respectively.

\medskip\noindent
The leading order 1PE potential is given by  
\beq
\label{opep}
W_{T}^{(0)} (q) = -\biggl(\frac{g_A}{2F_\pi}\biggr)^2 \, 
\frac{\vec{\sigma}_1 \cdot\vec{q}\,\vec{\sigma}_2\cdot\vec{q}}
{q^2 + M_\pi^2}~.
\eeq
At NLO one has to take into account various corrections  
which result from one--loop diagrams with the leading vertices and 
tree graphs with one insertion of the $d_{16}$-- and $\tilde d_{28}$--vertices 
(in the notation of ref.~\cite{Fet98}) 
from the dimension three Lagrangian $\mathcal{L}_{\pi N}$ and $l_{3,4}$--vertices
from the dimension four Lagrangian $\mathcal{L}_{\pi}$. 
All these graphs lead just to renormalization of the LECs $g_A$, $F_\pi$ and 
the pion mass $M_\pi$, see \cite{CHlim} for more 
details. In addition, one has a contribution from the 1PE graphs with 
one $g_A$--vertex replaced by the $d_{18}$--vertex  from the dimension three 
$\pi N$ Lagrangian. This correction leads to the so-called
Goldberger--Treiman discrepancy and can be accounted for by the replacement 
%*EE
\beq
\label{d18}
g_A \rightarrow g_A - 2 d_{18}M_\pi^2 
\eeq
in  Eq.~(\ref{opep}).
The corrections at NNLO arise from one--loop diagrams with one subleading $\pi NN$ vertex
and lead to renormalization of the LEC $g_A$ \cite{norb}. 
The corrections to 1PE at N$^3$LO are due to two--loop diagrams with all vertices of the 
lowest chiral dimension, one--loop graphs with one subleading vertex from $\mathcal{L}_{\pi N}$
or $\mathcal{L}_{\pi}$ and tree graphs with two subleading vertices or one sub--subleading vertex.
After performing renormalization of the LECs, one finds the N$^3$LO contribution to the 1PE 
potential to be proportional to \cite{Norbert04}:
\beq
 \label{opep2}
W_{T}^{(4)} (q) \propto 
\frac{1}{q^2 + M_\pi^2} F (q^2) \,,
\eeq
where the function
%*EE
$F (q^2) = \alpha_1 M_\pi^4 + \alpha_2 M_\pi^2 q^2 + \alpha_3 q^4$ can be viewed as the pion--nucleon form--factor.  
The latter does  not represent an observable quantity. 
Expressing the function $F(q)$ as 
\beq
F (q^2) = (\alpha_1 - \alpha_2 + \alpha_3) M_\pi^4+
(q^2 + M_\pi^2) (\alpha_2 M_\pi^2 + \alpha_3 (q^2 - M_\pi^2))
\eeq
the N$^3$LO contribution reduces to a renormalization of the 1PE potential (\ref{opep})
and contact interactions. Notice that one also encounters an additional
correction to the Goldberger--Treiman discrepancy. 
In addition, one has to take into account relativistic $1/m^2$--corrections to the 
static 1PE potential, 
which depend on a particular choice of the unitary transformation 
in the NN system and the form of the scattering (or bound state) equation, see ref.~\cite{Friar99} for details. 
The final expression for the 1PE potential 
adopted in the present work 
takes  the form
\beq
\label{opep_full}
V_{1\pi} (q) = -\biggl(\frac{g_A}{2F_\pi}\biggr)^2 \, 
\left( 1 - \frac{p^2 + p'{}^2}{2 m^2} \right) \vec{\tau}_1 \cdot
\vec{\tau}_2 \, \frac{\vec{\sigma}_1 \cdot\vec{q}\,\vec{\sigma}_2\cdot\vec{q}}
{q^2 + M_\pi^2}~,
\eeq
where, as in our previous work \cite{EGMs2},  
we take the larger value $g_A = 1.29$ instead of  $g_A = 1.26$ in order to 
account for the Goldberger--Treiman discrepancy. This corresponds to the
pion-nucleon coupling constant $g_{\pi N} =13.1$.

\medskip\noindent
We now turn to the 2PE contributions. The 2PE potential $V_{2\pi}^{(2)} +  V_{2\pi}^{(3)}$
is discussed in \cite{fr94,norb,EGM1,EGM2} and in \cite{ubi} using an energy--dependent formalism.
While dimensional regularization or equivalent schemes have been used in \cite{fr94,norb,EGM2}
to calculate matrix elements of the potential, a finite momentum cut--off approach has been 
applied in \cite{ubi}. The N$^3$LO corrections have been recently obtained by Kaiser 
using dimensional regularization \cite{NK21}.
In the following, we will adopt the SFR method to obtain the non--polynomial contributions 
to the 2PE potential with the short--range components being explicitly excluded, see \cite{EGMs1} 
for more details. The expressions for the 2PE potential in the SFR scheme up to NNLO have 
already been given in \cite{EGMs2}. To keep the presentation self--contained, we 
give below the corresponding (non--polynomial) terms at NLO
\beqa
\label{2PE_nlo}
W_C^{(2)} (q) &=& - \frac{1}{384 \pi^2 F_\pi^4}\,
L^{\tilde \Lambda} (q) \, \biggl\{4M_\pi^2 (5g_A^4 - 4g_A^2 -1)  + q^2(23g_A^4 - 10g_A^2 -1)
+ \frac{48 g_A^4 M_\pi^4}{4 M_\pi^2 + q^2} \biggr\}\nn
V_T^{(2)} (q) &=& -\frac{1}{q^2} V_S^{(2)} (q)  = - \frac{3 g_A^4}{64 \pi^2 F_\pi^4} \,L^{\tilde \Lambda} (q)
\eeqa
and at NNLO: 
\beqa
\label{2PE_nnlo}
V_C^{(3)} (q) &=& -\frac{3g_A^2}{16\pi F_\pi^4}  \biggl\{2M_\pi^2(2c_1 -c_3) -c_3 q^2 \biggr\} 
(2M_\pi^2+q^2) A^{\tilde \Lambda} (q) \nn
W_T^{(3)} (q) &=& -\frac{1}{q^2} W_S^{(3)} (q) = - \frac{g_A^2}{32\pi F_\pi^4} \,  c_4 (4M_\pi^2 + q^2) 
A^{\tilde \Lambda}(q)\,,
\eeqa
where the NLO and NNLO loop functions $L^{\tilde \Lambda} (q)$ and $A^{\tilde \Lambda} (q)$
are given by 
\beq
\label{def_LA}
L^{\tilde \Lambda} (q) = \theta (\tilde \Lambda - 2 M_\pi ) \, \frac{\omega}{2 q} \, 
\ln \frac{\tilde \Lambda^2 \omega^2 + q^2 s^2 + 2 \tilde \Lambda q 
\omega s}{4 M_\pi^2 ( \tilde \Lambda^2 + q^2)}~, \quad \quad
\omega = \sqrt{ q^2 + 4 M_\pi^2}~,  \quad \quad
s = \sqrt{\tilde \Lambda^2 - 4 M_\pi^2}\,.
\eeq
and
\beq
A^{\tilde \Lambda} (q) = \theta (\tilde \Lambda - 2 M_\pi ) \, \frac{1}{2 q} \, 
\arctan \frac{q ( \tilde \Lambda - 2 M_\pi )}{q^2 + 2 \tilde \Lambda M_\pi}\,.
\eeq
The N$^3$LO corrections to the 2PE potential $V_{2\pi}^{(4)}$ have been 
recently calculated by Kaiser \cite{NK21}.  They arise from the one--loop
``bubble'' diagrams with both dimension two $\pi \pi NN$ vertices of the 
$c_{1,\ldots , 4}$--type and
from the diagrams which contain the third order  pion--nucleon amplitude and lead to 
one--loop and two--loop graphs. We begin with the first group 
of corrections, for which one finds:
\beqa
\label{2PE_nnnlo}
V_C^{(4)} (q) &=& \frac{3}{16 \pi^2  F_\pi^4} \, L^{\tilde \Lambda} (q) \, 
\left\{ \left[ \frac{c_2}{6} \omega^2 + c_3 (2 M_\pi^2 + q^2 ) - 4 c_1 M_\pi^2 
\right]^2 + \frac{c_2^2}{45} \omega^4 \right\} 
\nonumber \\
W_T^{(4)} (q) &=& -\frac{1}{q^2} W_S^{(4)} (q) = \frac{c_4^2}{96 \pi^2 F_\pi^4} \omega^2  
\, L^{\tilde \Lambda} (q)\,.
\eeqa
No closed expressions can be given for some of the corrections from 
the second group.\footnote{Entem and Machleidt were able to 
calculate most of the integrals in eqs.~(\ref{TPE2loop}) analytically (for $\tilde \Lambda = \infty$)
and to express the corresponding contributions to the potential in terms
%*EE
of the loop functions $L^\infty (q)$ and $A^\infty (q)$ \cite{EM2}.}
It appears to be convenient to give the contributions to the 
potential using the (subtracted) spectral function representation:
\beqa
V_{C,S} (q) &=& - \frac{2 q^6}{\pi} \int_{2M_\pi}^\infty \, d \mu 
\frac{\rho_{C,S} (\mu )}{\mu^5 ( \mu^2 + q^2 )}\,, \quad \quad
V_T (q) = \frac{2 q^4}{\pi} \int_{2M_\pi}^\infty \, d \mu 
\frac{\rho_{T} (\mu )}{\mu^3 ( \mu^2 + q^2 )}\,, \nn
W_{C,S} (q) &=& - \frac{2 q^6}{\pi} \int_{2M_\pi}^\infty \, d \mu 
\frac{\eta_{C,S} (\mu )}{\mu^5 ( \mu^2 + q^2 )}\,, \quad \quad
W_T (q) = \frac{2 q^4}{\pi} \int_{2M_\pi}^\infty \, d \mu 
\frac{\eta_{T} (\mu )}{\mu^3 ( \mu^2 + q^2 )}\,. 
\eeqa
For the spectral functions $\rho_i (\mu)$ ($\eta_i (\mu)$) one finds:
\cite{NK21}:
\beqa
\label{TPE2loop}
\rho_C^{(4)} (\mu ) &=& - \frac{3 g_A^4 (\mu^2 - 2 M_\pi^2 )}{\pi \mu (4 F_\pi)^6}
\, \theta ( \tilde \Lambda - \mu ) \, 
\bigg\{ (M_\pi^2 - 2 \mu^2 ) \bigg[ 2 M_\pi + \frac{2 M_\pi^2 - \mu^2}{2 \mu} 
\ln \frac{\mu + 2 M_\pi}{\mu - 2 M_\pi } \bigg] \nn
&& \mbox{\hskip 5 true cm} + 4 g_A^2 M_\pi (2 M_\pi^2 - \mu^2 )
\bigg\}\,, \nn
\eta_S^{(4)} (\mu ) &=& \mu^2 \eta_T^{(4)} (\mu ) = - 
\frac{g_A^4 (\mu^2 - 4 M_\pi^2 )}{\pi (4 F_\pi)^6}
\, \theta ( \tilde \Lambda - \mu ) \, 
\left\{ \left(M_\pi^2 - \frac{\mu^2}{4} \right) \ln \frac{\mu +  2 M_\pi}{\mu - 2 M_\pi }
+ (1 + 2 g_A^2 ) \mu M_\pi \right\}\,, \nn
\rho_S^{(4)} (\mu ) &=& \mu^2 \rho_T^{(4)} (\mu ) = - 
\theta ( \tilde \Lambda - \mu ) \,  \left\{
\frac{g_A^2 r^3 \mu}{8 F_\pi^4 \pi} 
(\bar d_{14} - \bar d_{15} ) - 
\frac{2 g_A^6 \mu r^3}{(8 \pi F_\pi^2)^3} \left[ \frac{1}{9} - J_1  + J_2 \right] \right\}\,, \nn
 \eta_C^{(4)} (\mu ) &=&  \theta ( \tilde \Lambda - \mu ) \, \Bigg\{
\frac{r t^2}{24 F_\pi^4 \mu \pi} \left[ 2 (g_A^2 - 1) r^2 - 3 g_A^2 t^2 \right] (\bar d_1 + \bar d_2 ) \nn
&& {}+ \frac{r^3}{60 F_\pi^4 \mu \pi} \left[ 6 (g_A^2 - 1) r^2 - 5 g_A^2 t^2 \right] \bar d_3
- \frac{r M_\pi^2}{6 F_\pi^4 \mu \pi} \left[ 2 (g_A^2 - 1) r^2 - 3 g_A^2 t^2 \right] \bar d_5 \nn
&& {} - \frac{1}{92160 F_\pi^6 \mu^2 \pi^3} \Big[ - 320 (1 + 2 g_A^2 )^2 M_\pi^6 + 
240 (1 + 6 g_A^2 + 8 g_A^4 ) M_\pi^4 \mu^2 \nn
&& {}   \mbox{\hskip 3 true cm} - 60 g_A^2 (8 + 15 g_A^2 ) M_\pi^2  \mu^4
+ (-4 + 29 g_A^2 + 122 g_A^4 + 3 g_A^6 ) \mu^6 \Big] \ln \frac{2 r + \mu}{2 M_\pi} \nn
&& {} - \frac{r}{2700 \mu ( 8 \pi F_\pi^2 )^3} \Big[ -16 ( 171 + 2 g_A^2 ( 1 + g_A^2) 
(327 + 49 g_A^2)) M_\pi^4 \nn
&& {}   \mbox{\hskip 3 true cm} + 4 (-73 + 1748 g_A^2 + 2549 g_A^4 + 726 g_A^6 ) M_\pi^2 \mu^2 \nn
&& {}   \mbox{\hskip 3 true cm} - (- 64 + 389 g_A^2 + 1782 g_A^4 + 1093 g_A^6 ) \mu^4  \Big] \nn
&& {} + \frac{2 r}{3 \mu ( 8 \pi F_\pi^2 )^3} \Big[ 
g_A^6 t^4 J_1 - 2 g_A^4 (2 g_A^2 -1 ) r^2 t^2 J_2 \Big] \Bigg\}\,,
\eeqa
where we have introduced the abbreviations
\beq
r = \frac{1}{2} \sqrt{ \mu^2  - 4 M_\pi^2}\,, \quad \quad \quad
t= \sqrt{\mu^2 - 2 M_\pi^2}\,,
\eeq
and 
\beqa
J_1 &=& \int_0^1 \, dx \, \bigg\{ \frac{M_\pi^2}{r^2 x^2} - \bigg( 1 + \frac{M_\pi^2 }{r^2 x^2} \bigg)^{3/2}
\ln \frac{ r x + \sqrt{ M_\pi^2 + r^2 x^2}}{M_\pi} \bigg\}\,,\nonumber \\
J_2 &=& \int_0^1 \, dx \, x^2 \bigg\{ \frac{M_\pi^2}{r^2 x^2} - \bigg( 1 + \frac{M_\pi^2 }{r^2 x^2} \bigg)^{3/2}
\ln \frac{ r x + \sqrt{ M_\pi^2 + r^2 x^2}}{M_\pi} \bigg\}\,.
\eeqa
We use the scale--independent LECs $\bar d_{1}, \; \bar d_{2}, \; \bar d_{3}, \; \bar d_{5}, \; \bar d_{14}$ and 
$\bar d_{15}$ defined in \cite{Fet98}.
In addition, one has to take into account the leading relativistic $1/m$--corrections
to the 2PE potential:
\beqa
\label{tpe1m}
V_C^{(4)} (q) &=& \frac{3 g_A^4}{512  \pi m F_\pi^4} \bigg\{ \frac{2 M_\pi^5}{\omega^2}-
 3  ( 4 M_\pi^4 - q^4 )  A^{\tilde \Lambda} (q) \bigg\}\,,\nonumber \\
W_C^{(4)} (q) &=& \frac{g_A^2}{128 \pi m F_\pi^4} \Bigg\{ \frac{3 g_A^2 M_\pi^5}{\omega^2} 
- \bigg[ 4 M_\pi^2 + 2 q^2 - g_A^2 \left( 7 M_\pi^2 + \frac{9}{2} q^2 \right) \bigg] (2 M_\pi^2 + q^2 )
  A^{\tilde \Lambda} (q) \Bigg\}\,, \nn
V_T^{(4)} (q) &=& -\frac{1}{q^2} V_S^{(4)} (q) =  \frac{9 g_A^2}{512 \pi m F_\pi^4}  \left(
4 M_\pi^2 + \frac{3}{2} q^2 \right) A^{\tilde \Lambda} (q) \,, \nn
W_T^{(4)} (q) &=& -\frac{1}{q^2} W_S^{(4)} (q) = - \frac{g_A^2}{256 \pi m F_\pi^4}  \left[
8 M_\pi^2 + 2 q^2 - g_A^2 \left( 8 M_\pi^2 + \frac{5}{2} q^2 \right) \right]  A^{\tilde \Lambda} (q) \,, \nn
V_{LS}^{(4)} (q) &=& - \frac{3 g_A^4}{64 \pi m F_\pi^4} (2 M_\pi^2 + q^2 ) A^{\tilde \Lambda} (q) \,, \nn
W_{LS}^{(4)} (q) &=& - \frac{g_A^2 ( 1 - g_A^2)}{64 \pi m F_\pi^4} 
(4 M_\pi^2 + q^2 ) A^{\tilde \Lambda} (q) \,.
\eeqa
Notice that these relativistic corrections differ from the ones given in  
ref.~\cite{norb}. In fact, the specific form of the terms in eq.~(\ref{tpe1m})
depends on the form of the Schr\"odinger (or Lippmann--Schwinger) equation,
see \cite{Friar99} for more details. The relativistic corrections given in 
eq.~(\ref{tpe1m}) are consistent with the relativistic Schr\"odinger equation 
(\ref{schroed_rel}) and with the $1/m^2$--corrections to the 1PE potential in eq.~(\ref{opep_full}).

\medskip\noindent
Three--pion exchange starts to contribute at N$^3$LO. The corresponding expressions 
for the spectral functions and the potential (obtained using dimensional regularization)
have been given by Kaiser in \cite{NK31,NK32,NK33}. It has been pointed out in these 
references that the 3PE potential is much weaker than the N$^3$LO 2PE contributions
at physically interesting distances $r > 1$ fm. Having the explicit expressions for the 
3PE spectral functions, it is easy to calculate the potential in the SFR scheme.
It is obvious even without performing the explicit calculations that the finite--range 
part of the 3PE potential in the SFR scheme is strongly suppressed at intermediate 
and short distances compared to the result obtained using DR. This is because the 
short range components which dominate the 3PE spectrum are explicitly excluded 
in this approach. To illustrate that let us consider the isoscalar spin--spin  
contribution proportional to $g_A^4$, which has been found in \cite{NK32} to provide
the strongest 3PE potential for $0.6$ fm $< r <$ $1.4$ fm. The corresponding DR spectral functions 
$\rho_{S, \; 3 \pi}^{\rm DR} (\mu )$ and $\rho_{T, \; 3\pi}^{\rm DR} (\mu )$ are 
given by \cite{NK32}:
\beqa
\label{3pe_spect}
\rho_{S, \;3 \pi}^{\rm DR} (\mu ) &=& - \frac{g_A^4 (\mu - 3 M_\pi )^2}{35 \pi (32 F_\pi^3)^2}
\bigg[ 2 M_\pi^2 - 12 \mu M_\pi - 2 \mu^2 + 15 \frac{M_\pi^3}{\mu} 
+2 \frac{M_\pi^4}{\mu^2} + 3 \frac{M_\pi^5}{\mu^3} \bigg] \,, \nn
 \rho_{T, \; 3 \pi}^{\rm DR} (\mu ) &=& - \frac{g_A^4 (\mu - 3 M_\pi )}{35 \pi (32 \mu F_\pi^3)^2}
\bigg[ \mu^3 + 3 \mu^2 M_\pi + 2 \mu M_\pi^2 + 6 M_\pi^3 + 18 \frac{M_\pi^4}{\mu} 
-9 \frac{M_\pi^5}{\mu^2} -27 \frac{M_\pi^6}{\mu^3} \bigg]\,.
\eeqa
The finite--range part of the spin--spin potential $V_{S} (r)$ can be obtained 
from the corresponding spectral functions via
\beq
V_{S} (r) = - \frac{1}{6 \pi^2 r} \int_{3 M_\pi}^\infty d \mu \, \mu \,
e^{-\mu r} \left( \mu^2 \rho_T (\mu ) - 3 \rho_S (\mu ) \right)\,.
\eeq
Using the spectral functions in eq.~(\ref{3pe_spect}) one then gets
for $V_{S, \; 3\pi}^{\rm DR} (r)$:
\beq
V_{S, \; 3 \pi}^{\rm DR} (r) = \frac{g_A^4}{2 (8 \pi F_\pi^2)^3} \frac{e^{-3 M_\pi r}}{r^7} 
(1 + M_\pi r)^2 (2 + M_\pi r)^2\,.
\eeq
Introducing the cut--off $\tilde \Lambda$ in the spectral functions via
$\rho_{S, \; 3 \pi}^{\rm SFR} (\mu ) = \rho_{S, \; 3 \pi}^{\rm DR} (\mu ) \, \theta 
( \tilde \Lambda - \mu )$ and $\rho_{T, \; 3 \pi}^{\rm SFR} (\mu ) = 
\rho_{T, \; 3 \pi}^{\rm DR} (\mu ) \, \theta ( \tilde \Lambda - \mu )$
we obtain for the SFR potential $V_{S, \; 3\pi}^{\rm SFR} (r)$:
\beqa
V_{S, \; 3\pi}^{\rm SFR} (r) &=& V_{S, \; 3 \pi}^{\rm DR} (r) - \frac{g_A^4}{30 (32 \pi F_\pi^2)^3} 
\frac{e^{-3 y}}{r^7} \bigg[ 120 + 120 y + 60 y^2 + 20 y^3 + 5 y^4 + y^5 \\
&-& {} 25 M_\pi^2 r^2 (6 + 6 y + 3 y^2 + y^3 ) + 45 M_\pi^3 r^3 (2 + 2 y + y^2)
+ 30 M_\pi^4 r^4 (1 + y) - 63 M_\pi^5 r^5 \bigg]\,,
\nonumber  
\eeqa
where we have introduced the abbreviation $y = \tilde \Lambda r$.
In Fig.~\ref{fig:pic1} we plot $V_{S, \; 3 \pi}^{\rm DR} (r)$ and $V_{S, \; 3 \pi}^{\rm SFR} (r)$ 
for $r$ from $0.5$ to $2$ fm. The potential calculated using 
the spectral function regularization is much smaller in magnitude
compared to the one obtained using dimensional regularization. 
Clearly, such a suppression does not take place at very large $r$,
where $V_{S, \; 3 \pi}^{\rm SFR} (r)$ approaches $V_{S, \; 3\pi}^{\rm DR} (r)$. 
At such distances, however, the 3PE potential becomes negligibly small 
compared to the 1PE and 2PE contributions simply due to its shorter range. 
As a consequence, the 3PE potential can be neglected everywhere except 
the region of very small $r$, where it anyhow becomes unreliable. This 
is further exemplified in Fig.~\ref{fig:pic2}, where we show the ratio of the 
N$^3$LO isoscalar spin--spin contributions of 3PE and 2PE
using both regularization schemes for a wide range of $r$.
It turns out that the 3PE contribution reaches for $r > 0.5$ fm at most $2\%-8\%$
of the corresponding N$^3$LO 2PE contribution depending on the 
choice of the spectral function cut--off.
We therefore neglect all 3PE contributions in the present analysis.

\medskip\noindent
Although we have regularized the 2PE contributions by cutting off the large--mass 
components in the spectrum (or, equivalently, by explicitly shifting the 
corresponding short--distance components to contact terms), 
the resulting potential still behaves incorrectly
at large momenta (or equivalently at short distances).
The effective potential is valid for small values 
of the momentum transfer $q$ and becomes meaningless for momenta $q \gtrsim \Lambda_\chi$.
Moreover, since the potential $V$ grows with increasing momenta $q$, 
the scattering equation is ultraviolet divergent and needs to be regularized.
Following the standard procedure, see e.g.~\cite{EGM2},  
we  introduce an additional cut--off in the LS equation by 
multiplying the potential $V (\vec p, \; \vec p \, ')$ with a regulator function 
$f^\Lambda$,
\beq
\label{pot_reg}
V (\vec p, \; \vec p \, ') \rightarrow f^\Lambda ( p ) \, 
V (\vec p, \; \vec p \, ')\, f^\Lambda (p ' )\,.
\eeq 
In what follows, we use the exponential regulator function 
\beq
\label{reg_fun}
f^\Lambda (p ) = \exp [- p^6/\Lambda^6 ]~.
\eeq 
We will specify the values of the cut--offs below.

\medskip\noindent
It should be understood that our treatment of the effective potential is 
based on the heavy baryon formalism. As demonstrated  in \cite{rob00,higa03}, 
heavy baryon expansion becomes formally invalid for certain two--pion exchange contributions 
at very large distances. This problem with the heavy baryon 
formalism has been first observed in the single--nucleon sector and can be dealt with 
using e.g. the Lorentz invariant scheme proposed by Becher and Leutwyler \cite{bech99}. 
It is clear, however, that the NN interaction due to two--pion exchange becomes very weak at 
large distances, so that the problem with the formal inconsistency of the heavy baryon   
approach is expected to have little relevance for practical applications.

\medskip\noindent
Last but not least, we would like to comment on some key features of the SFR 
scheme adopted in the present work. First of all, it is crucial to understand that 
this approach does not affect the ``chiral features'' of the NN potential. 
We remind the reader that the resulting effective NN potential consists of the long--
and short--range pieces. Spontaneously broken approximate chiral symmetry of QCD leads 
to highly non--trivial constraints for the the long--range part of the potential, 
which is given by the terms nonpolynomial in momenta. The short--range part of the 
potential given by a series of the most general contact interactions with increasing 
power of momenta
is not affected by chiral symmetry with the exception of the quark--mass dependence 
of the corresponding LECs, which is not relevant for the present analysis. In other words,
only the long--distance asymptotics of the potential is constrained by chiral symmetry. 
The SFR scheme does, per constraction, not affect the long--distance asymptotics of the 
potential and leads to the same result as obtained using DR, see \cite{EGMs1} for more details. 
The only difference to the DR result is given by a series of the short--range interactions. 
It is, therefore, obvious, that the the SFR method does not affect the constraints of the 
chiral symmetry implemented in the NN potential. Further, we point out that the equivalence of the SFR 
and the finite cut--off regularization has only been established at a one--loop level 
and does not hold true for both loop integrals of the N$^3$LO 2PE contribution. The prominent 
feature of the applied regularization scheme is given by the fact, that it only affects 
the two--nucleon interaction. One can, therefore, directly adopt the values for 
various LECs resulting from the single--nucleon sector analyses, where dimensional 
regularization has been used. On the contrary, if a finite momentum cut--off regularization would 
be applied to both loop integrals entering the N$^3$LO 2PE contribution, 
one would need to re--extract the values of the corresponding LECs from 
pion--nucleon scattering and the process $\pi N \rightarrow \pi \pi N$ 
using the same regularization scheme.

%%%%%%%%%%%%%%%%%%%%%%%%%%%%%%%%%%%%%%%%%%%%%%%%%%%%%%%%%%%%%%%%%%%%%%%%%%%%%%%%%
\subsection{Isospin--breaking effects}
\def\theequation{\arabic{section}.\arabic{equation}}
\label{sec:isosp}

Isospin--breaking nuclear forces have been extensively studied within effective field
theory approaches, see e.g.~\cite{kolck,vKNij,kolck96,Ep99,WME,friar03}, as well as using more
phenomenological methods, see e.g.~\cite{coon96,nis02} for some recent references. 
In  the Standard Model,  isospin--violating effects have their origin 
in both strong (i.e. due to the different masses of the up and down quarks) 
and electromagnetic interactions (due to different charges of the up and down quarks).
The electromagnetic effects can be separated
into the ones due to soft and hard photons. While effects of hard photons 
are incorporated in effective field theory  
by inclusion of electromagnetic short distance  operators in the 
effective Lagrangian, soft photons have to be taken into account explicitly.

\medskip\noindent
In the present analysis we are rather limited in the treatment of isospin--violating interaction, which have
to be included precisely in the way it is done by the Nijmegen group \cite{nijpwa}. This is due to the fact that we 
are using the Nijmegen phase shifts instead of the real data as an input to fit the unknown LECs. 
Let us explain this point in more detail.  
With the only exception of the $^1S_0$ partial wave, the {\it np} isovector phase shifts in the Nijmegen PWA are not 
obtained independently from {\it np} data, but rather extracted from the 
proton--proton ({\it pp}) phase shifts using the assumption 
that the differences in the phase shifts result entirely due 
to isospin--breaking effects associated with $m_p \neq m_n$ and $M_{\pi^\pm} \neq M_{\pi^0}$ in 
the 1PE potential as well as due to electromagnetic interactions.  In order to be consistent with 
the Nijmegen phase shift analysis, we therefore have to neglect various isospin--breaking corrections
and adopt the same isospin--breaking and electromagnetic interactions as in \cite{nijpwa}.
Nevertheless, we have decided to overview the dominant  isospin--breaking contributions 
and to remind the reader on their relative size
following mainly the lines of reference \cite{WME} but extending the consideration 
to higher orders. For a detailed review of charge--symmetry breaking in the nucleon--nucleon interaction 
the reader is referred to \cite{friar03}. 

\medskip\noindent
Consider first isospin breaking in the strong interaction. The QCD quark mass term can be expressed as 
\begin{equation}
\label{eq1}
\mathcal{L}_{\rm mass}^{\rm QCD} = -\frac{1}{2}\bar{q} \, 
(m_{\rm u}+m_{\rm d})(1-\epsilon\tau_{3})\,q~,
\end{equation}
where 
\beq\label{epsdef}
\epsilon \equiv {m_d-m_u \over m_d+m_u} \sim {1 \over 3}~.
\eeq 
The above numerical estimation is based on the light quark mass values
utilizing a modified  $\overline{\rm MS}$ subtraction scheme
at a renormalization scale of 1~GeV. 
The isoscalar term in eq.~(\ref{eq1}) breaks chiral but 
preserves isospin symmetry. It leads to the nonvanishing pion mass, 
$M_\pi^2 = (m_u + m_d ) B \neq 0$, 
where $B$  is a low--energy constant that describes the strength of the  bilinear light quark condensates. 
All chiral symmetry 
breaking interactions in the effective Lagrangian are proportional to positive powers of $M_\pi^2$.
The isovector term ($\propto \tau_3$) in eq.~(\ref{eq1}) breaks isospin symmetry and generates 
a series of isospin--breaking effective interactions $\propto (\epsilon  M_\pi^2)^n$
 with $n \geq 1$.
It therefore appears to be natural to count strong isospin violation in terms of $\epsilon  M_\pi^2$. However, we note already here that isospin-breaking effects are
in general much smaller than indicated by the numerical value of $\epsilon$, because the
relevant scale for the isospin--conserving contributions is  $\Lambda_\chi$ rather than
$m_u+m_d$.

\medskip\noindent
Electromagnetic terms in the effective Lagrangian can be generated using the method of external sources, 
see e.g. \cite{urech95,MS,MM} for more details. All such terms are proportional to the nucleon charge matrix 
$Q_{\rm ch}= e \, (1 + \tau_3 )/2$, where $e$ denotes the electric charge.\footnote{Or equivalently, one can use 
the quark charge matrix $e \, (1/3 + \tau_3 )/2$.}
More precisely, the vertices which contain (do not contain) the photon fields are proportional to $Q^{n}_{\rm ch}$
($Q^{2n}_{\rm ch}$), where $n=1,2,\ldots$. Since we are interested here in nucleon--nucleon scattering in the absence of 
external fields, so that no photon can leave a Feynman diagram, it is convenient to introduce the 
small parameter $e^2 \sim 1/10$ for isospin--violating effects caused by the electromagnetic interactions. 

\medskip\noindent
Due to its perturbative nature induced by the small parameters $\epsilon M_\pi^2$ and $e^2$, we
treat the strong and electromagnetic
isospin violation in addition to the power counting of the isospin symmetric potential 
mentioned in section~\ref{sec:GenRem}. Although not necessary, in practical applications it often appears to be more 
convenient to have a single expansion parameter. Thus, one has to relate the quantities $\epsilon$, 
$e$ to the generic low--momentum scale $Q$ related to external three--momenta of nucleons and the pion mass
($p  \sim p' \sim M_\pi \sim Q$) and introduced before.
Here and below, we
will make use of the following simple counting rules\footnote{This suggests a 
slightly different counting of the strong isospin--breaking effects as compared to \cite{WME}. 
Most of the conclusions of 
\cite{WME} remain, however, unchanged. The important difference is that the leading isospin--violating short--range interaction
is now proportional to the quark mass difference, while electromagnetic contact terms are shifted to higher orders.}:
\beq\label{CountRules1}
\epsilon \sim e \sim \frac{Q}{\Lambda_\chi}\,.
\eeq
The counting of the electric charge is consistent with the one commonly used in the pion and pion--nucleon 
sectors, see e.g. \cite{urech95,fet99,fet00,fet01} 
%*EE
(it differs, however, from what is commonly used
in the description of extremely non-relativistic hadronic bound state, see e.g. \cite{Bern}).
In addition to the above mentioned counting rules, we need to deal with the extra $1/(4 \pi)^2$--factors,
which typically arise when calculating loop integrals. For pion loops, such factors are naturally
incorporated in the chiral power counting through the relation $\Lambda_\chi \sim 4 \pi F_\pi$. 
For photon loops we will further assume, that 
\beq\label{CountRules2}
\frac{e^2}{(4 \pi )^2}  \sim \frac{Q^4}{\Lambda_\chi^4}\,,
\eeq
which simply means that the factors $1/(4 \pi)^2$ provide two additional powers of the small parameter.
In the following, we will denote the order of various isospin--violating interactions by ``$\LOI$'',
``N$\LOI$'', $\ldots$ in order to distinguish the above mentioned phenomenological extension of the counting rules
from the usual chiral power counting in the isospin--conserved case.
Certainly, one has always the option to discard this generalization of the chiral counting rules  
and to perform separate expansions in $\epsilon$, $e$ and $Q/\Lambda_\chi$.
Notice further that eqs.~(\ref{CountRules1}) and (\ref{CountRules2}) suggest  a 
different counting of the strong isospin--breaking effects compared to ref.~\cite{WME}.
In that work strong and electromagnetic effects have been classified using a separate expansion 
without introducing a unified expansion scheme. 
The $\LOI$, N$\LOI$, $\ldots$  contributions in the present work should therefore not be confused with 
the corresponding terms in \cite{WME}. Last but not least, the above counting scheme is similar 
to the one adopted e.g.~in \cite{kolck,kolck96}, where effects $\sim \alpha /\pi$ were also
considered as being one order suppressed compared to the ones $\sim \epsilon M_\pi^2/\Lambda_\chi^2$.

\medskip\noindent
Let us now apply the power counting rules to estimate isospin--violating corrections to hadronic masses,
see also \cite{friar03} for a similar estimation. 
We begin with the pion mass. It is well known that the pion mass does not receive contributions linear in 
the quark mass difference and the strong contribution to the pion mass starts at the second order in $m_d - m_u$.
Consequently, the leading strong term can be estimated as $(\Delta M_\pi^2)_{\rm str}  \equiv 
(M_{\pi^\pm}^2 - M_{\pi^0}^2 )_{\rm str}
\propto (\epsilon M_\pi^2 )^2 \Lambda_\chi^{-2}$, which is of the order $\nu =6$  to be compared to $\nu=2$
for the isospin--symmetric term $\propto M_\pi^2$. One thus expects the strong contribution to the pion mass difference 
$(\Delta M_\pi )_{\rm str} \equiv (M_{\pi^\pm} - M_{\pi^0} )_{\rm str}$ to have 
the size $\sim 0.1 \ldots 0.3$ MeV depending on whether 
one substitutes in the numerical estimation $M_\rho$ or $4 \pi F_\pi$ for $\Lambda_\chi$. 
Clearly, we cannot predict whether the shift is positive or negative.  
The leading electromagnetic contribution to $\Delta M_\pi^2$  is of the order $(\Delta M_\pi^2)_{\rm em}
\sim e^2 (4 \pi)^{-2} \Lambda_\chi^2$,
which is a $\nu = 4$--effect according to our counting rules. Numerically, one estimates the size of 
the pion mass difference to be 
$( \Delta M_\pi )_{\rm em}\sim 1 \ldots 3$ MeV. We see that both power counting arguments and 
numerical estimations suggest that the pion   
mass difference is mainly of electromagnetic origin. 
%This agrees well with the empirical observations, see \cite{don93}
%for unrelated discussion on the size of electromagnetic effects. 
Furthermore, the estimated size of the electromagnetic 
shift agrees well with the observed value $M_{\pi^\pm} - M_{\pi^0} = 4.6$ MeV.
All these statements can also be backed by hard calculations, for a classical review see \cite{GLmass}.
%*EE
For the nucleon mass difference, the strong contribution is linear in the quark mass differences and can be estimated as:
$(\Delta m)_{\rm str} \equiv (m_n - m_p )_{\rm str} \sim (\epsilon M_\pi^2)  \Lambda_\chi^{-1} \sim 6 \ldots 9$ MeV.\footnote{This too large value reflects our earlier statement about the use
of the parameter $\epsilon$ to estimate isospin-breaking corrections.} 
According to the counting rules, this is the $\nu = 3$--effect. Electromagnetic shift appears at $\nu = 4$ and 
is expected to be of the order $(\Delta m)_{\rm em}
\sim e^2 (4 \pi)^{-2} \Lambda_\chi \sim 0.5 \ldots 0.7$ MeV. In reality, the effects are of opposite sign and the 
difference between them is less pronounced.  
One observes $(m_n-m_p)_{\rm str} \simeq 2.1\,$MeV
and $(m_n-m_p)_{\rm em} \simeq - 0.8\,$MeV, leading to the physical value of
$m_n - m_p = 1.3\,$MeV (see again \cite{GLmass} for more details).

\medskip\noindent
We are now in the position to discuss various isospin--breaking contributions to the two--nucleon force. 
As explained in \cite{WME}, the leading--order (i.e. $\LOI$) isospin--breaking interactions are due to the pion
mass difference in the 1PE potential and the static Coulomb interaction. The latter is clearly of the order\footnote{The
factor  $F_\pi^2$ results from the common normalization of the isospin--symmetric part of the 
two--nucleon potential adopted in this work. 
This factor can be understood e.g. from looking at the 1PE potential in eq.~(\ref{opep}).} 
$\sim e^2 Q^{-2} F_\pi^2$, while the former is 
\beq 
\sim \frac{\Delta M_\pi^2}{M_\pi^{2}} \sim 
\left( \frac{e^2}{(4 \pi)^{2}} \Lambda_\chi^2 \right) \frac{1}{M_\pi^{2}}  = \mathcal{O} 
\left[ \frac{Q^2}{\Lambda_\chi^2} \right] \,,
\eeq
where we used of 
the counting rules (\ref{CountRules1}) and (\ref{CountRules2}) together with  $M_\pi \sim Q$.
Thus, the $\LOI$ isospin--breaking force is of the order $\nu = 2$.
Consider now N$\LOI$ corrections to this result, which appear at $\nu = 3$. 
The pion--nucleon coupling constant 
receives strong isospin--violating contributions of the order $\epsilon M_\pi^2 / \Lambda_\chi^2$.
The corresponding LECs in the pion--nucleon Lagrangian are denoted by $d_{17}$, $d_{18}$ and $d_{19}$ in 
the notation of Ref.~\cite{Fet98}. This charge dependence of the pion--nucleon coupling constant
leads to isospin--violating 1PE of the order $\nu = 3$. In addition, one has to take into account 
strong isospin--breaking contact interaction of the kind
\beq
\label{cont_str}
\epsilon M_\pi^2\, (N^\dagger  \tau_3 N ) \, (N^\dagger N)\,,
\eeq
which leads to charge symmetry breaking. 
We can  check the accuracy of our estimation  numerically using the values for the LECs found in \cite{WME}.
According to eq.~(\ref{cont_str}), we expect the ratio of the isospin--breaking terms to isospin--conserving ones
to be typically of the size: $\epsilon M_\pi^2 /\Lambda_\chi^2 \sim 0.5\% \ldots 1.1\%$, where the uncertainty results again from 
using two different estimations for $\Lambda_\chi$. Picking up the numbers from Table 2 in \cite{WME} we find for 
this ratio the values $0.8\%$, $0.8\%$ and $3.1\%$ for three different 
values of the (sharp) cut--off $\Lambda$ in the Lippmann--Schwinger equation: $\Lambda=300$, $\Lambda=400$ and $\Lambda=500$ MeV.
Thus, our numerical estimation is consistent with the results of \cite{WME}.
The NN$\LOI$ corrections are of the order $\nu = 4$ and arise from various sources. 
First, one has to take into account isospin--breaking in the 2PE potential due to electromagnetic 
corrections to the pion--nucleon coupling, see e.g.\cite{kolck96}. 
The correction due to the pion mass difference in the leading 2PE potential can be estimated as:
\beq
\sim   \frac{\Delta M_\pi^2 }{M_\pi^{2}} \, \frac{Q^2}{ \Lambda_\chi^{2}}   = \mathcal{O} 
\left[ \frac{Q^4}{\Lambda_\chi^4} \right] \,.
\eeq
Another isospin--violating  two--pion exchange interaction at this order is generated
by the triangle and football diagrams with one insertion of the isospin--breaking 
$\pi \pi NN$ vertex with the LEC $c_5$ (using again the notation of \cite{Fet98}).
This vertex is proportional to $\epsilon M_\pi^2$ and is thus formally of the lower order than 
the electromagnetic $\pi \pi NN$ vertices $\propto e^2/(4 \pi)^2$. 
As we have seen on an example of the nucleon mass difference, in practice, both effects might be 
of a comparable size. For a recent work on this kind of isospin--breaking forces see \cite{nis02}.
The remaining contributions are given by the 
static $\pi \gamma$--exchange of the order $\sim  e^2/(4 \pi )^2$ and by two independent contact interactions  
\beq
\label{cont_em}
\frac{e^2}{(4 \pi )^2} \, (N^\dagger  \tau_3 N )\, (N^\dagger  \tau_3 N) \quad \quad \mbox{and}
\quad \quad  \frac{e^2}{(4 \pi )^2} \, (N^\dagger  \tau_3 N )\, (N^\dagger N)\,
\eeq
which lead to both charge independence and charge symmetry breaking.
Notice that, in practice, the effect of the second interaction cannot be disentangled from the effect of 
term in eq.~(\ref{cont_str}).

\medskip\noindent
Let us now estimate the size of isospin violation in the NN scattering due to the nucleon mass difference.
We first note that the first relativistic corrections to the isospin--symmetric part of the two--nucleon force 
appear at at N$^3$LO ($\nu = 4$) and are given by $1/m^2$--corrections to the 1PE and 
$1/m$--corrections to the leading 2PE potential.
Consequently, the size of the corresponding isospin--violating terms can be estimated as
\beq
\frac{\Delta m}{m} \, \frac{Q^4}{\Lambda_\chi^4} \sim \frac{\epsilon M_\pi^2}{m \Lambda_\chi} \, \frac{Q^4}{\Lambda_\chi^4} 
= \mathcal{O} \left[ \frac{Q^8}{\Lambda_\chi^8} \right] \,,
\eeq
Such terms therefore contribute only at the order $\nu=8$. In addition to the above mentioned corrections, 
one has to account for the fact that the neutron--proton mass difference leads to energy shifts 
of virtual states when calculating two--pion exchange diagrams. This can also easily been understood in 
the language of the heavy baryon formalism: factoring out the exponential factor $\exp ( i m_p v \cdot x)$
from the proton and neutron fields, where $v$ and $x$ denote the proton velocity and position, the neutron 
propagator receives a shift in the denominator $\propto (m_n - m_p)$  after integrating out the small field components.
It is then easy to see that the isospin--violating 2PE is suppressed against its isospin--conserving part
by a factor:
\beq
\sim \frac{\Delta m}{Q} \sim \frac{\epsilon M_\pi^2}{\Lambda_\chi Q} = \mathcal{O} \left[\frac{Q^2}{\Lambda_\chi^2} \right]\,. 
\eeq
Therefore, neutron--proton mass difference in 2PE starts to contribute at $\nu = 4$. 
This sort of charge symmetry breaking corrections has been studied recently in \cite{coon96,friar03}.
Notice further that, as pointed out in \cite{friar03}, certain loop integrals in the 2PE contributions 
give only one power of $(4 \pi)$ instead of expected two powers and are, therefore,  enhanced.   
We will not take this enhancement  into account in the present work. 
Apart from the above mentioned corrections to the nucleon-nucleon force, the
neutron--proton mass difference 
has to be taken into account in kinematical relations as discussed in appendix \ref{sec:kinem}, as well as 
in the expression for the kinetic energy of the nucleons. Let us consider this last effect. Its contribution 
to the scattering amplitude can be estimated by looking at the Lippmann--Schwinger equation
\beq
T = V + V \, G_0 \, T\,,
\eeq
where $G_0$ refers to the free propagator of two nucleons. Both terms on the right-hand-side of the above 
equation are of the same order $\nu = 0$. Taking into account the nucleon mass difference in $G_0$ leads 
therefore to a correction to the T--matrix of the order 
\beq
\sim  \frac{\Delta m}{m} \sim \frac{\epsilon M_\pi^2}{m \Lambda_\chi}  = \mathcal{O} 
\left[ \frac{Q^4}{\Lambda_\chi^4} \right] \,,
\eeq
and thus contribute at $\nu = 4$. Notice that this estimation is valid for both relativistic 
and nonrelativistic expressions for the two--nucleon propagator $G_0$.

\medskip\noindent
All other isospin--violating corrections are suppressed by further powers of the small parameter.
We would like, however, to point out an important limitation of our estimation due to the fact
that we do not explicitly account for the long--range nature of electromagnetic forces. Consider, for example,
the leading one--pion and one--photon exchange forces. For simplicity, we will restrict ourselves to the 
$^1S_0$ proton--proton channel, where the 1PE potential takes the form:
\beq
V_{1 \pi}^{(0)} (q) = \biggl(\frac{g_A}{2F_\pi}\biggr)^2 \, \frac{q^2}
{q^2 + M_\pi^2}~.
\eeq
The static Coulomb interaction, 
\beq
\label{coul}
V_{\rm Coulomb} (q) = \frac{e^2}{q^2}\,,
\eeq
is suppressed compared to $V_{1 \pi}^{(0)} (q)$ by two powers of the small 
parameter $Q/\Lambda_\chi$ according to the power counting.
Such an estimation works fairly well for momenta $q$ of the order  $q \sim M_\pi$,
for which we get $V_{1 \pi}^{(0)} (q = M_\pi) \sim  23$ GeV$^{-2}$ and 
$V_{\rm Coulomb} (q= M_\pi) \sim 5$ GeV$^{-2}$. The power counting, however, breaks down for 
small momenta $q \ll M_\pi$ due to the long--range nature of the Coulomb interaction. 
For example, for $q = M_\pi/4$ one gets: $V_{1 \pi}^{(0)} (q = M_\pi/4) \sim  3$ GeV$^{-2}$ while 
$V_{\rm Coulomb} (q= M_\pi/4) \sim 82$ GeV$^{-2}$. Consequently, the Coulomb interaction provides the 
dominant contribution to the potential for small momenta and requires 
a nonperturbative treatment at low energy.
A possible way out of the above mentioned inconsistency 
would be to develop separate and systematic power counting for momenta 
much smaller than the pion mass. This is, however, beyond the scope of the present work.
Notice that a similar idea with two different power counting regimes has been applied recently to 
the nucleon Compton scattering in order to extend the region of applicability of the effective field theory in the 
$\Delta$--region \cite{pasc03}. In the present analysis, we will simply take 
into account higher--order corrections to the 
long--range electromagnetic interactions when determining the values of the LECs 
in order to correct for the low--momentum behavior of the NN potential.
The first long--range corrections beyond the ones considered above result from two--photon exchange, whose size can be estimated  
as 
\beq
\sim \frac{e^2}{Q^2} \frac{e^2}{(4 \pi)^2} F_\pi^2 = \mathcal{O} 
\left[ \frac{Q^6}{\Lambda_\chi^6} \right]\,.  
\eeq
It thus formally appears at the order $\nu = 6$. 
In addition, at the same order  $\nu = 6$ one has to take into account relativistic $1/m$ corrections to the 
static one--photon exchange, which provide a contribution of the following size:
\beq
\sim \frac{e^2}{Q^2} \, F_\pi^2 \, \frac{Q^2}{m^2}  = \frac{e^2}{m^2} \, F_\pi^2 = \mathcal{O} 
\left[ \frac{Q^6}{\Lambda_\chi^6} \right]\,.
\eeq
The relative sizes of various isospin--breaking contributions discussed above
are summarized in Table~\ref{tab:isosp}.
In what follows, we will give explicit expressions for the above mentioned interactions.

%%%%%%%%%%%%%%%%%%%%%%%%%%%%%%%%%%%%%%%%%%%%%%%%%%%%%%%%%%%%%%%%%%%%%%%%%%%%%%%%%
\subsubsection{Finite--range isospin--breaking forces}
\def\theequation{\arabic{section}.\arabic{equation}}
\label{sec:isospFR}

Let us now give the explicit expressions for the finite--range isospin--violating interactions 
up to N$\LOI$. 
The dominant $\nu = 2$ contribution ($\LOI$) due to $M_{\pi^\pm} \neq M_{\pi^0}$ can be 
taken into account by replacing the isospin--conserving expression 
%*EE
$V_{1\pi} (q)$ in eq.~(\ref{opep_full}) by: 
\beqa
\label{ope}
V_{1\pi, \; pp} (q) &=& V_{1\pi, \; nn}  (q)= -\biggl(\frac{g_A}{2F_\pi}\biggr)^2 \, \vec{\tau}_1 \cdot
\vec{\tau}_2 \, \frac{\vec{\sigma}_1 \cdot\vec{q}\,\vec{\sigma}_2\cdot\vec{q}}
{q^2 + M_{\pi^0}^2} \, \left( 1 - \frac{p^2 + p'{}^2}{2 m^2} \right) ~, \nn
V_{1\pi, \; np, \; {T=1}} (q) &=& -\biggl(\frac{g_A}{2F_\pi}\biggr)^2 \, \vec{\tau}_1 \cdot
\vec{\tau}_2 \, \vec{\sigma}_1 \cdot\vec{q}\,\vec{\sigma}_2\cdot\vec{q}
\left( \frac{2}{q^2 + M_{\pi^\pm}^2} - \frac{1}{q^2 + M_{\pi^0}^2} \right)
\, \left( 1 - \frac{p^2 + p'{}^2}{2 m^2} \right)\,, \nn
V_{1\pi, \; np, \; {T=0}} (q) &=& -\biggl(\frac{g_A}{2F_\pi}\biggr)^2 \, \vec{\tau}_1 \cdot
\vec{\tau}_2 \, \frac{\vec{\sigma}_1 \cdot\vec{q}\,\vec{\sigma}_2\cdot\vec{q}}{q^2 + M_{\pi}^2} 
\, \left( 1 - \frac{p^2 + p'{}^2}{2 m^2} \right)\,,
\eeqa
where $T$ denotes the total isospin, $M_{\pi^\pm}$ and $M_{\pi^0}$ are the masses of the charged and neutral pions, respectively,
and 
\beq\label{avmass}
\Mp = \frac{2}{3} \Mpp +  \frac{1}{3} \Mpn = 138.03~{\rm MeV}~.
\eeq
The 1PE potential gets further charge independence and charge symmetry breaking 
contributions at N$\LOI$ and NN$\LOI$ due to isospin violating 
pion--nucleon couplings. The final expression for the 1PE potential is then of the kind: 
$\eta_i V_{1 \pi}^{(0)}$, where 
$\eta_i$ are the channel--dependent constants: $\eta_{nn} \neq \eta_{np} \neq  \eta_{pp}$.
Unfortunately, the actual size of isospin--violating corrections to the pion--nucleon coupling is
not well determined at presence \cite{kolck96}.

%Furtehr corrections result from pion mass difference in the NLO part of the two--pion exchange 
%potential \cite{friar99}, $\pi \gamma$--exchange \cite{bira98} as well as two charge--dependent 
%contact interactions without derivatives. The letters only contribute to S--waves and enable 
%an accurate fit of all three $^1S_0$ scattering length $a_{pp}$,  $a_{np}$, and  $a_{nn}$.
%In the present work we however have to take into account only this short--range terms and to omite 
%the two remaining subleading contributions in order to be consistent with the results of Nijmegen partial 
%wave analysis which are used in our fits instead of real data. 
%Indeed, taking into account e.g. isospin violation in the two--pion exchange potential would lead to charge 
%independence breaking in the {\it np} and {\it pp} triplet P--wave phase shifts (which comes out parameter--free
%at this order) not consistent with the Nijmegen PWA. For the sake of completeness, we will nevertheless 
%briefly discuss the above mentioned corrections. 
\medskip\noindent
The pion mass difference
in the 2PE can be incorporated as outlined in ref.\cite{FvK}.
It is most convenient to consider the isoscalar and isovector 2PE piece
separately,
\beq
V_{2\pi} = V_{2\pi}^0 + V_{2\pi}^1 \, \vec{\tau}_1 \cdot
\vec{\tau}_2~.
\eeq
The isoscalar part $V_{2\pi}^0$ can be expressed as \cite{FvK} 
\beq
 V_{2\pi}^0 = \frac{2}{3} V_{2\pi}^0 (\Mpp, \Mpp ) +
 \frac{1}{3} V_{2\pi}^0 (\Mpn, \Mpn ) = V_{2\pi}^0 (\Mp, \Mp )   + {\cal O}\left[ \left(\frac{\Mpp - \Mpn}{\Mpp}
\right)^2 \right]~,
\eeq
where the arguments of $V_{2\pi}^0$ denote the masses of exchanged pions.
%*EE
% and 
%\beq\label{avmass}
%\Mp = \frac{2}{3} \Mpp +  \frac{1}{3} \Mpn = 138.03~{\rm MeV}~.
%\eeq
For the isovector 2PE $V_{2\pi}^1$, one has the general structure
\beqa\label{TPE1}
V_{2\pi}^1 &=& \tau_1^3 \, \tau_2^3 \, V_{2\pi}^1 (\Mpp, \Mpp )
+ (\vec{\tau}_1 \cdot \vec{\tau}_2 -\tau_1^3 \, \tau_2^3 ) \,
V_{2\pi}^1 (\Mpp, \Mpn )  \nonumber \\
&=& \left\{ \begin{array}{ll}
V_{2\pi}^1 (\Mpp , \Mpp ) & {\rm for}~~pp~~{\rm and}~~nn~, \\
2V_{2\pi}^1 (\Mpp , \Mpn ) - V_{2\pi}^1 (\Mpp , \Mpp ) \sim V_{2\pi}^1 (\Mpn , \Mpn ) 
& {\rm for}~~np, T = 1~. \end{array}\right. \quad 
\eeqa
The result in the last line of the above equation is valid modulo 
$(( \Mpp - \Mpn )/ \Mpp)^2$--corrections.
For the $T=0$ case the 2PE potential reads  $V_{2 \pi} (M_\pi, \, M_\pi )$.

\medskip\noindent
The $\pi\gamma$ exchange diagrams have been calculated in
ref.\cite{vKNij} and we give below the results obtained in that paper
omitting all computational details. Due to isospin, only charged pion
exchange can contribute to the $\pi\gamma$ potential $V_{\pi\gamma}$
and thus it only affects the $np$ system. The potential has the form
\beqa\label{Vpiga}
V_{\pi\gamma} (\vec{q} \, ) &=& -{g_A^2 \over 4F_\pi^2 \Mppz} \,
(\vec{\tau}_1 \cdot \vec{\tau}_2 -\tau_1^3 \, \tau_2^3 ) \,
\vec{\sigma}_1 \cdot \vec{q} \, \vec{\sigma}_2 \cdot \vec{q} \,\, V_{\pi\gamma}
(\beta)~, \nonumber \\
V_{\pi\gamma}(\beta ) &=& {\alpha \over \pi} \biggl[ -{ (1-\beta^2)^2
  \over 2\beta^4 (1+\beta^2) } \ln(1+\beta^2) + {1\over 2\beta^2} 
- {2\bar{\gamma} \over 1+\beta^2} \biggr]~.
\eeqa 
Here, $\beta = |\vec{q} \,|/\Mpp$ and $\bar{\gamma}$ is a
regularization scheme dependent constant. 
The analytical form of
$V_{\pi\gamma}$ is similar to the one of the 1PE potential,  but it differs in
strength by the factor $\alpha/\pi \simeq 1/400$. 

\medskip\noindent
Finally, the expressions for the remaining isospin--violating 2PE contributions at NN$\LOI$
have been given in \cite{friar03}.

\subsubsection{Long--range (soft) isospin--breaking forces}
\def\theequation{\arabic{section}.\arabic{equation}}
\label{sec:isospIR}

We now discuss long--range isospin--breaking interactions
which are often referred to in the literature as 
``electromagnetic forces''.
The static Coulomb force in eq.~(\ref{coul})  does certainly not  completely represent 
the electromagnetic interaction between two nucleons but only its leading 
contribution. The first long--range corrections to the static Coulomb force are either suppressed
by $m^{-2}$ (relativistic corrections to the static one--photon exchange)
or by an additional power of the fine--structure constant $\alpha$ (two--photon exchange).
Although all these effects are formally of  higher order,
we nevertheless prefer to take them into account explicitly for the following reasons.
First of all, the effects of these interactions are magnified at low energy due to their long--range nature.
Further, as explained above, in our analysis we have to take into account isospin--breaking effects in the same
way as it is done in  \cite{nijpwa}.
The electromagnetic interaction for the {\it pp} and {\it np} case is given by
\beqa
\label{vc1vc2}
V_{\rm EM} (pp) &=& V_{\rm C1} +  V_{\rm C2} + V_{\rm VP} + V_{\rm MM} (pp)\,, \nn
V_{\rm EM} (np) &=& V_{\rm MM} (np)\,, \nn
V_{\rm EM} (nn) &=& V_{\rm MM} (nn)\,,
\eeqa
where $V_{\rm C1}$ and $V_{\rm C2}$ are usually referred to as ``improved Coulomb potential''.
They take into account  the  relativistic $1/m^2$--corrections to the static 
Coulomb potential and include contributions of the two--photon--exchange diagrams \cite{austin}.
The explicit coordinate--space expressions read: 
\beqa
\label{modCoul}
V_{\rm C1} &=& \frac{\alpha '}{r} \,, \nn
V_{\rm C2} &=& - \frac{1}{2 m_p^2} \left[ (\Delta + k^2 ) \frac{\alpha}{r} + 
\frac{\alpha}{r} (\Delta + k^2 ) \right] \sim - \frac{\alpha \alpha '}{m_p r^2}\,,
\eeqa
where $\Delta$ denotes the Laplacian.
The energy--dependent constant 
$\alpha '$ is given by 
\beq
\alpha ' = \alpha 
\frac{m_p^2 + 2 k^2}{m_p \sqrt{m_p^2 + k^2}}\,.
\eeq
Here $k$ is the c.m.s.~scattering momentum. 
The term $V_{\rm C2}$ is chosen in such a way that it leads to an exact cancellation between the proper 
two--photon and the iterated one--photon exchange, see \cite{austin} for more details. 
The approximation made in the second line of eq.~(\ref{modCoul}) is based upon using Coulomb distorted--wave
Born approximation (CDWBA), see \cite{Berg88} for more details. The modified
Coulomb potential $V_{\rm C1}$ in eq.~(\ref{modCoul}) can be treated in momentum space in the 
same way as the usual static Coulomb potential as described  
in appendix \ref{sec:coul}.\footnote{Clearly, one has to use the appropriately adjusted regular and 
irregular Coulomb functions $F_l (r)$ and $G_l (r)$.} 
The magnetic moment interaction $V_{\rm MM}$ in eq.~(\ref{vc1vc2}) is given by \cite{stoks90}:
\beqa
\label{vmm}
V_{\rm MM} (pp) &=& - \frac{\alpha}{4 m_p^2 r^3} \left[ \mu_p^2 S_{12} + (6 + 8 \kappa_p ) 
   \vec L \cdot \vec S \right]\,, \nn
V_{\rm MM} (np) &=& - \frac{\alpha \kappa_n}{2 m_n r^3} \left[ \frac{\mu_p}{2 m_p}  S_{12} + \frac{1}{m}
   (\vec L \cdot \vec S  + \vec L \cdot \vec A ) \right] \,, \nn
V_{\rm MM} (nn) &=& - \frac{\alpha \mu_n^2}{4 m_n^2 r^3}  S_{12} \,, 
\eeqa
where $\mu_p=2.793$ and  $\mu_n=-1.913$ are the proton and neutron magnetic moments and 
$\kappa_p= \mu_p -1$, $\kappa_n= \mu_n$ their anomalous magnetic moments. Further, 
$\vec L$ is the orbital angular momentum, $\vec A = ( \vec \sigma_1 - \vec \sigma_2 )/2$  
and $S_{12} = ( \vec \sigma_1 \cdot \vec r ) 
( \vec \sigma_2 \cdot \vec r )/r^2 - \vec \sigma_1 \cdot \vec \sigma_2/3$.
The corresponding expressions in momentum space can be found e.g. in \cite{stoksPhD}.
Finally, the vacuum polarization potential $V_{\rm VP}$ derived by Ueling \cite{ueling}, 
see also \cite{durand}, reads:
\beq
\label{vp}
V_{\rm VP} = \frac{2 \alpha}{3 \pi} \frac{\alpha '}{r} \int_1^\infty dx \, 
e^{-2 m_e r x} \left( 1 + \frac{1}{2 x^2} \right) \frac{(x^2 -1)^{1/2}}{x^2}\,,
\eeq
where $m_e$ is the electron mass. Clearly, the vacuum polarization potential is not of an infinitely long range. 
Its range is governed by the electron mass, which is still tiny compared to the relevant mass 
scales in the nucleon--nucleon problem. This is similar to the treatment of vacuum polarization
in EFT approaches for hadronic bound states, see e.g. \cite{ES}.

\medskip\noindent
It is important to realize that the expressions (\ref{vc1vc2}) refer 
to point--like nucleons and only define the long--distance 
asymptotics of the corresponding electromagnetic interaction. The short--distance structure is more 
complicated and not shown explicitly. In particular, we do not include zero--range (for point--like nucleons) 
terms as well as electromagnetic  form factors which can, in principle, be calculated consistently in EFT.
Such short--range terms with the nucleon form factors of a dipole form are, for example, 
included in the Argonne V18 potential.
Last but not least, we note that the above consideration of the electromagnetic effects is based on the 
``nonrelativistic'' Schr\"odinger equation (\ref{Schroed_nonrel}), which will be defined in the next section.
To close this section let us point out some well known  
practical complications which arise due to the presence of the long--range electromagnetic forces.
\begin{itemize}
\item
Asymptotic states are affected by electromagnetic interactions. The S-matrix has to be formulated in terms of 
asymptotic Coulomb states.
\item
The formally suppressed (as compared to the strong nuclear force) electromagnetic interactions are enhanced at low energy. 
The Coulomb interaction requires a nonperturbative treatment. Even the effects due to magnetic moment interaction 
might be large for certain observables under specific  kinematical conditions. For example, 
in the  {\it np} system, it gives rise to a forward--angle dip structure for the 
analyzing power.
\item
The expansion of the scattering amplitude in partial waves converges very slowly in the presence 
of magnetic moment interactions. 
\end{itemize}

%%%%%%%%%%%%%%%%%%%%%%%%%%%%%%%%%%%%%%%%%%%%%%%%%%%%%%%%%%%%%%%%%%%%%%%%%%%%%%%%%
\section{Scattering equations}
\def\theequation{\arabic{section}.\arabic{equation}}
\setcounter{equation}{0}
\label{sec:LS}

We start with the relativistic Schr\"odinger equation (\ref{schroed_rel})
and assume the potential to be of a finite range. The treatment of the nucleon--nucleon scattering  
problem in the presence of the long--range Coulomb interaction is relegated to
appendix~\ref{sec:coul}.
The scattering states are described by the Lippmann--Schwinger equation
corresponding to the Schr\"odinger equation (\ref{schroed_rel}).
The LS equation (for the $T$--matrix) projected onto states with orbital angular momentum
$l$, total spin $s$ and total angular momentum $j$ is 
\beq\label{LSeq}
T^{sj}_{ll'} (p,p') = V^{sj}_{ll'} (p,p') + \sum_{l''} \,
\int_0^\infty \frac{dp'' \, {p''}^2}{(2 \pi )^3} \,  V^{sj}_{ll''} (p,p'')
\frac{1}{2 \sqrt{{p'}\,^2+m^2}- 2\sqrt{{p''}^2 + m^2} +i\eta} T^{sj}_{l''l'} (p'',p')~,
\eeq
with $\eta \to 0^+$.
In the uncoupled case, $l$ is conserved. The partial wave projected
potential $V^{sj}_{l',l} (p',p)$ can be obtained using the formulae collected in appendix~\ref{sec:pw}.
The relation between the  $S$-- and on--the--energy shell $T$--matrix is given by 
\beq
\label{Sdef}
S_{l l'}^{s j} (p) = \delta_{l l'} - \frac{i}{8 \pi^2} 
\, p \, \sqrt{p^2 + m^2} \,  T_{l l'}^{s j} (p)~.
\eeq
The phase shifts in the uncoupled cases can be obtained from the
$S$--matrix via
\beq
S_{jj}^{0j} = e^{ 2 i \delta_{j}^{0j} } \; , \quad 
S_{jj}^{1j} = e^{  2 i \delta_{j}^{1j} } \;,
\eeq
where we have used the notation $\delta^{sj}_l$.
Throughout, we use the so--called Stapp parametrization~\cite{stapp}
of the $S$--matrix in the coupled channels ($j>0$):
\beqa
S &=& \left( \begin{array}{cc} S_{j-1 \, j-1}^{1j} &  S_{j-1 \, j+1}^{1j} \\
S_{j+1 \, j-1}^{1j} &  S_{j+1 \, j+1}^{1j} \end{array} \right) \nn
&=& 
\left( \begin{array}{cc} \cos{(2 \epsilon)} \exp{(2 i \delta^{1j}_{j-1})} &
i \sin{(2 \epsilon)} \exp{(i \delta^{1j}_{j-1} +i \delta^{1j}_{j+1})} \\
i \sin{(2 \epsilon)} \exp{(i \delta^{1j}_{j-1} +i \delta^{1j}_{j+1})} &
\cos{(2 \epsilon)} \exp{(2 i \delta^{1j}_{j+1})} \end{array} \right)~.
\eeqa
For the discussion of the effective range expansion for the $^3S_1$ partial wave
we will use the different parametrization of the $S$--matrix, namely the one due to 
Blatt and Biedenharn \cite{Bl52}. The connection between these two sets of parameter 
is given by the following equations:
\beqa
\label{blattb}
\delta_{j-1} + \delta_{j+1} &=&  \hat{\delta}_{j-1} + \hat{\delta}_{j+1}\;, \nn
\sin ( \delta_{j-1} - \delta_{j+1} ) &=& \frac{\tan ( 2 \epsilon)}{\tan (2 \hat{\epsilon})}\;, \nn
\sin (\hat{\delta}_{j-1} - \hat{\delta}_{j+1}) &=& \frac{\sin ( 2 \epsilon)}{\sin (2 \hat{\epsilon})}\;,
\eeqa
where $\hat{\delta}$ and $\hat{\epsilon}$ denote the quantities in the 
Blatt--Biedenharn parametrization 
and we have omitted the superscripts for $\delta$'s.

\medskip\noindent
To close this section we would like to remind the reader that the Schr\"odinger 
and Lippmann--Schwinger equations (\ref{schroed_rel}) and (\ref{LSeq})
may be cast into a nonrelativistic form. 
One way to do that is using the Kamada--Gl\"ockle transformation \cite{KGT},
which relates the relativistic and nonrelativistic c.m.s.~momenta $\vec p$ and $\vec q$
via:
\beq
\label{momenta}
T_{\rm kin} = 2 \sqrt{p^2 + m^2} -2 m  =  \frac{q^2}{m}\,.
\eeq
The potential $\tilde V$ to be used in the nonrelativistic 
Schr\"odinger equation
\beq
\label{schr_nr1}
\left[ \frac{q^2}{m} + \tilde V \right] \phi
= E \phi\,,
\eeq
is defined in the partial--wave projected representation as 
\beq
\label{Vnonrel_true}
\tilde V^{sj}_{ll'} (q, \,q ') = \sqrt{\left(1+\frac{q^2}{2 m^2} \right)
\sqrt{1+\frac{q^2}{4 m^2}}} \;V^{sj}_{ll'}  \left( \sqrt{q^2+\frac{q^4}{4 m^2}},
\sqrt{q' \,^2+\frac{q ' \,^4}{4 m^2}} \right) \sqrt{\left(1+\frac{q ' \,^2}{2 m^2} \right)
\sqrt{1+\frac{q ' \, ^2}{4 m^2}}}\,,
\eeq
where $V^{sj}_{ll'} (p, \, p' )$ is the potential entering the relativistic Schr\"odinger 
equation (\ref{schroed_rel}). The wave--function $\phi$ is related to $\Psi$ in eq.~(\ref{schroed_rel})
via
\beq
\label{phipsi}
\phi ( q) = \sqrt{\left(1+\frac{q^2}{2 m^2} \right)
\sqrt{1+\frac{q^2}{4 m^2}}} \; \Psi\left( \sqrt{q^2+\frac{q^4}{4 m^2}} \right)\,.
\eeq
The S--matrix is defined via
\beq
\label{Sdef1}
\tilde 
S_{l l'}^{s j} (q) = \delta_{l l'} - \frac{i}{8 \pi^2} 
\, q \,  m \,  \tilde T_{l l'}^{s j} (q)~,
\eeq
where the T--matrix $\tilde T_{l l'}^{s j}$ satisfies the usual nonrelativistic Lippmann--Schwinger 
equation
\beq
\label{LSeq_nonrel}
\tilde T^{sj}_{ll'} (q,q') = \tilde V^{sj}_{ll'} (q,q') + \sum_{l''} \,
\int_0^\infty \frac{dq'' \, {q''}^2}{(2 \pi )^3} \,  \tilde V^{sj}_{ll''} (q,q'')
\frac{m}{q' \, ^2- {q''}\,^2 +i\eta} \tilde T^{sj}_{l''l'} (q'',q')~.
\eeq
It can be demonstrated \cite{KGT} that the S--matrix $ \tilde S^{sj}_{ll'}$
equals for any given energy the S--matrix $S^{sj}_{ll'}$ defined in eq.~(\ref{Sdef}),
that is $ \tilde S^{sj}_{ll'} (q) = S^{sj}_{ll'} (p)$. 
Another commonly used way to cast the relativistic Schr\"odinger equation (\ref{schroed_rel})
into a nonrelativistic--like form is based upon the algebraic manipulations with this 
equation, see  \cite{Friar99}.  
More precisely, adding $2 m$ to both sides in eq.~(\ref{schroed_rel}) with subsequent squaring them, 
subtracting $4 m^2$ and dividing both sides by $4 m$ leads to
\beq\label{Schroed_nonrel}
\left[ \frac{p^2}{m} + \bar V \right] \Psi
= \frac{k^2}{m} \Psi\,,
\eeq
where the momentum $k$ is related to the energy $E$ in eq.~(\ref{schroed_rel}) via 
\beq
\label{rel_kin}
E = 2 \sqrt{k^2 + m^2} - 2 m\,,
\eeq 
and the potential operator $\bar V$ is given by 
\beq
\label{pot_nonrel_symb}
\bar V = \left\{ \frac{\sqrt{p^2 + m^2}}{2 m} , \; V \right\} + \frac{V^2}{4 m}\,,
\eeq
or, in the partial--wave projected basis, by
\beq
\label{pot_nonrel}
\bar V^{sj}_{ll'} (p,p') = \left( \frac{\sqrt{p^2 + m^2}}{2 m}+\frac{\sqrt{p'\,^2 + m^2}}{2 m} \right)
V^{sj}_{ll'} (p,p')
+\frac{1}{4 m} \sum_{l''} \,
\int_0^\infty \frac{dp'' \, {p''}^2}{(2 \pi )^3} \,  V^{sj}_{ll''} (p,p'')
V^{sj}_{l''l'} (p'',p')\,.
\eeq
The curly bracket in eq.~(\ref{pot_nonrel_symb}) denote an anticommutator.
Notice that contrary to the previously described approach, the ``nonrelativistic''
Schr\"odinger equation (\ref{Schroed_nonrel}) still requires relativistic kinematics 
in relating the energy and momentum, see eq.~(\ref{rel_kin}).
The S-- and T--matrices $\bar S_{l l'}^{s j}$ and $\bar T_{l l'}^{s j}$ are defined via 
eqs.(\ref{Sdef1}) and (\ref{LSeq_nonrel}), respectively (with $\tilde S$, $\tilde T$, $q$, $q '$  being replaced 
by $\bar S$, $\bar T$, $p$, $p'$).
At any given momentum $p$ one has $\bar S_{l l'}^{s j} (p) = S_{l l'}^{s j} (p)$.
We have also checked numerically that both equations (\ref{schr_nr1}) and  (\ref{Schroed_nonrel}) 
lead to identical results. 
It should be understood that both ways to cast the relativistic 
Schr\"odinger equation (\ref{schroed_rel}) into a nonrelativistic form discussed in this section 
are limited to the two--nucleon problem. To the best of our knowledge, no extension  to different systems 
has yet been offered. 
Consequently, three-- and more--nucleon observables calculated using a nonrelativistic approach 
with the NN potential $\tilde V$ or $\bar V$ will lead to different results. 
One should therefore use the relativistic Schr\"odinger equation (or Faddeev--Yakubovsky equations) 
with the potential $V$ 
in such cases. The same applies for processes with external probes.

\section{Bound state}
\setcounter{equation}{0}
\label{sec:bs}

We now turn to bound state (i.e.~deuteron) properties.
The deuteron binding energy $E_{\rm d}$ and wave function $\Psi_l^{\rm d} (p)$ 
can be obtained from the homogeneous part of eq.(\ref{LSeq}):
\beq\label{LSb1}
\Psi_l^{\rm d} (p) = \frac{1}{E_{\rm d} - (2 \sqrt{p^2 + m^2} - 2 m )} \,  \sum_{l'} \, \int_0^\infty \frac{dp'
\, {p'}^2}{(2 \pi )^3} \,  V^{sj}_{l,l'} (p,p') \, \Psi_{l'}^{\rm d} (p')~,
\eeq
with $s=j=1$ and $l=l'=0,2$, or, alternatively, from the nonrelativistic--like equation
\beq\label{LSb2}
\Psi_l^{\rm d} (p) = \frac{1}{E_{\rm d} + E_{\rm d}^2/(4 m) -  p^2/m} \,  \sum_{l'} \, \int_0^\infty \frac{dp'
\, {p'}^2}{(2 \pi )^3} \,  {\bar V}^{sj}_{l,l'} (p,p') \, \Psi_{l'}^{\rm d} (p')~,
\eeq
where ${\bar V}^{sj}_{l,l'} (p,p')$ is related to ${V}^{sj}_{l,l'} (p,p')$ via eq.~(\ref{pot_nonrel}).
Here we have used the relation (\ref{rel_kin}) between the binding energy and momentum.
In addition, one can also use the nonrelativistic Schr\"odinger approach as described in the 
previous section, which leads to:
\beq\label{LSb3}
\phi_l^{\rm d} (p) = \frac{1}{E_{\rm d} -  p^2/m} \,  \sum_{l'} \, \int_0^\infty \frac{dp'
\, {p'}^2}{(2 \pi )^3} \,  {\tilde V}^{sj}_{l,l'} (p,p') \, \phi_{l'}^{\rm d} (p')~,
\eeq
where ${\tilde V}^{sj}_{l,l'} (p,p')$ is defined via eq.~(\ref{Vnonrel_true}). 
The wave functions $\phi_l^{\rm d} (p)$ and $\Psi_l^{\rm d} (p)$ are related via eq.~(\ref{phipsi}).

\medskip\noindent
We will now regard the so--called static properties of the deuteron using the nonrelativistic--like equation (\ref{LSb2}).
The latter is fully equivalent to the relativistic equation (\ref{LSb1}) and leads to the same wave function,
but has the advantage that one can apply the standard nonrelativistic formulae to study various deuteron properties. 
We denote by $u(r)$ and $w(r)$ the S-- and D--wave components of the coordinate
space wave function $\Psi_{l}^{\rm d} (r)$ and by $u(p)$ and $w(p)$ the 
momentum space representations of $u(r)/r$ and $w(r)/r$:
\beq
u(p) = \frac{2}{\pi} \int_0^\infty u(r) j_0(pr) r dr \quad \quad \mbox{and} \quad \quad 
w(p) = \frac{2}{\pi} \int_0^\infty w(r) j_0(pr) r dr\,.
\eeq
The wave functions $u$ and $w$ are normalized according to:
\beq
\int_0^\infty dp \, p^2 \, [{u}(p)^2
+ {w}(p)^2] = \int_0^\infty dr  \, [{u}(r)^2+{w}(r)^2] =
1\,.
\eeq
The probability  $P_{\rm d}$ to find the nucleons inside of the deuteron in a D--state can be calculated via
\beq
\label{Pd}
P_{\rm d}=\int_0^\infty dp \, p^2 \,
{w}(p)^2 = \int_0^\infty dr \, w(r)^2\,.
\eeq
Further, one can compute the deuteron quadrupole moment $Q_{\rm d}$ and the matter root--mean--square (rms) radius 
$\sqrt{\langle r^2 \rangle^{\rm d}_m}$
through the following equations:
\beqa
\label{quadr}
Q_{\rm d} &=& \frac{1}{20} \int_0^\infty dr \, r^2
\, w(r) \, [\sqrt{8} u(r) - w(r) ] \nn
&=& -\frac{1}{20}\int_0^\infty dp \,
\biggl\{ \sqrt{8} \biggl[ p^2 \frac{d {u}(p)}{dp}
\frac{d {w}(p)}{dp} + 3p {w}(p) \frac{d {u}(p)}{dp}
\biggr] +
p^2\biggl(\frac{d {w}(p)}{dp}\biggr)^2 +6 {w}(p)^2
\biggr\}
\eeqa
and 
\beq
\sqrt{\langle r^2 \rangle^{\rm d}_m} \label{rd}
 = \frac{1}{2}\,\biggl[ \,
\int_0^\infty dr \, r^2 \,  [{u}(r)^2+{w}(r)^2] \, \biggr]^{1/2} \,.
\eeq
The wave functions $u(r)$ and $w(r)$ behave at large $r$ as:
\beq
u(r) \sim A_S \, {\rm e}^{-\gamma\, r}\,, \quad  \quad  \quad  \quad 
w(r) \sim A_D \,  {\rm
  e}^{-\gamma \, r} \, \biggl(
1+ \frac{3}{\gamma r} + \frac{3}{(\gamma r)^2} \biggr)\,,
\eeq
where $A_S$ and $A_D$ are the asymptotic normalization factors of the S-- and D--states, respectively,
and $\gamma = \sqrt{| m  E_{\rm d}  + E_{\rm d}^2/4 |}$. Instead of the quantities  $A_S$ and $A_D$, one often introduces 
the deuteron normalization $N_{\rm d}$ and the asymptotic D/S ratio $\eta_{\rm d}$ according to:
\beq
N_{\rm d}^2 = A_S^2 + A_D^2\,, \quad \quad \quad \quad
\eta_{\rm d} = \frac{A_D}{A_S}\,.
\eeq
Not all of the above mentioned deuteron properties are observable and can be measured experimentally.
The D--state probability $P_{\rm d}$ is well known to be unobservable
\cite{friar79}. The deuteron electric quadrupole moment corresponds to 
the quadrupole form factor at $| \vec q \,| = 0$, where $\vec q$ 
denotes the momentum transfer.
Clearly, the expression (\ref{quadr}), which gives just the deuteron expectation value 
of the quadrupole operator $Q_{ij}$
\beq
Q_{ij} \equiv \frac{1}{4} (3 r_i r_j - \delta_{ij} r^2 )\,,
\eeq
is only an approximation to the experimentally measured value for the quadrupole moment, which i.e.~does not take into 
account two--nucleon currents and relativistic corrections, see e.g. \cite{kohno83} for more details. 
A related discussion in the framework of EFT can be found in \cite{CP}.
The situation is similar with the deuteron matter rms--radius $\sqrt{\langle r^2 \rangle^{\rm d}_m}$, which 
is related to the experimentally measured deuteron charge rms--radius $\sqrt{\langle r^2 \rangle^{\rm d}_{ch}}$ via
\cite{klar86,mart95,friar97}:
\beq
\langle r^2\rangle^{\rm d}_{ch} = 
\langle r^2\rangle^{\rm d}_{pt} +\langle r^2\rangle^{p}_{ch} + \langle r^2\rangle^{n}_{ch}\,,
\eeq
where  
$\sqrt{\langle r^2\rangle^{p}_{ch}}= 0.886(11)$ fm (taking the mean of the three recent values form
Refs.\cite{RR,MR,IS} and adding the errors in quadrature) 
and  $\langle r^2\rangle^{n}_{ch}=-0.113(5)$ fm$^2$ 
\cite{kop95} are the proton and neutron ms--radii, respectively, and 
the ``point--nucleon'' radius of the deuteron $ \langle r^2\rangle^{\rm d}_{pt}$ is given by 
\beq
\langle r^2\rangle^{\rm d}_{pt} = \langle r^2\rangle^{\rm d}_{m} + \langle r^2\rangle^{\rm d}_{B}\,.
\eeq
Here $\langle r^2\rangle^{\rm d}_{B}$ subsumes the ``nuclear'' effects due to 
two--body currents as well as relativistic corrections.
Notice that while the ``point--nucleon'' deuteron radius is measurable, the matter
radius $\langle r^2\rangle^{\rm d}_{m}$
is clearly not an observable quantity. In particular, the separate contributions 
$\langle r^2\rangle^{\rm d}_{m}$  and $\langle r^2\rangle^{\rm d}_{B}$
change by a unitary transformation in the two--nucleon system, see \cite{friar97} for more details. 
For one specific choice of such a transformation, the effects due to two--nucleon currents 
in $\langle r^2\rangle^{\rm d}_{B}$ are estimated to be of the order 
$\sim 0.016$ fm$^2$ \cite{friar97}.
Contrary to the previously discussed deuteron quadrupole moment and rms--radius, the asymptotic quantities
$A_S$ and $A_D$ (or, equivalently, $N_{\rm d}$ and $\eta_{\rm d}$) as well as the deuteron binding energy are 
observables related to the ``pure'' nucleon--nucleon system. In particular, the binding energy gives the position of the 
NN S--matrix pole, while  the normalization $N_{\rm d}$ is related to the residue of the pole in the following way, see 
e.g.~\cite{stoksPhD,jen00}:
\beq
\label{nd}
N_{\rm d}^2 = \lim_{p \to k_{\rm d}}  \frac{(p - k_{\rm d} )}{8 \pi^2} p m \hat{\bar T}{}_{00}^{11} ( p )\,.
\eeq
Here $k_{\rm d} \equiv i \gamma$ and $\hat{\bar T}{}_{00}^{11} ( p )$ is the $l=l'=0$--component of the diagonalized T--matrix 
$\hat{\bar T}{}^{11} \equiv U \bar{T}{}^{11} U^{-1}$, where 
\beq
U = \left( \begin{array}{cc} \cos \hat \epsilon  & \sin \hat\epsilon \\ - \sin \hat\epsilon & \cos 
\hat\epsilon \end{array} \right)\,,
\eeq
and $\hat \epsilon$ is the Blatt and Biedenharn mixing angle \cite{Bl52}. Alternatively, one can rewrite 
equation (\ref{nd}) in terms of the Blatt and Biedenharn eigenphase shift $\hat \delta_0 (p)$ as:
\beq
\label{nd2}
N_{\rm d}^2 = \lim_{p \to k_{\rm d}} \frac{ 2 i (p - k_{\rm d} )}{1 - i \tan \big[\hat \delta_0 (p) \big]}\,.
\eeq
Notice that the T--matrix becomes real and the phase shift $\delta_0 (p)$ imaginary at negative energies.
Finally, we point out that the asymptotic D/S ratio $\eta_{\rm d}$ is given by the negative of the 
Blatt and Biedenharn mixing angle at the deuteron pole:
\cite{stoksPhD}:
\beq
\label{etaD}
\eta_{\rm d} = - \tan \hat \epsilon ( k_{\rm d} )\,.
\eeq

\medskip\noindent
Up to now we have discussed the deuteron properties in the context of the nonrelativistic--like 
equation (\ref{LSb2}). As already pointed out before, one could alternatively use the nonrelativistic 
equation (\ref{LSb3}). Both schemes are completely equivalent for the two--nucleon system and lead 
to the same phase shifts and the deuteron binding energy. It is also clear from eq.~(\ref{etaD})
that the asymptotic D/S ratio $\eta_{\rm d}$ does not change when one uses eq.~(\ref{LSb3})
instead of eq.~(\ref{LSb2}). On the other hand, the normalization 
$N_{\rm d}$ or, equivalently, the asymptotic normalization $A_S$ will change. Eq.~(\ref{nd2}) 
takes the form 
\beq
\label{nd3}
\tilde N_{\rm d}^2 = \lim_{\tilde p \to \tilde k_{\rm d}} 
\frac{ 2 i (\tilde p - \tilde k_{\rm d} )}{1 - i \tan \big[\hat{\tilde \delta}_0 (\tilde p) \big]}\,.
\eeq
if one uses the nonrelativistic Schr\"odinger equation (\ref{LSb3}). Here, $\tilde k_{\rm d} = i \sqrt{|m E_{\rm d}|}$
is the nonrelativistic deuteron binding momentum and $\hat{\tilde \delta}_0$ is the S--wave eigenphase shift  
calculated using eq.~(\ref{LSeq_nonrel}). Since both schemes are phase equivalent, one has:
\beq
\hat{\tilde \delta}_0 (\tilde p) = \hat \delta_0 ( \tilde p \sqrt{1 + \tilde p^2/(4 m^2)} )\,. 
\eeq
Here we made use of the relation (\ref{momenta}) between the relativistic and nonrelativistic momenta. 
We have therefore:
\beqa
\label{nd4}
\tilde N_{\rm d}^2 &=& \lim_{\tilde p \to \tilde k_{\rm d}} 
\frac{ 2 i (\tilde p - \tilde k_{\rm d} )}{1 - i \tan \big[\hat{\delta}_0 ( \tilde p 
\sqrt{1 + \tilde p^2/(4 m^2)}  ) \big]} \nn
&=& \lim_{p \to k_{\rm d}} 
\frac{ 2 i \left( \sqrt{2 m^2 ( \sqrt{1 + p^2/m^2} -1)}  -  \sqrt{2 m^2 ( \sqrt{1 + k_{\rm d}^2/m^2} -1)} \right)}
{1 - i \tan \big[\hat{\delta}_0 (  p  ) \big]} \nn
&\cong & N_{\rm d}^2 \left( 1 + \frac{3 E_{\rm d}}{8 m} \right)\,. 
\eeqa
Here we have again used the relation (\ref{momenta}) between the relativistic and nonrelativistic momenta.
The result in the third line of eq.~(\ref{nd4}) is valid up to corrections of the order $E_{\rm d}^2/m^2$.  
It can be rewritten in terms of the asymptotic S--wave normalization as follows:
\beq
\label{ASfin}
\tilde A_S = A_S \left[ 1 + (1 + \eta_{\rm d}^2 ) \frac{3 E_{\rm d}}{16 m} \right]\,.
\eeq
To end this section, we note that the other deuteron properties such as $P_{\rm d}$ in eq.~(\ref{Pd}), 
$Q_{\rm d}$ in eq.~(\ref{quadr})\footnote{It would be more appropriate to introduce a special notation 
for the quadrupole moment defined in eq.~(\ref{quadr}) using nonrelativistic impulse approximation in 
a way similar to the deuteron rms--radius. Unfortunately, no such notation appears in the literature,
which might lead to a confusion. It should be understood that while the deuteron quadrupole moment 
represents  the response of the deuteron to an external electromagnetic field and is certainly measurable,
Eq.~(\ref{quadr}) gives only an approximation, which is model-dependent and not observable.} 
and $\sqrt{\langle r^2\rangle^{\rm d}_{m}}$  in eq.~(\ref{rd}), 
which are not observable, are expected to change when calculated using the nonrelativistic wave function 
from eq.~(\ref{LSb3}). For a recent reviews on the deuteron the reader is referred to 
refs.~\cite{vanorden,GG}.

%%%%%%%%%%%%%%%%%%%%%%%%%%%%%%%%%%%%%%%%%%%%%%%%%%%%%%%%%%%%%%%%%%%%%%%%%%%%%%%%%
\section{The fits}
\def\theequation{\arabic{section}.\arabic{equation}}
\setcounter{equation}{0}
\label{sec:fit}

In this section we discuss the determination 
and specify the values
of the various LECs adopted in the present analysis. 
Throughout this work, we use the following values 
for the pion decay constant $F_\pi$,
the pion masses $M_{\pi^\pm}$, $M_{\pi^0}$ and the proton and neutron  masses $m_p$ and $m_n$:
$F_\pi =92.4$ MeV, $M_{\pi^\pm} = 139.5702$ MeV,  $M_{\pi^0} = 134.9766$ MeV,
$m_p=938.2720$ MeV, $m_n=939.5653$ MeV.

\medskip\noindent
For the 1PE potential, we use the expression (\ref{opep_full}) with $g_A = 1.29$. This larger value of 
the LEC $g_A$ as compared to the standard one $g_A = 1.26$ is in order to account for the Goldberger--Treiman 
discrepancy as discussed in section \ref{sec:pion_pot}. 
Notice further that we take into account the pion mass difference as given in eq.~(\ref{ope}).
The leading 2PE potential given in eq.~(\ref{2PE_nlo}) is parameter--free. The NNLO and N$^3$LO 2PE contributions
in eqs.~(\ref{2PE_nnlo}), (\ref{2PE_nnnlo})--(\ref{TPE2loop})
depend on the LECs $c_1$,  $c_2$,  $c_3$ and  $c_4$ from the second--order $\pi N$ Lagrangian as well as on 
$\bar d_1 + \bar d_2$, $\bar d_3$,   $\bar d_5$ and $\bar d_{14} - \bar d_{15}$ from the third--order $\pi N$ Lagrangian.
For the LECs $c_{1,4}$ we adopt the central values  
from the $Q^3$--analysis of the $\pi N$ system  \cite{Paul}:
$c_1=-0.81$ GeV$^{-1}$, $c_4=3.40$ GeV$^{-1}$.  For the constant $c_3$ the 
value $c_3=-3.40$ GeV$^{-1}$ is used, which is on the lower side but still
consistent with the results from  reference~\cite{Paul}: $c_3=-4.69 \pm 1.34$ GeV$^{-1}$.
The same value for $c_3$ has been adopted in our NNLO analysis \cite{EGMs2}.
Further, this value was found in Ref.~\cite{EM2} to be consistent with empirical 
NN phase shifts as well as the results from dispersion and conventional 
meson theories. 
Notice, however, that it is about 25\% smaller in magnitude than the 
value extracted from the partial wave analysis of the {\it pp} and {\it np} data \cite{rentm03}.
The LEC $c_2$ could not be fixed accurately analyzing pion--nucleon scattering
inside the Mandelstam triangle in \cite{Paul}. We therefore adopt the central value found in the 
third--order analysis \cite{Fet98}: $c_2 = 3.28$ GeV$^{-1}$. For the combinations of $d_i$'s, we 
again use the values found in \cite{Fet98}: $\bar d_1 + \bar d_2 = 3.06$ GeV$^{-2}$,
$\bar d_3  = - 3.27$ GeV$^{-2}$, $\bar d_5 = 0.45$ GeV$^{-2}$ and 
$\bar d_{14} - \bar d_{15} = -5.65$ GeV$^{-2}$. 
 
\medskip\noindent
We now turn to short--range contact interactions. 
The two $\nu=0$ --LECs $C_{S,T}$, seven $\nu=2$ --LECs $C_{1 \ldots 7}$ as well as 
fifteen $\nu=4$ --LECs $D_{1 \ldots 15}$ in eq.~(\ref{Vcon}) are unknown and have to be 
fixed from a fit to data (i.e.~to Nijmegen phase shifts). Contributions of the contact interactions
to various partial waves are given in eq.~(\ref{VC}). Thus, we have to determine 8 LECs in the 
$^3S_1 - ^3D_1$ channel, 4 LECs in the $^1S_0$ channel,  3 LECs in the $^3P_2 - ^3F_2$ channel, 
2 LECs in each of the $^1P_1$,  $^3P_1$,  $^3P_0$ partial waves and 1 LEC in each of the 
$^1D_2$, $^3D_2$, $^3D_3 - ^3G_3$ channels.  
In addition to the above mentioned isospin--conserving  contact interactions, we have two
isospin--violating contact terms with unknown coefficients, see eqs.~(\ref{cont_str}) and (\ref{cont_em}). 
Both terms contribute to the 
$^1S_0$ partial wave and provide charge--dependent contributions to the LEC $\tilde C_{1S0}$.  In the following,
we will therefore distinguish between $\tilde C_{1S0}^{pp}$, $\tilde C_{1S0}^{np}$ and $\tilde C_{1S0}^{nn}$.
We also note that we always use the proper kinematics as given in appendix \ref{sec:kinem}.
Let us now specify precisely our way of fixing the LECs. The LECs contributing to isovector channels 
$^3P_2 - ^3F_2$, $^3P_1$,  $^3P_0$ and  $^1D_2$ have been fixed from a fit to Nijmegen {\it pp} 
phase shifts \cite{nijpwa}, which are much more precise than the corresponding {\it np} phase shifts.
The isovector {\it np} phase shifts are then extracted from the {\it pp} ones in a  parameter--free way
by taking into account the proper 1PE potential and switching off the electromagnetic interaction. 
This is precisely the same procedure as used in the Nijmegen PWA \cite{nijpwa}. 
In the $^1S_0$ partial wave we have to take into account isospin--violating contact interactions as discussed above.
We determine the LECs $\tilde C_{1S0}^{pp}$, $\tilde C_{1S0}^{np}$, $C_{1S0}$, $D_{1S0}^1$ and  $D_{1S0}^2$
from a combined fit in the $^1S_0$ {\it pp} and {\it np} channels. The LEC $\tilde C_{1S0}^{nn}$ is then 
obtained from the requirement to reproduce the experimental value \cite{how98,gonz99}
$a_{\rm nn} =-18.9$ fm for the {\it nn} scattering length. 
All remaining LECs are fixed from a fit to {\it np} phases from Nijmegen PWA \cite{nijpwa}.
We notice that contrary to our NLO and NNLO analysis \cite{EGM2,EGMs2},
we had to use here a large energy interval, i.e. up to $E_{\rm lab} = 200$ MeV, 
in order to fix the LECs.
This is because of two reasons: first, the phase shifts in the $^1S_0$ and in the $^3S_1$--$^3D_1$ channels simply 
do not show enough structure beyond $E_{\rm lab} = 100$ MeV in order to fix reliably 4 and 8 parameters, respectively. 
Secondly, phase shifts at low energy are not very sensitive to higher--order contact interactions
except maybe in the two S--waves. 

\medskip\noindent
It remains to specify the values for the cut--offs $\Lambda$ and $\tilde \Lambda$ which enter the 
Lippmann--Schwinger equation and the spectral--function representation of the two--pion exchange 
potential, respectively. 
Certainly, both cut--offs are introduced in order to remove high--momentum
components of the interacting nucleon and pion fields, which are beyond the range of applicability of 
the chiral EFT. We remind the reader that from the formal point of view, one 
can choose any value for the SFR cut--off which is large enough so that the relevant physics is still 
present. Even the choice $\tilde \Lambda = \infty$, which is equivalent 
to dimensional regularization, is formally possible since all terms with 
positive powers of $\tilde \Lambda$ (and $\propto \ln \tilde \Lambda$)  can be absorbed by redefinition 
of the corresponding LECs. It has been argued in \cite{EGMs2}, however, that 
the choice $\tilde \Lambda = 500 \ldots 700$ MeV  leads to a natural separation of the 
long-- and short--range parts of the nuclear force 
and allows to improve the convergence of the low--momentum expansion. 
In the present analysis we use this range for $\tilde \Lambda$. 

\medskip\noindent
While $\tilde \Lambda$ is related to perturbative renormalization of 
the pion loop integrals, the cut--off $\Lambda$ specifies the way of nonperturbative renormalization 
of the Lippmann--Schwinger equation. 
Contrary to the SFR cut--off  $\tilde \Lambda$, one, in general, cannot 
arbitrarily increase the value of $\Lambda$ \cite{Lepage,Lepage_INT,Geg01,Geg04}. This is because one needs 
an infinite number of counter terms in order to absorb  all divergences arising through iteration 
of the potential in the Lippmann--Schwinger equation.\footnote{It has been shown in \cite{beane00}
that $1/r^n$ singular potentials, which arise e.g.~from pion 
exchange contributions, can be renormalized by a one--parameter square--well counterterm, see 
\cite{Betal} for a related work. Although 
the authors of \cite{beane00} have demonstrated that the low-energy NN observables can be made independent 
of the square-well width by adjusting the square-well strength, the power counting scheme adopted in
the present work is not consistent with such an approach.} 
Keeping $\Lambda$ finite and of the order of the separation scale in the NN problem, one 
expects the contribution of the higher--order counter terms to be suppressed by powers of the generic
low--momentum scale provided that the corresponding LECs are of natural size \cite{Lepage,Lepage_INT}. 
In our previous NLO and NNLO analyses \cite{ep2002} based on the 
dimensionally regularized expressions for the potential we have used $\Lambda = 500 \ldots 600$ MeV
with the regulator function being defined as $f^\Lambda (p ) = \exp [- p^4/\Lambda^4 ]$.
In the more recent study \cite{EGMs2} based on the SFR approach we have 
increased this range to  $\Lambda = 450 \ldots 650$ MeV using  $f^\Lambda (p ) = \exp [- p^6/\Lambda^6 ]$.
We have, however, found  in \cite{EGMs2}, that the upper values 
of $\Lambda$ are already rather close to its critical value $\Lambda^c$, 
above which one encounters spurious deeply--bound states. Notice that the values of various LECs 
start to strongly vary for $\Lambda \sim \Lambda^c$ leaving the natural range. 
In order to avoid such a situation we slightly reduce  the range of variation of $\Lambda$
to $450 \ldots 600$ MeV in the present analysis. To be more specific, we will use the following cut--off combinations 
(all values in MeV):
\beq
\label{cutoffs}
\{ \Lambda, \; \tilde \Lambda \} = \{ 450, \; 500 \},  \; \{ 600, \; 600 \},  \; 
\{ 450, \; 700 \},  \; \{ 600, \; 700 \}\,.
\eeq
For $\tilde \Lambda = 500$ MeV the value $\Lambda = 600$ MeV is already found to be close to $\Lambda^c$.
We therefore replace the cut--off combination $\{ 600, \; 500 \}$ by $\{ 600, \; 600 \}$.
%Clearly, the precise value of $\Lambda^c$ depends on the SFR cut--off $\tilde \Lambda$,
%which defines the (shortest) range of the two--pion exchange potential included explicitely in our calculation.
%One therefore expects the lower values $\Lambda^c$ for smaller values of $\tilde \Lambda$. 
%While $\Lambda = 600$ MeV is found to be far enough from  $\Lambda^c$ for $\tilde \Lambda = 700$ MeV, 
%this is no more the case for $\tilde \Lambda = 500$ MeV. We will comment more on that later on.
Notice that further reducing of the $\Lambda$--values beyond $\Lambda = 450$ MeV is, in principle, possible 
but leads to a strong increase of the theoretical uncertainty. We therefore refrain from doing that. 
Finally, we notice that a more elegant regularization 
prescriptions, like e.g. lattice regularization, would allow to regularize
pion loop integrals and the Lippmann--Schwinger equation in the same way without
introducing two independent scales $\Lambda$ and $\tilde \Lambda$. 
For a related recent discussion of the role and optimal choice of the cut--off  $\Lambda$ in the 
LS equation the reader is referred to Refs.~\cite{Lepage_INT,Geg04}.

\medskip\noindent
Let us now give the precise definition of the phase shifts considered in the present work and remind
the reader on the type of phase shifts used in the Nijmegen PWA \cite{nijpwa}.
We will adopt here the notation of ref.~\cite{Berg88} and denote by $\delta^{\rm V}_{\rm W}$  the phase shift 
generated by the potential $W$ with respect to the solution with $V$ as the interaction. 
\begin{itemize}
\item{{\it pp} phases.}

The full phase shifts $\delta_{\rm EM+N}$ of electromagnetic plus strong interaction
can be expressed as:
\beq
\label{ph0}
\delta_{\rm EM+N} = \delta_{\rm EM} + \delta^{\rm EM}_{\rm EM + N}\,.
\eeq
The above expression applies to uncoupled channels. For 
coupled channels one has to translate the addition law for the phase shifts into a multiplication 
law for the corresponding S--matrices 
\beq
\label{Smul}
S_{\rm EM + N} = (S_{\rm EM} )^{1/2} S_{\rm  EM + N}^{EM}  (S_{\rm EM} )^{1/2}\,, 
\eeq
%since the magnetic moment interaction (\ref{vmm}) contains 
%a tensor part, 
see \cite{stoks90,Berg90} for further details.
Such a modification for coupled channels will, however, not change the conclusions of this 
section. We will therefore not consider the coupled case in detail in what follows. 
The last term in eq.~(\ref{ph0}) corresponds to the phase shifts 
of the electromagnetic plus nuclear interaction with respect to electromagnetic wave functions. 
These phase shifts $\delta^{\rm EM}_{\rm EM + N}$ are the ones which are given in the Nijmegen PWA \cite{nijpwa}.
Notice that the electromagnetic phase shift $\delta_{\rm EM}$ can be represented as
\beq
\label{ph1}
\delta_{\rm EM} = \delta_{\rm C1} +  \delta_{\rm C1+ C2}^{\rm C1} + \delta_{\rm C1+ C2+MM}^{\rm C1+C2} +
\delta_{\rm C1+ C2+MM+VP}^{\rm C1+C2+MM} =  \delta_{\rm C1}  + \rho + \phi + \tau \,, 
\eeq
where we have introduced the abbreviations $\rho \equiv  \delta_{\rm C1+ C2}^{\rm C1}$, 
$\phi \equiv  \delta_{\rm C1+ C2+MM}^{\rm C1+C2}$ and 
$\tau \equiv \delta_{\rm C1+ C2+MM+VP}^{\rm C1+C2+MM}$. In practice, the quantities 
$\rho$, $\phi$  and $\tau$ are usually calculated using the CDWBA. 
%(XXX spell out?)
This is justified 
due to the smallness of the corresponding interactions $V_{\rm C2}$, $V_{\rm MM}$ and $V_{\rm VP}$. In this case 
one has approximately:
\beq
\phi \sim \delta_{\rm C1 + MM}^{\rm C1}\,, \quad \quad \quad
\tau \sim \delta_{\rm C1 + VP}^{\rm C1}\,.
\eeq
For more details the reader is
referred to refs.~\cite{Berg88,stoks90}.

\medskip\noindent
The phase shifts $\delta^{\rm EM}_{\rm EM + N}$ obtained in the Nijmegen PWA do, however, not correspond 
to the type of phase shifts, which is usually considered in practical calculations, namely the phase shifts
$\delta^{\rm C1}_{\rm C1 + N}$ of the modified Coulomb plus strong interactions with respect to the 
phase shifts of the modified Coulomb potential. These phase shifts can easily be calculated for any 
given nuclear potential using 
e.g. the method described in appendix \ref{sec:coul}. The {\it pp} phases considered in the 
present work are of that type. We therefore need to relate $\delta^{\rm C1}_{\rm C1 + N}$ to the previously
discussed phase shifts $\delta^{\rm EM}_{\rm EM + N}$. 
This can be done by noting  that the total phase shift $\delta_{\rm EM+N}$ can be expressed in the form:
\beq
\label{ph2}
\delta_{\rm EM+N} = \delta_{\rm C1} + \delta^{\rm C1}_{\rm C1 + C2 + MM +VP  + N}\,.
\eeq
In the coupled case one has to modify this relation in a way analogous to eq.~(\ref{Smul}).
Again, due to the smallness of the potentials $V_{\rm C2}$, $V_{\rm MM}$ and $V_{\rm VP}$  one can make use of the DWBA to relate 
$\delta^{\rm C1}_{\rm C1 + C2 + MM+ VP + N}$ and $\delta^{\rm C1}_{\rm C1 + N}$,
which leads to \cite{Berg88}:
\beqa
\label{ph3}
(\delta^{\rm C1}_{\rm C1 + C2 + MM+ VP + N})_l - (\delta^{\rm C1}_{\rm C1 + N})_l &\equiv & \tilde \Delta_l  \\
&=&  (\delta^{\rm C1+N }_{\rm C1 + N+ C2 + MM+ VP} )_l \nn
&\sim& \frac{-m_p}{k} \int_0^\infty dr \, \chi_l (r) \left[ V_{\rm C2} (r) + V_{\rm MM} (r) +V_{\rm VP} (r) \right] \chi_l (r)\,,
\nonumber
\eeqa
where $l$ is the angular momentum and $\chi_l (r)$ is the wave function for the potential $V_{\rm C1}+ V_{\rm N}$. 
Combining now eqs.~(\ref{ph0})--(\ref{ph3}) we end up with the following formula which relates the phase shifts 
$\delta^{\rm C1}_{\rm C1 + N}$ we are calculating to the ones $\delta^{\rm EM}_{\rm EM + N}$ of the Nijmegen PWA:
\beq
\label{phfin}
(\delta^{\rm C1}_{\rm C1 + N})_l \sim (\delta^{\rm EM}_{\rm EM + N})_l  - \tilde \Delta_l + \rho_l + \phi_l + \tau_l\,.
\eeq
This formula can be further simplified if one notes that for all $l \geq 1$ the wave functions $\chi_l (r)$ 
near the threshold are almost not affected by the nuclear interaction \cite{Berg88}. Thus, one can approximately
replace in eq.~(\ref{ph3}) the wave functions $\chi_l (r)$ by the regular Coulomb wave functions, which leads to 
$\tilde \Delta_l \sim \rho_l + \phi_l + \tau_l $. Therefore, for all partial waves except $^1S_0$ one has:
$(\delta^{\rm C1}_{\rm C1 + N})_l \sim (\delta^{\rm EM}_{\rm EM + N})_l$. In the case of the  $^1S_0$ 
partial wave, it has been argued based on explicit calculations 
that the quantity $\tilde \Delta_0$ is ``sufficiently model--independent''
for a wide range of nuclear forces at least in the case $V_{\rm C2}=0$. Notice that the magnetic 
moment interaction in eq.~(\ref{vmm}) does not contribute in that and all other spin--singlet channels.  The values 
of $\tilde \Delta_0$ based on the Nijmegen N78 potential \cite{nag78} as well as for the 
$\tau_0$ and $\rho_0$ are given in \cite{Berg88} at various energies.
In the present analysis, we will use these 
values for the above mentioned quantities in order to relate our $^1S_0$ phase shift to the one of the Nijmegen PWA via
eq.~(\ref{phfin}).\footnote{The quantity $\tilde \Delta_l$ 
in \cite{Berg88} does not contain the contribution due to magnetic moment interaction, which has been neglected in 
that work. In our analysis we only need to know  $\tilde \Delta_l$ explicitly for $l=0$ and therefore can still 
use the result of \cite{Berg88}.} 
Notice that a more accurate way to determine the phase shifts $(\delta^{\rm C1}_{\rm C1 + N})_l$ would be 
to define the regularized expressions for electromagnetic interactions and to calculate the 
quantities $\rho$, $\phi$,  $\tau$ and $\tilde \Delta_l$ explicitly (one can still use DWBA). We, however,
believe that there is no need for such a refinement at the level of accuracy of N$^3$LO. 

\medskip\noindent
Let us now summarize our way of calculating phase shifts in the {\it pp} system. We compute the phase shifts
$\delta^{\rm C1}_{\rm C1 + N}$ of the modified Coulomb plus nuclear potential $V_{\rm C1} + V_{\rm N}$ with respect 
to wave functions of the $V_{\rm C1}$--potential. The strong interaction part of the chiral nuclear force 
at order N$^3$LO is discussed  in section \ref{sec:pot}. 
It has to be used in the relativistic Schr\"odinger equation (\ref{schroed_rel}), while 
the electromagnetic interactions in eqs.(\ref{vc1vc2}),  
including the modified Coulomb force $V_{\rm C1}$, are to be used in 
the nonrelativistic--like equation  (\ref{Schroed_nonrel}). We therefore first apply 
eq.~(\ref{pot_nonrel}) to derive the modified strong potential for use in 
eq.~(\ref{Schroed_nonrel}).\footnote{Notice that in principle, the modified potential should contain contributions
due to electromagnetic interactions. Equation (\ref{pot_nonrel}) should actually be applied to the sum of the 
strong interaction and electromagnetic potentials $V_{\rm N}$ and $V_{\rm em}$. The modified potential would 
then contain pieces $\propto V_{\rm N}$, $V_{\rm em}$, $(1/m) \, V_{\rm N} V_{\rm N}$, 
$(1/m) \, V_{\rm N} V_{\rm em}$ and $(1/m) \,V_{\rm em} V_{\rm em}$. Applying eq.~(\ref{pot_nonrel}) only to 
the strong potential $V_{\rm N}$ and adding the appropriate (modified) electromagnetic contributions, we thus 
miss terms $\propto (1/m) \, V_{\rm N} V_{\rm em}$. Such contributions are suppressed by a factor $Q/m$ compared to the 
$\pi \gamma$--exchange in eq.~(\ref{Vpiga}) and beyond the accuracy of the present 
calculation.}  
We then calculate phase shifts $\delta^{\rm C1}_{\rm C1 + N}$ in momentum space as described
in appendix \ref{sec:coul}. In the fitting procedure the calculated phase shifts $\delta^{\rm C1}_{\rm C1 + N}$ are 
compared with the phases $\delta^{\rm EM}_{\rm EM + N}$ of the Nijmegen PWA \cite{nijpwa}. For all partial 
waves except $^1S_0$ we use the approximation $(\delta^{\rm C1}_{\rm C1 + N} )_l \sim ( \delta^{\rm EM}_{\rm EM + N} )_l$.
For the $^1S_0$ phase, we make use of equation (\ref{phfin}), where the quantities $\tilde \Delta_l$, $\rho_l$
and $\tau_l$ are taken from reference \cite{Berg88}.

\item{{\it np} phases.}

As already pointed out in section \ref{sec:isospIR}, the electromagnetic interaction 
in the {\it np} case is given entirely in terms of magnetic moment interaction
$V_{\rm MM} (np)$. Consequently, eq.~(\ref{ph0}) takes the form
\beq
\label{phase_np}
\delta_{\rm MM+N} = \delta_{\rm MM} + \delta^{\rm MM}_{\rm MM + N}\,.
\eeq
The {\it np} phase shifts of the Nijmegen PWA as well as in our analysis correspond to the phase 
shifts $\delta_{\rm MM + N}^{\rm MM}$ of nuclear plus magnetic moment interactions with respect to magnetic 
moment interaction wave functions. Notice that the term in $V_{\rm MM} (np)$ in eq.~(\ref{vmm}) 
proportional to $\vec A$ gives rise to  the so--called 
``class IV'' isospin--breaking force \cite{Hen79}, which mixes spin--singlet and spin--triplet states.   
The contribution of this term is very small and usually only taken into account when constructing the 
magnetic moment scattering amplitude, see e.g.~\cite{V18}. In our analysis we  
make use of the standard approximation \cite{V18}
\beq
\delta^{\rm MM}_{\rm MM + N} \sim \delta_{\rm N},
\eeq
for all $l \neq 0$--states. Therefore and because of the fact that the 
magnetic moment interaction does not contribute to the $^1S_0$ channel, we have to take into account $V_{\rm MM} (np)$
explicitly only in the $^3S_1$--$^3D_1$ partial wave. In that case the phase shift $\delta^{\rm MM}_{\rm MM + N}$
is calculated by subtracting $\delta_{\rm MM}$ from $\delta_{\rm MM+N}$, where the phase shifts 
$\delta_{\rm MM}$ are obtained using the Born approximation. 

%Further, it is crucial to realize that the isospin--breaking nuclear force 
%is parameter--free for all $l \geq 1$ at the order we are working. Up to N$\LOI$ in the notation of 
%\cite{WME}, it is given by  the neutral--to--charged pion mass difference in the one-- and the leading two--pion 
%exchange as well as by the $\pi \gamma$--exchange potential. In our work we follow the strategy of the Nijmegen group
%\cite{nijpwa} for isovector phase shifts with $l \neq 0$ and fix the unknown isospin--invariant contact 
%interactions by fit to the {\it pp} phases of \cite{nijpwa}, which are much more precise than the corresponding 
%{\it np} phase shifts. The letters are then obtained by taking into account the appropriate nuclear isospin--breaking
%effects and by switching off the Coulomb interaction. It is therefore clear that we have to use the same isospin--breaking 
%effects as used by the Nijmegen group (i.e. to neglect the $\pi \pi$ and $\pi \gamma$ contributions)
%in order to be consistent. The contact terms which contribute to 
%isoscalar phase shifts are fixed from fit to {\it np} phases. 

\item{{\it nn} phases.}

As in the previously considered case of neutron--proton scattering, in the {\it nn} system 
one has to take into account only magnetic 
moment interaction. Decomposing the phase shifts as in eq.~(\ref{phase_np})
one can make use of the approximation $\delta^{\rm MM}_{\rm MM + N} \sim
\delta_{\rm N}$ in all 
{\it nn} partial waves. 
As already stated before, this approximation is accurate for partial waves with  $l \neq 0$.
In the case of the $^1S_0$ partial wave, the phase shift is still given by $\delta_{\rm N}$
since the long--range magnetic moment interaction does not contribute to this channel.
Therefore, {\it nn} phase shifts in all partial waves correspond to $\delta_{\rm N}$.
\end{itemize}

%%%%%%%%%%%%%%%%%%%%%%%%%%%%%%%%%%%%%%%%%%%%%%%%%%%%%%%%%%%%%%%%%%%%%%%%%%%%%%%%%
\section{Results and discussion}
\def\theequation{\arabic{section}.\arabic{equation}}
\setcounter{equation}{0}
\label{sec:res}

\subsection{Phase shifts}
\label{sec:phases}

In the following sections, we will show the results for the {\it np} phase shifts. 
Before showing the results of our analysis, let us make a simple estimate for the 
expected theoretical uncertainty at N$^3$LO. Following the reasoning of ref.~\cite{EGMs2},
we expect for the uncertainty of 
a scattering observable at c.m.s.~momentum $k$ at N$^3$LO to be of
the order $\sim (\max [ k, \, M_\pi] /\lambda)^{5}$. To provide a fair estimate, 
we identify the hard scale $\lambda$ with the smallest value of the ultraviolet 
cut--off, i.e.~we adopt $\lambda \sim 450$ MeV. This results in the following 
estimations for the theoretical (maximal) uncertainty:
\bigskip 
\hskip 3 true cm
\begin{minipage}{10cm}
\begin{itemize}
\item[]
$\sim$ 0.5\%  at $E_{\rm lab} \sim 50$ MeV and below,
\item[]
$\sim$ 7\%  at $E_{\rm lab} \sim 150$ MeV,
\item[]
$\sim$ 25\%  at $E_{\rm lab} \sim 250$ MeV.
\end{itemize}
\end{minipage}

\medskip\noindent
One should keep in mind that the above estimations are fairly rough. 
For a detailed discussion on the theoretical uncertainty, especially at NLO and NNLO, 
the reader is referred to \cite{EGMs2}.

\subsubsection{S--waves}
\label{sec:SW}

The phase shifts in the $^1S_0$ and $^3S_1$ partial waves are shown in Fig.~\ref{fig1}.
Both are visibly improved compared to the NNLO result.
For $E_{\rm lab} = 50$ MeV, 150 MeV and 250 MeV we find the phase shift in the 
$^1S_0$ partial wave in the ranges $40.42^\circ \ldots 40.72^\circ$, 
$16.04^\circ \ldots 17.03^\circ$ and $2.22^\circ \ldots 4.76^\circ$, respectively.
These values agree well with the ones from the Nijmegen PSA: $\delta = 40.54^\circ$, 
$\delta = 16.94^\circ$ and $\delta = 1.96^\circ$. The relative uncertainty of our results 
is in agreement with the above estimations except for  $E_{\rm lab} = 250$ MeV, 
where the phase shift is close to $0$. The results for the $^3S_1$ partial wave are similar to the 
ones in the $^1S_0$ channel. The uncertainty due to the cut--off variation 
is found to be smaller in this case.

\medskip\noindent
The pertinent S--wave LECs are tabulated in table~\ref{tab:LEC1} for the four pairs of cut-offs 
(\ref{cutoffs}). Here, several remarks are in order. First, we note that, in general, 
one has to expect multiple solutions for the LECs. This problem has already been discussed 
in \cite{EGM2} at NLO and NNLO. For the $^1S_0$ channel, we have to fix
five LECs $\tilde C_{1S0}^{pp}$, $\tilde C_{1S0}^{np}$, $C_{1S0}$, $D_{1S0}^1$, $D_{1S0}^2$, from a fit to 
Nijmegen {\it pp} and {\it np} phase shifts. We did find multiple solutions for LECs which describe the data equally well
if we neglect  isospin breaking and fix 
$\tilde C_{1S0}^{np}$, $C_{1S0}$, $D_{1S0}^1$, $D_{1S0}^2$ from a fit to the Nijmegen {\it np} phase shift.
Taking into account isospin breaking effects and performing a combined fit 
to both {\it pp} and {\it np} phase shifts turns out to improve the situation and help to 
sort out the true solution. We found a single solution for the LECs for the cut--off
combinations $\{ \Lambda, \; \tilde \Lambda \} = \{ 450, \; 700 \}$ and   
$\{ 600, \; 700 \}$. For the two other cut--off combinations
several local minima in the $\chi^2$--plot have been observed. We then adopted the values for the 
LECs corresponding to the global minimum. We have checked that these values result
from the ones for different  $\{ \Lambda, \; \tilde \Lambda \}$ by a continuous change of 
the cut--offs. 

\medskip\noindent
The situation in the $^3S_1$--$^3D_1$--channel is even more complex since one has to determine eight LECs. 
Our results for the LECs are shown  in tables~\ref{tab:LEC1}, \ref{tab:LEC2} and \ref{tab:LEC3}.
Due to the large dimension of the parameter space, we cannot definitely claim that the 
found values for the LECs correspond to a true global minimum of the $\chi^2$. 

\medskip\noindent
Let us now comment on the naturalness of the determined LECs. 
In general, the natural size for the LECs can be (roughly) estimated as follows:
\beq
\tilde C_i \sim \frac{4 \pi}{F_\pi^2}\,, \quad \quad 
C_i \sim \frac{4 \pi}{F_\pi^2 \Lambda_{\rm LEC}^2}\,, \quad \quad 
D_i \sim \frac{4 \pi}{F_\pi^2 \Lambda_{\rm LEC}^4}\,,
\eeq
where $\Lambda_{\rm LEC}$ is the scale entering the values of the LECs and 
the factor $4 \pi$ results from the angular integration in the partial wave decomposition,
see appendix~\ref{sec:pw} for the details. The S--wave LECs shown in table~\ref{tab:LEC1}
are of the natural size except the LECs $D_{1S0}^1$ and $D_{3S1}^1$, which are  
somewhat large in magnitude. Indeed, estimating the scale $\Lambda_{\rm LEC}$ as 
$\Lambda_{\rm LEC} \sim 500$ MeV leads to $| D_i | \sim 2.4$ in the same units as used in 
table~\ref{tab:LEC1}. Still, the higher--order contact interactions are suppressed compared 
to the lower--order operators at low momenta. 
For example, for the cut--off combination $\{ 450, \; 500 \}$ and 
$p = p' = M_\pi$ the contributions of the contact operators at various orders are given by:
\beqa
\langle ^1S_0 | V_{\rm cont}^{\rm np} (p, \; p' \,)| ^1S_0 \rangle \bigg|_{p = p' = M_\pi}
&=& \bigg[\tilde C_{1S0}^{\rm np} + C_{1S0} ( p^2 + p '^2) +
\Big( D_{1S0}^1 \, p^2 \, {p'}^2 + D_{1S0}^2 \, ({p}^4+{p}'^4) \Big) \bigg]_{p = p' = M_\pi} \nn
&=& \Big[-0.091 + 0.057 +(-0.010+0.003) \Big] \times 10^4 \mbox{ GeV}^{-2}\,.
\eeqa

\subsubsection{P--waves} 
\label{sec:PW}

Our results for the {\it np} P--waves and the mixing angle $\epsilon_1$ are shown in Fig.~\ref{fig2}.
All phase shifts are visibly improved compared to the NLO and NNLO results. One has, however, 
to keep in mind that two independent parameters appear now in each of these channels 
instead of one parameter at both NLO and NNLO. The results of the Nijmegen PWA are reproduced 
in our N$^3$LO analysis within the theoretical uncertainty in all phase shifts with exception 
of $^1P_1$ at larger energies. 
%*EE
In the case of the $^3P_2$ partial wave, the band is dominated 
by variation of the SFR cut--off. In particular, lower values for this cut--off lead to a better 
agreement with the data at higher energies.
At $E_{\rm lab} = 250$ MeV, the N$^3$LO phase shifts 
deviate from the data by an amount of up to $\sim 8^\circ$. The typical size of the P--wave phase shifts
at this energy is of the order $\sim 25^\circ$. The uncertainty in the calculated phase shifts due 
to the cut--off variation agrees therefore fairly well with the estimation in section \ref{sec:phases}.
We remind the reader that the theoretical bands at NLO and NNLO are expected to have a similar 
width, since the effective potential at these orders contains the same 
set of contact interactions (counter terms). As explained in detail in \cite{EGMs2}, the 
uncertainty resulting from the cut--off variation at NLO is smaller than the actual theoretical 
uncertainty at this order.

\medskip\noindent
The  pertinent P--wave LECs are tabulated in table~\ref{tab:LEC2} four these four pairs of cut-offs. 
As in the previously discussed case of the S--waves, we found multiple solutions for the LECs. 
The physical solution can easily be determined due
to the smaller number of parameters (two unknown LECs in each P--wave).
To illustrate this point consider the $^3P_1$ partial wave with the cut--offs 
$\{ \Lambda, \; \tilde \Lambda \} = \{ 450, \; 500 \}$. We find 
two solutions for the LECs $C_{3P1}$ and $D_{3P1}$ fitting to the Nijmegen {\it pp}
phase shifts:
\beqa
\label{LECs_3P1}
&& 
C_{3P1} = -0.6334\,, \mbox{\hskip 1.5 true cm} D_{3P1} = 4.2359\,, \nn
\mbox{and} && C_{3P1} = 5.9620\,,  \mbox{\hskip 1.8 true cm} D_{3P1} = -20.6154 \,, 
\eeqa
where we used the same units as in Table~\ref{tab:LEC2}. Both sets of parameters lead to 
an accurate description of the data, which is shown in Fig.~\ref{fig:3P1}. The solution 
in the first line of eq.~(\ref{LECs_3P1}) satisfies the naturalness assumption for the 
LECs and has been adopted in the present analysis. Notice further that the value 
$C_{3P1} = -0.6334 \times 10^4$ GeV$^{-2}$ is close to the NLO and NNLO values for these LECs
(for the same cut--off combination) $C_{3P1} = -0.4932 \times 10^4$ GeV$^{-2}$ and 
$C_{3P1} = -0.7234\times 10^4$ GeV$^{-2}$, respectively. The results for other partial waves 
are similar. All LECs $C_i$ in the P--waves are found to be of  natural size and take the values 
which are close to the ones at NLO and NNLO. The P--wave LECs $D_i$ are natural as well.

\medskip\noindent
Let us now comment on isospin--breaking. 
As already explained before, the LECs in the isovector partial waves are fitted to 
the {\it pp} phase shifts. To calculate the corresponding {\it np} phase shifts, we
switch off the electromagnetic interaction and adjust for the proper pion mass in the 
1PE potential. The differences  between 
the corresponding {\it pp} and {\it np} phase shifts at three different energies $E_{\rm lab} = 10$, $25$ and 
$50$ MeV are shown in Tables \ref{tab1a},  \ref{tab1b} and \ref{tab1c}, respectively. 
In general, we see that the effects due to removal of the Coulomb interaction, $\Delta_i^\gamma$, agree very well with the 
ones of Nijmegen PWA. The uncertainty due to the cut--offs variation becomes larger at higher energies. 
The effects due to including the pion mass difference, $\Delta_i^\pi$, show typically somewhat larger deviations from 
the Nijmegen PWA. This is presumably to a large extent due to a different treatment of the 1PE force: while
we use the potential in momentum space with high momenta being cut off, the 
Nijmegen group performs calculations in coordinate space,
and chooses to cut--off the long--range potential at $1.4$ fm. 
Although both methods certainly lead to the same long--distance 
asymptotics of the 1PE potential, they differ significantly in the treatment of its shorter--range part. 
The largest deviations from $\sim 15\%$ at $E_{\rm lab} = 10$ MeV to  $\sim 20\%$--$30\%$ at 
$E_{\rm lab} = 50$ MeV for $\Delta_i^\pi$ from the Nijmegen PWA are observed in the $^3P_2$ partial wave. 
It is comforting to see that both isospin--violating effects (i.e.~due to the Coulomb force and the pion mass difference
in 1PE) are  in most cases of the same size, as it is also expected from power counting arguments, see section 
\ref{sec:isosp} for more details. We also note that these two effects have often opposite sign and  tend to  
cancel. For example, one observes at $E_{\rm lab}=10$ MeV \cite{nijpwa}: $\Delta_{\rm 3P0} \equiv
\Delta_{\rm 3P0}^\gamma + \Delta_{\rm 3P0}^\pi = 0.371^\circ  -0.447^\circ = -0.076^\circ$.

\subsubsection{D-- and higher partial waves} 
\label{sec:DW}

The results for D--, F-- and G--waves are shown in Figs.~\ref{fig3}, \ref{fig4} and \ref{fig5},
respectively. We remind the reader that at N$^3$LO there is one adjustable constant $D_i$ in each of the D--waves,
while F-- and higher partial waves are parameter--free. As depicted in Fig.~\ref{fig3}, the shape 
of the $^3D_3$ partial wave is still not properly reproduced at N$^3$LO, although it is greatly improved
compared to NLO and NNLO predictions. This phase shift is, however, rather small as compared to other 
D--wave phase shifts, and thus one expects relatively small effect of this phase shift on the NN scattering 
observables. We also note that the absolute deviation from the data in this channel is not larger than in the 
other D--waves. 
Most of the F-- and G--waves are at N$^3$LO in agreement with the data. 
One observes that the theoretical bands  do not get thinner at N$^3$LO, which 
might at first sight appear strange. This, however, is naturally explained by the fact
that there are no short--range contact terms in these channels. Such terms start to contribute to 
F--waves at N$^5$LO ($Q^6$) and to G--waves at N$^7$LO ($Q^8$).  Consequently, one should expect 
the uncertainty due to the cut--off variation to be of the same size for calculations up to these high 
orders in the chiral expansion. 
Clearly, peripheral partial waves are strongly dominated by the 1PE potential, which represents the 
longest--range part of the strong nuclear force. Indeed, one observes that the  phase shifts are 
mostly well reproduced already at NLO, while NNLO and N$^3$LO corrections only produce minor changes.   
Notice further that due to the smallness of the phase shifts, the Born approximation works very well in 
high partial waves and the phase shifts are essentially given by the diagonal (in momentum space) matrix 
elements of the two--nucleon potential. It is then clear that the bands arise almost completely due 
to multiplying the potential by the regulator function.  
%*EE
The only exception from this rule is given by the $^1D_2$ partial wave,
where the SFR cut--off has a larger impact on the phase shift at higher energy than the cut--off in the 
Lippmann--Schwinger equation. In particular, the 
lower values of this cut--off lead to larger values of the phase shift.

\medskip\noindent
The determined LECs $D_i$ are tabulated in Table~\ref{tab:LEC3}. All of them are of natural size.

\subsection{S--wave effective range expansion}
\label{sec:ERE}

We now regard the  S--wave effective range parameters and begin with the {\it np} system. 
In that case one can make use of the usual effective range expansion for finite--range potentials,
eq.~(\ref{ere1}). The reason is that the long--range magnetic moment interaction does not 
contribute to states with $l=0$. Notice, however, that one should not use the standard effective range 
expansion for the $^3D_1$ partial wave and mixing angle $\epsilon_1$, 
which are modified in the presence of the  
long--range ($\sim 1/r^3$) magnetic moment interaction. 
Our results for the $^3S_1$ and $^1S_0$  scattering length, effective range and 
shape coefficients $v_{2,3,4}$ are summarized in Tables \ref{tab1} and \ref{tab2}.  
The results for the $^1S_0$ scattering length and effective range are improved compared 
to the NLO and NNLO predictions of \cite{EGMs2}. The N$^3$LO result for the scattering length fills a 
small gap between the NNLO prediction and the value of the Nijmegen PWA. The uncertainty 
due to the cut--off variation for all effective range parameters turns out to be smaller at N$^3$LO compared to NNLO, as 
it should.  We observe a minor discrepancy for the shape coefficient $v_2$, which might however simply reflect the 
lack of numerical accuracy, with which this quantity is calculated. The description of the  $^3S_1$  effective range 
parameters is similar to the one in the $^1S_0$ channel. 

\medskip\noindent
Next, we consider the {\it pp} system. This case is much more complex since one has to account for 
electromagnetic interaction. Ideally, one should use the phase shifts $\delta_{\rm EM +N}^{\rm EM}$
and the expression for the effective range function given in  \cite{Berg88} to obtain 
the effective range expansion for the nuclear force in presence of the long--range electromagnetic interactions,
which  in the $^1S_0$ channel  are given by the improved Coulomb and vacuum polarization potentials 
in eqs.(\ref{modCoul}) and (\ref{vp}). In the present analysis we have used a simplified description for the 
{\it pp} phase shift as explained in section \ref{sec:fit}. We do not calculate explicitly the 
phase shifts $\delta_{\rm EM +N}^{\rm EM}$ but rather the ones $\delta_{\rm C1 +N}^{\rm C1}$ 
of nuclear plus modified Coulomb potential with respect to Coulomb wave functions, 
adjusting for the difference as explained in section  \ref{sec:fit}. We have therefore made use of eq.~(\ref{ere2})
to calculate the {\it pp} scattering length and effective range. We obtain the following values:
\beq
\label{ere_pp}
a_{pp} = -7.795 \ldots -7.812 \; \mbox{fm}\,, 
\quad \quad \quad \quad 
r_{pp} = 2.73 \ldots 2.76 \; \mbox{fm}\,,
\eeq
where the uncertainty is due to the cut--off variation.
These values agree nicely with the experimental ones \cite{Berg90}:
\beq
\label{app}
 a_{pp}^{\rm exp} = -7.8149 \pm 0.0029\; \mbox{fm}\,, 
\quad \quad \quad \quad 
r_{pp}^{\rm exp} = 2.769 \pm 0.014 \; \mbox{fm}\,.
\eeq
One should, however, keep in mind that we made an approximation and neglected the 
effects due to the long--range part of the vacuum polarization potential and the 
interaction in the second line of eq.~(\ref{modCoul}). As found in \cite{Berg88}
neglecting these electromagnetic interactions affects the values of $a_{pp}$
and $r_{pp}$ by an amount smaller than $0.01$ fm, which is within the theoretical
uncertainty of the present analysis.

\medskip\noindent
Finally, we  consider the {\it nn} system. Since no long--range electromagnetic 
interactions contribute to the $^1S_0$ partial wave, one can use the effective range 
expansion (\ref{ere1}). Since we have used the 
``standard value''  for the {\it nn} scattering length $a_{nn}$
\beq
a_{nn}^{\rm std} = -18.9 \pm 0.4 \, \mbox{fm}
\eeq
as an input to fix the LEC of the leading isospin--violating short--range interaction, 
we can only make predictions for the effective range $r_{nn}$:
\beq
r_{nn} = 2.76 \ldots 2.80 \; \mbox{fm}\,.
\eeq
This agrees with the experimental number \cite{Mill90}
\beq
r_{nn}^{\rm exp} = 2.75 \pm 0.11 \; \mbox{fm}\,.
\eeq
Notice that there is still some controversy about the experimental value of 
the {\it nn} scattering length extracted using different reactions. For example, 
the value $a_{nn}^{\rm exp} = - 18.50 \pm 0.53$ fm has been reported from 
studying the $^2$H($\pi^-$, {\it n}$\gamma$){\it{n}} process \cite{how98}, while measurements 
of the neutron  deuteron breakup reaction lead on one hand to 
$a_{nn}^{\rm exp} = - 18.7 \pm 0.6$ fm \cite{Gonz99} and on the other hand to
different values,  $a_{nn}^{\rm exp} = - 16.1 \pm 0.4$ fm 
and $a_{nn}^{\rm exp} = - 16.3 \pm 0.4$ fm \cite{Hu00}.

\medskip\noindent
Last but not least, we would like to point out that the relation between the values of the 
effective range parameters, which would result if there would be no electromagnetic interaction,
and the observed ones is highly nontrivial. Neglecting  electromagnetic nucleon mass shifts, we can, for example, 
switch off the Coulomb interaction in the {\it pp} system and recalculate 
the effective range coefficients using eq.~(\ref{ere1}). This leads to:
\beq
\label{ere_pp_str}
\tilde a_{pp} = -16.00 \ldots -16.63 \; \mbox{fm}\,, 
\quad \quad \quad \quad 
r_{pp} = 2.81 \ldots 2.86 \; \mbox{fm}\,.
\eeq
Although the value for $\tilde a_{pp}$ is fairly close to the one for the {\it nn} and {\it np}
scattering lengths, as one would expect from the approximate isospin invariance of the strong interaction, 
it should be understood that effects due to electromagnetic interaction are not completely removed from these quantities.
To clarify this point let us take a look at the leading 
{\it nn},  {\it np} and {\it pp} short--range interactions:
\beqa
\label{betas}
\tilde C_{1S0}^{nn} &=& (\tilde C_{1S0}^{nn} )_{\rm str} + \beta_{nn} \frac{e^2}{(4 \pi)^2}\,, 
\nonumber\\
\tilde C_{1S0}^{np} &=& (\tilde C_{1S0}^{np} )_{\rm str} + \beta_{np} \frac{e^2}{(4 \pi)^2}\,, \nn
\tilde C_{1S0}^{pp} &=& (\tilde C_{1S0}^{pp} )_{\rm str} + \beta_{pp} \frac{e^2}{(4 \pi)^2}\,. 
\eeqa
Here the LECs $(\tilde C_{1S0}^{i} )_{\rm str}$ are entirely due to the strong interaction. 
If only linear terms in the quark mass difference are included, see eq.~(\ref{cont_str}), these LECs are related 
with each other as $(\tilde C_{1S0}^{pp} )_{\rm str} + (\tilde C_{1S0}^{nn} )_{\rm str} = 2 (\tilde C_{1S0}^{np} )_{\rm str}$
and the difference $(\tilde C_{1S0}^{pp} )_{\rm str} - (\tilde C_{1S0}^{nn} )_{\rm str}$ is proportional to $\epsilon M_\pi^2$.
The terms $\propto \beta_i$ in eq.~(\ref{betas}) are due to the short--range electromagnetic interactions,
see eq.~(\ref{cont_em}).\footnote{The fact that we have three independent LECs $\beta_{nn}$,   $\beta_{np}$
and  $\beta_{pp}$ and only two terms in eq.~(\ref{cont_em}) might appear confusing. 
In fact, we have only shown explicitly electromagnetic isospin--breaking  and omitted isospin 
conserving terms in eq.~(\ref{cont_em}). One of  the two electromagnetic isospin conserving contact interactions
contribute to $^1S_0$ and one to $^3S_1$ NN scattering. Therefore, three and not two independent 
electromagnetic terms contribute to $^1S_0$  NN scattering.} Since we do not know the values of the LECs
$\beta_i$ in eq.~(\ref{betas}) and it is not possible to disentangle them from $\tilde C_{1S0}^{i}$ in 
the two--nucleon system, we cannot extract the values for NN observables due to the strong interaction out of the 
experimentally measured quantities. Notice that the LECs $\beta_i$ might (at least in principle) be determined 
from processes with external pions.
%(XXX check the above statement. Not sure whether  it is true...)
Notice further that the {\it pp} scattering length with  the long--range 
Coulomb interaction being switched off is even not a well--defined quantity in an effective field theory approach since  
it is sensitive to details of the strong interaction at short distances. Indeed, the extracted scattering length 
$\tilde a_{pp}$ in eq.~(\ref{ere_pp_str}) shows a significant cut--off dependence. Clearly, the scattering length 
due to pure strong interaction is perfectly well defined and the cut--off dependence is (largely) 
absorbed by the appropriate ``running'' of $\beta_{pp}$. For related discussion on the proton--proton
scattering length in context of effective field theory see \cite{kong00,gegelia03}.
Furthermore, a useful approximation for the 
quantity $\tilde a_{pp}$ based on the short--range nature of the strong interaction and 
the fact that the scattering length is large can be found in \cite{jackson50}.

\subsection{Two--nucleon scattering observables}
\label{sec:NNobs}

Once the NN phase shifts are calculated, all two--nucleon scattering observables
can be obtained in a straightforward way using e.g.~the formulae 
collected in \cite{stoksPhD}.
In Figs.~\ref{fig6}, \ref{fig7}, \ref{fig8} and \ref{fig9} we show the {\it np} 
differential cross section and vector analyzing power at $E_{\rm lab} =$25, 
50, 96 and 143 MeV at NNLO and N$^3$LO in comparison with the data and the Nijmegen PWA results.
In this calculation, we have included all {\it np} partial waves up to $j \leq 8$
and did not take into account the magnetic moment interaction. 
At the lowest energy we have considered, $E_{\rm lab} = 25$ MeV, both NNLO and N$^3$LO 
results are consistent with the ones of the Nijmegen PWA. The small disagreement with the 
Nijmegen PWA in the analyzing power at forward direction is due to the neglected magnetic moment 
interaction.  At higher energies the NNLO predictions become less precise.  At N$^3$LO  
the uncertainty  in the cross--section due to the cut--off variation at the largest energy we have calculated,
$E_{\rm lab} = 143$ MeV, is less than 10\%. It is comforting to see that NNLO and N$^3$LO
results overlap in most cases and are both in agreement with the Nijmegen PWA. 
We further notice that the small but visible deviations
of our N$^3$LO result for the differential cross section from the Nijmegen PWA curve at 
forward and backward angles and higher energies is most probably due to the lack of partial waves with 
$j>8$ in our calculations. The convergence of the partial wave expansion is well known to be 
slow in these particular cases. For example, it has been found in \cite{Fachr01} that 
a sum up to $j = 16$ is needed to obtain convergence for the cross section at $E_{\rm lab} = 300$ MeV
within 1\%.

\subsection{Deuteron properties}
\label{sec:deut}

We now turn to the bound state properties.  We stress that we do not use the deuteron
binding energy as a fit parameter as it is frequently done but  rather adopt the same
LECs as obtained in the fit to the low phases. In Table~\ref{tab3} we collect the
resulting deuteron properties in comparison to the NLO and NNLO results from \cite{EGMs2}.
All results for the deuteron properties in this table have been calculated using the formulae given in 
section \ref{sec:bs}  based on the relativistic wave function $\Psi^{\rm d} (p)$.
%%Notice that we did not use the deuteron binding energy in the determination of the LECs. 
First, we note a clear improvement at N$^3$LO in the chiral expansion. The predicted 
binding energy at N$^3$LO is within 0.4\% of the experimental value. This has to be 
compared with 1\%--1.5\%  ($\sim$2\%--2.5\%) deviation at NNLO (NLO). 
Also visibly improved is the asymptotic S--wave normalization strength $A_S$, which 
now deviates from the experimental (central) value by 0.3\% as compared to $\sim$1.1\% ($\sim$1.9\%)
at NNLO (NLO). Our predictions for the asymptotic D/S--ratio have a tendency to slightly reduce 
its value when going from NLO to NNLO to N$^3$LO. The results at all orders are in
agreement with the data within the experimental uncertainty. Further, our N$^3$LO result for $\eta_{\rm d}$ 
agrees well with the one of the Nijmegen PWA \cite{Swart95}, $\eta_{\rm d} = 0.0253(2)$. 
We do not observe any improvement for the quadrupole momentum $Q_{\rm d}$ at N$^3$LO, which shows an even 
larger deviation from the data compared to NNLO (6\% - 8\% versus 4\% - 5\%). We, however, remind the 
reader that the present calculation of $Q_{\rm d}$ is based on formulae of section \ref{sec:bs}. It is 
incomplete and does, in particular, not take into account the contribution of the two--nucleon current.  
Notice that apart from the pion--exchange two--nucleon currents, there are contributions from 
two--nucleon contact current, where the corresponding LEC cannot be fixed from nucleon--nucleon 
scattering. Such current results from the operator in the effective Lagrangian with 
four nucleon fields, one photon field and two derivatives. 
It appears natural to fix the value of the accompanying LEC from the requirement to reproduce the 
value of the deuteron quadrupole moment. For the calculations of the various deuteron properties 
including the quadrupole moment as well as other two--nucleon observables in pionless EFT the reader 
might consult  Refs.~\cite{chen99_1,chen99_2}. Notice that the situation with the quadrupole moment 
is analogous to the one described in 
\cite{walzl01} for the deuteron magnetic moment. In that case the corresponding 
short--range two--nucleon current results from the operator with just one derivative and thus appears even at a 
lower order. 
The situation with  the deuteron rms-radius is similar to the one with the quadrupole moment:
we observe a larger deviation from the data at N$^3$LO as compared to the NLO and NNLO results.
Notice however that the deviations at N$^3$LO from the experimental number are still of the order 
of 0.5\% or less.
The above comment on the missing contributions in the quadrupole moment calculation 
applies to the  deuteron rms-radius  as well.
At N$^3$LO one should account for the contribution due to the short--range two--nucleon 
current, which results from the contact operator with four nucleon fields, 
one photon field and two derivatives, see \cite{chen99_1} for more details. 
It is remarkable that the cut--off dependence of $\sqrt{\langle r^2 \rangle^{\rm d}_m}$ at N$^3$LO is 
significantly larger compared to NLO and NNLO. This implicitly confirms our previous statement 
about the necessity to incorporate the short--range current at this order. The cut--off dependence 
of the corresponding LEC will compensate the cut--off dependence of  $\sqrt{\langle r^2 \rangle^{\rm d}_m}$ 
making  the deuteron ``point--nucleon'' radius cut--off independent up to higher--order corrections.
The complete N$^3$LO calculation of the quadrupole moment and the   ``point--nucleon'' 
electric charge radius of the deuteron will be presented in a separate publication.
As a numerical check, we have recalculated all deuteron properties using the nonrelativistic 
wave function $\phi^{\rm d} (p)$ in eq.~({\ref{LSb3}). As expected from the discussion in section \ref{sec:bs}, 
we reproduce the values for $E_{\rm d}$ and $\eta_{\rm d}$. The asymptotic S--wave normalization $A_S$
changes by $\tilde A_S - A_S = 0.00039$ fm$^{-1/2}$. This has to be compared
with the value $\tilde A_S - A_S = 0.000392$ fm$^{-1/2}$ from  eq.~(\ref{ASfin}).
The quadrupole moment and the rms--radius change by 0.4\% and 0.1\%, respectively.
Finally, we show the deuteron wave function in coordinate space in fig.~\ref{fig10} 
for a particular cut--off choice, together with results obtained 
at NLO and NNLO. One observes a stronger suppression of the S--wave component 
at short distances compared to NLO and NNLO, as well as the lower probability for the deuteron to be in the D--state.
The latter observation follows also from the smaller value of $P_{\rm d}$ quoted in Table~\ref{tab3}. 
The shape of the wave function changes for different cut--off choices.
We remind the reader that the deuteron wave function is not observable (except at very large distances).

%%%%%%%%%%%%%%%%%%%%%%%%%%%%%%%%%%%%%%%%%%%%%%%%%%%%%%%%%%%%%%%%%%%%%%%%%%%%%%%%%
\section{Summary}
\def\theequation{\arabic{section}.\arabic{equation}}
\setcounter{equation}{0}
\label{sec:summ}

In this paper, we have considered the interactions between two nucleons at N$^3$LO
in chiral effective field theory. The pertinent results of this study can be summarized as
follows:
\begin{itemize}
\item[i)]The two--nucleon potential at N$^3$LO consists of one-, two- and three-pion
exchanges and a set of contact interactions with zero, two and four derivatives, respectively, 
according to the chiral power counting, see also table~\ref{tab:isosp}. 
We have applied spectral function regularization to the
multi-pion exchange contributions. This allows for a better separation of the low and high momentum
components in the pion loop diagrams than dimensional regularization. Within this framework, 
we have shown that three-pion exchange can safely be neglected. The
corresponding cut--off is varied from 500 to 700 MeV. The LECs 
related to the dimension two and three $\bar NN\pi\pi$ 
vertices are taken consistently from studies of pion-nucleon scattering in
chiral perturbation theory, \cite{Fet98,Paul}.
In the isospin limit, there are 24 LECs related to four--nucleon 
interactions which feed into the
S--, P-- and D--waves and various mixing parameters, cf. Eq.~(\ref{VC}).
\item[ii)]We have reviewed
the various isospin breaking mechanisms and
proposed a novel ordering scheme, based on one small parameter that collects
strong as well as electromagnetic isospin violation,
cf. Eq.~(\ref{CountRules1}) accompanied by a particular counting rule for
photon loops, see Eq.~(\ref{CountRules2}). This differs from the scheme
proposed and applied in Ref.~\cite{WME}. In the actual calculations, we have
included the leading charge-independence and charge-symmetry breaking
four--nucleon operators, the pion mass difference in the 1PE, 
the kinematical effects due to the nucleon mass difference 
and the same electromagnetic corrections as done by
the Nijmegen group  (the static Coulomb potential and various corrections to
it, magnetic moment interactions and vacuum polarization). This is done because
we fit to the Nijmegen partial waves. In the future, it would be important to
also include isospin violation in the 2PE, $\pi\gamma$-exchange and the isospin
breaking corrections to the pion-nucleon scattering amplitude (which have been
consistently determined in \cite{fet01}).
\item[iii)]We have discussed in some detail the form of the scattering equation
that is used to iterate the potential and similar for the bound state. We use
the Lippmann-Schwinger equation with the relativistic form of the kinetic
energy. Such an approach can easily be extended to external probes or
few--nucleon systems. We have also discussed the reduction to a
nonrelativistic form which be might of easier use in some applications.
The LS equation is regulated in the standard way, cf. Eq.~(\ref{pot_reg}), 
with the cut-off varied from 450 to 600~MeV.
\item[iv)]The total of 26 four--nucleon LECs has been determined by
a combined fit  to some $np$ and $pp$ phase shifts from the Nijmegen analysis
together with the $nn$ scattering length value $a_{nn} = -18.9\,$fm, as 
detailed in section~\ref{sec:fit}. The resulting LECs are of natural
size except $D_{1S0}^1$ and $D_{3S1}^1$. Comparing to the fits at NLO and NNLO, we had to extend the
fit range to higher energies for the reasons discussed in
section~\ref{sec:fit}.
\item[v)] The description of the low phase shifts (S, P, D) is excellent, see
Figs.~\ref{fig1}-\ref{fig3}. In all cases, the N$^3$LO result is better 
than the NNLO one with a sizeably reduced theoretical uncertainty. This 
holds in particular for the problematic $^3P_0$ wave which was not well 
reproduced at NNLO. The peripheral waves (F, G, H, $\ldots$),
that are free of parameters, are also well described with the
expected theoretical uncertainty related to the cut--off variations,
see Figs.~\ref{fig4}-\ref{fig5}. We stress that the description of the
phases in general improves when going from LO to NLO to NNLO to N$^3$LO,
as it is expected in a converging EFT. 
\item[vi)] The resulting S-wave scattering lengths and range parameters in the $np$
(cf. tables \ref{tab1} and \ref{tab2}) and $pp$ systems (cf. Eq.~(\ref{ere_pp}))
are in good agreement with the ones obtained in the Nijmegen PWA. In addition,
we can give theoretical uncertainties for all these quantities, which are
mostly in the one percent range.
\item[vii)] The  scattering observables (differential cross sections,
analyzing powers) for the $np$ system displayed in 
Figs.~\ref{fig6}-\ref{fig9} are well described, with a small theoretical
uncertainty at the order considered here.
\item[viii)] The deuteron properties are further predictions. In particular,
we have not included the binding energy in the fits, the deviation from the
experimental value is in the range from 0.4 to 0.07$\%$. The asymptotic S-wave
normalization and the asymptotic $D/S$ are also well described. The remaining
discrepancies in the quadrupole moment and the rms matter radius are related
to the short-ranged two-nucleon current not considered here.
\end{itemize}

\noindent
In the future, these studies should be extended in various directions. In
particular, one should construct the electroweak current operators to the
same accuracy and work out the corresponding three--nucleon force, which
is of special interest since it does not contain any novel LECs. Furthermore,
a more systematic study of isospin violation in the two-- and three--nucleon
systems based on the formalism developed here should be pursued. Work along
these lines is under way.

%%%%%%%%%%%%%%%%%%%%%%%%%%%%%%%%%%%%%%%%%%%%%%%%%%%%%%%%%%%%%%%%%%%%%%%%%%%%%%%%%
\section*{Acknowledgments}

We are grateful to Franz Gross, Barry Holstein, Rocco Schiavilla and Bob Wiringa
for several discussions and comments. E.E.~also acknowledge the hospitality of the 
Institute for Nuclear Theory at the University of Washington, Seattle,
where part of this research was conducted.  This work has been supported by the 
U.S.~Department of Energy Contract No.~DE-AC05-84ER40150 under which the 
Southeastern Universities Research Association (SURA) operates the Thomas Jefferson 
National Accelerator Facility.

\bigskip

%%%%%%%%%%%%%%%%%%%%%%%%%%%%%%%%%%%%%%%%%%%%%%%%%%%%%%%
\appendix
\def\theequation{\Alph{section}.\arabic{equation}}
\setcounter{equation}{0}
\section{Kinematics}
\label{sec:kinem}

Consider two nucleons moving with momenta $\vec p_1$ and $\vec p_2$.
We use the relativistic kinematics for relating the energy $E_{\rm lab}$ of two nucleons
in the laboratory system to the square of the nucleon momentum $\vec p$ in the center--of--mass system, which
is defined by the condition $\vec p_1 + \vec p_2 =0$.\footnote{It would be more 
appropriate to call such a system center--of--momenta or rest--frame and not center--of--mass
as usually done in the literature.} The relation between  $E_{\rm lab}$ and $\vec{p} \, ^2$ reads
(here and in what follows: $p \equiv | \vec p |$):
\begin{itemize}
\item
Proton--proton case:
\beq
p^2 = \frac{1}{2} m_p E_{\rm lab}\,.
\eeq
\item
Neutron--neutron case:
\beq
p^2 = \frac{1}{2} m_n E_{\rm lab}\,.
\eeq
\item
Neutron--proton case:
\beq
p^2 = \frac{m_p^2 E_{\rm lab} (E_{\rm lab} + 2 m_n)}{(m_n+m_p)^2 + 2 E_{\rm lab} m_p}\,.
\eeq
\end{itemize}
\noindent
The relativistic Schr\"odinger equation for two protons or two neutrons 
in the c.m.~system reads:
\beq
\label{schroed_rel}
\left[ \left( 2 \sqrt{p^2 + m}  - 2 m \right) + V \right] \Psi
= E \Psi\,,
\eeq
where $m$ is the proton or neutron mass.
For the neutron--proton system it takes the form 
\beq
\left[ \left( \sqrt{p^2 + m_n^2} + \sqrt{p^2 + m_p^2} - m_n - m_p \right) + V \right] \Psi
= E \Psi\,.
\eeq
The free Hamiltonian $H_0$ can be expressed in terms of the mass $m$ defined as
\beq
m = \frac{ 2 m_p m_n}{m_p + m_n}\,,
\eeq
in the following way
\beq
\label{kinet_approx}
H_0 = \sqrt{p^2 + m_n^2} + \sqrt{p^2 + m_p^2} - m_n - m_p \simeq  2 \sqrt{p^2 + m^2} - 2 m  
% +(m_p - m_n ) \, \mathcal{O} 
%\left( \frac{m_p - m_n}{m_p + m_n} \right)
\,,
\eeq
modulo terms which are proportional to $(m_p - m_n )^2$. Taking into account such terms goes beyond the 
accuracy of the present analysis. We will therefore use the approximate expression
(\ref{kinet_approx}) in this work, which leads to the Schr\"odinger equation of the type
(\ref{schroed_rel}).

\section{Partial wave decomposition of the $NN$ potential}
\setcounter{equation}{0}
\label{sec:pw}

In this appendix we  describe the partial wave decomposition of the two--nucleon potential.
For that we first rewrite the potential $V$ in the form
\begin{eqnarray}
\label{pot_dec}
V &=& V_C + V_\sigma \; \vec{\sigma}_1 \cdot \vec{\sigma}_2 + V_{SL} \; i \; \frac{1}{2} 
(\vec{\sigma}_1 + \vec{\sigma}_2 )  \cdot
( \vec{k} \times \vec{q} )  + V_{\sigma L} \; \vec{\sigma}_1 \cdot ( 
\vec{q} \times
\vec{k} ) \; \vec{\sigma}_2 \cdot ( \vec{q} \times \vec{k} ) \nonumber \\
&& {} + V_{\sigma q} \; ( \vec{\sigma}_1 \cdot \vec{q} ) \; (\vec{\sigma}_2 \cdot
\vec{q} ) + V_{\sigma k} \; (\vec{\sigma}_1 \cdot \vec{k}) \; (\vec{\sigma}_2 \cdot \vec{k})~,
\end{eqnarray}
with six functions $V_C (p, p', z), \; \ldots , \; V_{\sigma k} (p, p', z)$
depending on $p \equiv | \vec{p}\,|$, $p' \equiv | \vec{p} \, '|$ and
the cosine of the angle between the two momenta is called $z$.
%$z \equiv \cos ( \widehat{\vec{p} \vec{p} \, '})$. 
These functions may depend on the isospin matrices $\fet{\tau}$ as well. 
To perform the partial wave decomposition of $V$, i.~e.~to express it in the 
standard $lsj$ representation, we have followed the steps of ref.~\cite{Er71}.
In particular, we start from the helicity state representation $|\hat{p} \, \lambda_1  \lambda_2 \rangle$,
where $\hat{p} = \vec{p}/p$ and $\lambda_1$ and $\lambda_2$ are 
the helicity quantum numbers corresponding to nucleons 1 and 2,
respectively. We then expressed the potential in the $| j m \lambda_1 \lambda_2 \rangle$ representation
using the transformation matrix 
$\langle  \hat{p} \, \lambda_1 \lambda_2  | j m \lambda_1 \lambda_2 \rangle$, given
in ref.~\cite{Er71}. The final step is to switch to the $| lsj \rangle$ representation.
The corresponding transformation matrix $\langle lsjm| jm\lambda_1 \lambda_2 \rangle$ is given in refs.~\cite{Ja59},
\cite{Er71}.

\medskip\noindent
For $j>0$, we obtain the following expressions for the non--vanishing matrix elements in the $| lsj\rangle$ 
representation:
\begin{eqnarray}
\langle j0j | V | j0j \rangle &=& 2 \pi \int_{-1}^{1} \, dz \, \left\{ V_C - 3 V_\sigma
+ {p '}^2 {p}^2 (z^2 -1) V_{\sigma L} - q^2 V_{\sigma q} - k^2 V_{\sigma k} 
\right\} P_j (z)~, \nonumber \\
\langle j1j | V | j1j \rangle 
&=& 2 \pi \int_{-1}^{1} \, dz \, \left\{ \left[ V_C + V_\sigma + 2 
p' p z V_{SL} - {p'}^2 p^2 (1 + 3 z^2) V_{\sigma L} + 4 k^2 V_{\sigma q} + 
\frac{1}{4} q^2 V_{\sigma k} \right]   \right. \nonumber \\
&&   {} \times P_j (z) + \left. \left[- p' p \, V_{SL} + 
2 {p'}^2 p^2 z V_{\sigma L} -2 p' p \, (V_{\sigma q} - \frac{1}{4} V_{\sigma k} )\right] \right. \nonumber \\
&& {} \times \left( P_{j-1} (z) +
P_{j+1} (z) \right) \bigg\}~,  \nonumber 
\end{eqnarray}
\begin{eqnarray}
\label{aa13}
\langle j\pm 1,1j | V | j \pm 1, 1j \rangle &=& 2 \pi \int_{-1}^{1} \, dz \, \left\{
p' p \left[ -V_{SL}  \pm \frac{2}{2j+1} \left( - p' p z V_{\sigma L} + V_{\sigma q} -
\frac{1}{4} V_{\sigma k} \right) \right] \right.  \nonumber \\
&&  {} \times P_j (z) + \bigg[ V_C + V_\sigma + p' p z V_{SL} + {p'}^2 p^2 (1-z^2) V_{\sigma L} \nn
&&  {} \left. \left. \pm\frac{1}{2j+1} \left( 2 {p'}^2 p^2 V_{\sigma L} - 
({p'}^2 + p^2)(V_{\sigma q} +
\frac{1}{4} V_{\sigma k} ) \right) \right] P_{j\pm 1} (z) \right\}  \nonumber \\
\langle j\pm 1,1j | V | j \mp 1, 1j \rangle &=& \frac{\sqrt{j(j+1)}}{2j+1} 2 \pi \int_{-1}^{1} \, dz \, 
\Bigg\{- p' p \, (4 V_{\sigma q} - V_{\sigma k} ) P_j (z)~,  \nonumber \\
&& {} + \left[\mp \frac{2 {p'}^2 p^2}{2j+1} V_{\sigma L} + {p'}^2 \Big(2 V_{\sigma q}+\frac{1}{2}
V_{\sigma k}\Big) \right] P_{j\mp 1} (z)  \nonumber \\
&& {} \left. + \left[\pm \frac{2 {p'}^2 p^2}{2j+1} V_{\sigma L} + {p}^2 \Big(2 V_{\sigma q}+\frac{1}{2}
V_{\sigma k}\Big) \right] P_{j\pm 1} (z) \right\}~.  
\end{eqnarray} 
Here, $P_j (z)$ are the conventional Legendre polynomials.
For $j=0$ the two non--vanishing matrix elements are
\begin{eqnarray} 
\label{aa14}
\langle 000 | V | 000 \rangle &=& 2 \pi \int_{-1}^{1} \, dz \, \left\{
V_C - 3 V_\sigma
+ {p '}^2 {p}^2 (z^2 -1) V_{\sigma L} - q^2 V_{\sigma q} - k^2 V_{\sigma k} 
\right\}~, \nonumber \\
\langle 110 | V | 110 \rangle &=& 2 \pi \int_{-1}^{1} \, dz \, \bigg\{ z V_C
+ z V_\sigma + p' p (z^2 -1) V_{SL} + {p'}^2 p^2 z (1-z^2) V_{\sigma L}  \nn
&& {}\left.  - \left( ( {p'}^2 + p^2 )
z - 2 p' p \right)  V_{\sigma q} - \frac{1}{4} 
\left( ( {p'}^2 + p^2 )
z + 2 p' p \right) V_{\sigma k}  \right\}~.
\end{eqnarray}
Note that sometimes another notation is used in which an additional overall  
minus sign enters the expressions for the 
off--diagonal matrix elements with $l=j+1, \, l'=j-1$ and $l=j-1, \, l'=j+1$.

\section{Momentum space treatment of the Coulomb interaction}
\setcounter{equation}{0}
\label{sec:coul}

In this appendix we would like to explain our way of treating the
nucleon--nucleon scattering problem in the presence of the 
Coulomb interaction in momentum space (following closely Ref.~\cite{WME}).
The starting point is the nonrelativistic Schr\"odinger equation of the 
form (\ref{schr_nr1}) or (\ref{Schroed_nonrel}), where the 
the potential consists of two pieces: the short--range one given 
by the strong interaction and the long--range one given by the Coulomb force.

\medskip\noindent
As the Coulomb potential is of infinite range, the S--matrix has 
to be formulated in terms of asymptotic Coulomb states. Therefore, the
phase shifts for a given angular momentum $l$
due to the strong potential in the presence of the
long--range electromagnetic interactions, denoted by $\delta_l^l$,
are defined in terms of a linear combination of (ir)regular 
Coulomb--functions $F(G)$ as
\beq\label{ucwave}
\chi_{l}^l (r) = F_{l}(r) + \tan(\delta_l^l) \, G_{l}(r)
\eeq
analogously to the expression for an arbitrary potential of short range 
(i.e. in the absence of the Coulomb force)
\beq\label{plainasym}
\chi_{l}^s (r) = F^0_{l}(r) + \tan(\delta_l^s) \, G^0_{l}(r)
\eeq
with $F^0$, $G^0$ denoting solutions of the Coulomb problem with zero
charge (conventionally expressed in terms of Bessel and Neumann
functions) and the corresponding phase shift is called $\delta_l^s$.
So far, we have restricted ourselves to uncoupled channels. We will
consider the coupled case later on.

\medskip\noindent
As eq.(\ref{ucwave}) exhibits  asymptotical Coulomb-states, 
we have to re-express our Lippmann-Schwinger-equation
in terms of them. A very convenient scheme for inclusion of the Coulomb 
force in momentum space was suggested long time ago by Vincent and Phatak 
\cite{VP} and  is
used in the present analysis. In what follows we will briefly   
describe this approach.

\medskip\noindent
The starting point of this technique is the observation, that for a potential of the form
\begin{equation}\label{VsC}
V = V_{\rm C} + V_{\rm S}
\end{equation}
with 
\begin{equation}\label{cond}
V_{\rm S}\,\psi_{l} = 0 \,\,\,(r\ge R)~,
\end{equation}
and $\psi_{l}$ the two--nucleon wave-function for a given angular momentum, 
two exact solutions for the wave-function can be given for every point
on a sphere with radius $R+\epsilon$. One is of the form as in eq.(\ref{ucwave}),
and another one according to eq.(\ref{plainasym}) with 
the phase shifts calculated for the following potential as in eq.(\ref{VsC}),
with V$_{\rm C}$, however,  being the Fourier-transformed Coulomb-potential
integrated to the radius $R$, 
\beqa\label{Coulombpot}
V_{\rm C}(\mid \vec{q}\,'- \vec{q}\mid) &=&  
\int_{0}^{R}d^{3}r \,  e^{i(\vec{q}\,'-\vec{q}\,) \cdot
\vec{r}}\,\,\frac{\alpha}{r} \nonumber\\
                               &=& 
\frac{4\pi \alpha}{\mid \vec{q}~'-\vec{q}\mid^{2}}
(1-\cos(\mid \vec{q}~'-\vec{q}\mid R))~.
\eeqa
Here, $\vec{q}, \vec{q}\,'$ are the cms momenta and $\alpha$ is the fine-structure
constant. On the above-defined sphere, both wave functions describe
the same system. 
Now we know how to obtain an expression for the strong phase shift
$\delta_l^l$ in the presence of the Coulomb interaction
in terms of the short-range shift $\delta_l^s$ in the absence of
electromagnetism: We only have to match the two solutions. 
This is most conveniently done by requiring the logarithmic derivative
of both solutions to be equal, what enables us to express the strong
shift in the presence of the Coulomb force in a Wronskian form:
\beq
\tan(\delta_l^l)\,=\,\frac{\tan(\delta_l^s)\lbrack 
F,G_0\rbrack + \lbrack F,F_0\rbrack}{\lbrack F_0,G\rbrack 
+ \tan(\delta_l^s)\lbrack G_0,G\rbrack}
\eeq
with
\beq
\lbrack F,G \rbrack\,=\,\biggl(G\frac{dF}{dr}-F\frac{dG}{dr}\biggr)_{r=R}
\eeq

\medskip\noindent
Let us now extend the previous consideration to the coupled case. For that we 
replace eq.~(\ref{ucwave}) by the matrix equation:
\beq\label{ucwave_coup}
\chi^l (r) = F(r) - m q K^l  G(r)\,,
\eeq
where $K^l$ is the K--matrix for the strong potential in the presence of the Coulomb interaction, 
$\chi^l (r)$ is the $2 \times 2$ matrix which contains the 
wave functions
\beq\label{coul_cop}
\chi^l (r) = \left( \begin{array}{cc} \chi^l_{j-1, \, j-1} (r)  &  \chi^l_{j-1, \, j+1} (r)\\
 \chi^l_{j+1, \, j-1} (r)  &  \chi^l_{j+1, \, j+1} (r) \end{array} \right)\,, 
\eeq
and $F(r)$ and $G(r)$ are the  $2 \times 2$ matrices which contain the 
Coulomb wave functions
\beq
F(r) = \left( \begin{array}{cc} F_{j-1} (r)  &  0\\
 0  &  F_{j+1} (r) \end{array} \right)\,, \quad \quad
G(r) = \left( \begin{array}{cc} G_{j-1} (r)  &  0\\
 0  &  G_{j+1} (r) \end{array} \right)\,.
\eeq
All subscripts in the above equations refer to the values of the angular momentum $l$.
Analogously, the equation (\ref{plainasym}) has to be replaced by 
\beq\label{plainasym_coup}
\chi^s (r) = F^0(r) - m q K^s  G^0(r)\,,
\eeq
where  $K^s$ is the K--matrix for the strong potential only. 

\medskip\noindent
Matching now the wave function $\chi^l (r)$ with $\chi^s (r)$, 
calculated from the potential $V$ defined in eqs.~(\ref{VsC}) and (\ref{Coulombpot}),
at some radius $R$ by equating the corresponding logarithmic derivatives 
as described above one obtains for the K--matrix $K^l$:
\beq
K^l = 
\frac{1}{mq} \left[ F (F_0 - m q K^s G_0 )^{-1} (F_0 ' - m q K^s G_0 ' ) - F ' \right]
\left[ G (F_0 - m q K^s G_0 )^{-1} (F_0 ' - m q K^s G_0 ' ) - G ' \right]^{-1}\,.
\eeq

%In the present analysis we have also incorporated relativistic corrections to 
%the Coulomb interaction in the same way as it is done by the Nijmegen group. As 
%already pointed out in text, this is acieved by replacing the usual nonrelativistic 
%Coulomb potential by the energy--dependent one of the form \cite{austin}
%\beq
%\label{impr_coul}
%V_C ' = \frac{\alpha '}{r}, \quad \quad \quad \mbox{where} \quad \alpha ' = \alpha 
%\frac{m_p^2 + 2 k^2}{m_p \sqrt{m_p^2 + k^2}}\,.
%\eeq
%Here $k$ is the absolute value of the c.m.~momentum of two protons.
%The modified
%Coulomb potential in eq.~(\ref{impr_coul}) can be treated in momentum space in the 
%same way as the standard Coulomb potential as described  
%above.\footnote{Clearly, one has to use the appropriately adjusted regular and 
%irregular Coulomb functions $F_l (r)$ and $G_l (r)$.} 
%Notice that the relativistic corrections given in eq.~(\ref{impr_coul}) are 
%designed to be used in eq.~(\ref{Schroed_nonrel}) which assumes relativistic 
%konematics in relating the c.m.s.~energy and momentum. In the present 
%analysis we therefore first compute the corresponding ``nonrelativistic'' strong potential 
%using eq.~(\ref{pot_nonrel}). The integration is performed numerically using
%Gauss--Legendre quadrature. We then add the modified Coulomb force  to the 
%resulting potential and solve the corresponding Lippmann--Schwinger equation in 
%momentum space following the lines of this appendix.  

\medskip \noindent
The only remaining difficulty is the determination of the
matching radius $R$, because the
given solution is wrong as long as (\ref{cond}) is not valid. On the
other hand, it is not possible to extend $R$ to arbitrarily 
large values, because the 
cosine in eq.(\ref{Coulombpot}) will cause rapid
oscillations. $R \sim 10$ fm turns out to be a good choice, see
\cite{WME}. We use the value $R=12$ fm in the present analysis.

\section{Effective range expansion}
\setcounter{equation}{0}
\label{sec:efr}

In this appendix we collect the formulae for the S--wave effective range expansion.
In the simplest case of the scattering with the finite--range potential, the 
quantity $k \cot ( \delta_0 )$, where $\delta_0$ is the S--wave phase shift and 
$k$ is the c.m.s.~momentum, is well known 
to have the low--momentum (or effective range)  expansion:
\beq
\label{ere1}
k \cot (\delta_0 )   = -\frac{1}{a} + \frac{1}{2} r k^2 + 
v_2 k^4 + v_3 k^6  + v_4 k^8 +\mathcal{O} (k^{10})\,.
\eeq
Here $a$ is the scattering length, $r$ the effective range and $v_{2,3,4}$
the shape parameters. 

\medskip\noindent
In the presence of the long--range potential the effective range expansion has to be 
modified. In that case one usually defines an effective range function
instead of the quantity $k \cot ( \delta_0 )$, 
in which the left--hand singularities due to the long--range interaction are 
removed, see reference \cite{Berg88} for more details. 
In the case of the modified Coulomb potential given in eq.~(\ref{modCoul}),
the effective range function $F_C$ takes the form \cite{Berg88}
\beq
F_C = C_0^2 (\eta ') \, k \, \cot (\delta_0^{C} ) + 2 k \, \eta ' \, h (\eta ' )\,,
\eeq
where the quantity $\eta '$ is given by
\beq
\eta ' = \frac{m_p}{2 k} \alpha ' \,,
\eeq
and the functions $C_0^2 (\eta ')$ (the Sommerfeld factor) and $h (\eta ' )$ read
\beq
C_0^2 (\eta ') = \frac{2 \pi \eta '}{e^{2 \pi \eta '} - 1} \,,  \quad  \quad  \mbox{and} \quad  \quad 
h (\eta ' ) = {\rm Re} \Big[ \Psi ( 1 + i \eta ' ) \Big] - \ln (\eta ' ) \,.
\eeq
Here, $\Psi$ denotes the digamma function. Notice that the phase shift $\delta_0^C$ is the 
S--wave phase shift of the finite--range plus Coulomb potential with respect to 
Coulomb wave functions. In the notation of section \ref{sec:fit}, $\delta_0^C$ should be written
as $\delta_{\rm C1 + N}^{\rm C1}$.
The effective range expansion for the function $F_C$ is
\beq
\label{ere2}
F_C =  -\frac{1}{a^C} + \frac{1}{2} r^C k^2 + 
v_2^C k^4 + v_3^C k^6  + v_4^C k^8 +\mathcal{O} (k^{10})\,.
\eeq

\medskip\noindent
In a general case of an arbitrary long--range interaction, the effective 
range function may be obtained e.g.~along the lines of Ref.~\cite{hink71}
provided that the long--range potential is weak enough to be treated perturbatively. 
For more discussion on the effective range expansion in presence of electromagnetic
interaction the reader is referred to \cite{Berg88}.

\bigskip\bigskip\bigskip
%%%%%%%%%%%%%%%% REFS %%%%%%%%%%%%%%%%%%%%%%%%%%%%%%%%%%%%%%%%%%%%%%%%

%\newpage

\pagebreak

\centerline {{\large \bf TABLES}}

\begin{table*}[htb] 
\vspace{1.cm}
\begin{center}
\begin{tabular}{||p{5.1cm}|p{5.1cm}|p{5.1cm}||}
\hline \hline 
{} & {} &  {} \\[-1.5ex]
\centerline{Isospin--symmetric}   &  Isospin--breaking, finite--range  & Isospin--breaking, long--range\\[-1ex]
\hline \hline 
{} & {} &  {} \\[-1.5ex]
{\bf LO} ($\nu = 0$): 

static 1PE, contact terms without derivatives 

& 

\vskip 0.3 true cm
\centerline{$-$} & 
\vskip 0.3 true cm
\centerline{$-$}
\\
{\bf NLO} ($\nu = 2$): 

leading 2PE, contact terms with 2 derivatives

&  
{$\fet \LOI$} ($\nu = 2$):

$M_{\pi^\pm} \neq M_{\pi^0}$ in 1PE

& 
{$\fet \LOI$} ($\nu = 2$):

static 1$\gamma$--exchange
\\
{\bf NNLO} ($\nu = 3$):

subleading 2PE

& 
{\bf N$\fet \LOI$} ($\nu = 3$):

isospin breaking in 1PE ($\propto \epsilon M_\pi^2$), 
contact term without derivatives $\propto \epsilon M_\pi^2$

&

\vskip 0.2 true cm
\centerline{$-$}\\
{\bf N$^3$LO} ($\nu = 4$):

sub--subleading 2PE, leading 3PE, 
$1/m^2$--corrections to 1PE, $1/m$--corrections to 2PE,
contact terms with 4 derivatives

& 

{\bf NN$\fet \LOI$} ($\nu = 4$):

isospin breaking in 1PE ($\propto e^2/(4 \pi)^2$), 
$M_{\pi^\pm} \neq M_{\pi^0}$ in 2PE, triangle and football 2PE diagrams $\propto c_5$, $\pi \gamma$--exchange,
$m_n \neq m_p$ in 2PE and in the LS equation,
contact terms without derivatives $\propto e^2/(4 \pi)^2$

& 
\vskip 1.5 true cm
\centerline{$-$}\\

\vskip 0.3 true cm
\centerline{$\ldots$}    & 
\vskip 0.3 true cm
\centerline{$\ldots$} & 
{\bf N$\fet{^4 \LOI}$} ($\nu = 6$):

$1/m^2$--corrections to the static 1$\gamma$--exchange,
2$\gamma$--exchange. \\[1ex]
\hline \hline
  \end{tabular}
\vskip 0.3 true cm
\parbox{16.5cm}{
\caption{Dominant contributions to the isospin--symmetric and isospin--breaking parts of the 
two--nucleon force. \label{tab:isosp}}
}
\end{center}
\end{table*}

\pagebreak

\begin{table*}[htb] 
\begin{center}
\begin{tabular}{||c||r|r|r|r||}
\hline \hline
  &  &  &  &  \\[-1.5ex]
LEC   &  $\{ 450, \; 500\}$   &  $\{ 600, \; 600\}$  &  $\{ 450, \; 700\}$  &  $\{ 600, \; 700\}$ \\[0.8ex]
\hline \hline 
  &  &  &  &  \\[-1.5ex]
$\tilde C_{1S0}^{\rm pp}$ &  $-0.0834$  & $-0.0800$  &  $-0.1247$  &  $-0.0436$ \\[0.2ex]
$\tilde C_{1S0}^{\rm np}$ &  $-0.0913$  & $-0.0892$  &  $-0.1289$  &  $-0.0544$ \\[0.2ex]
$\tilde C_{1S0}^{\rm nn}$ &  $-0.0880$  & $-0.0851$  &  $-0.1272$  &  $-0.0494$ \\[0.2ex]
$C_{1S0}$        &   $1.5007$  &  $1.8075$  &   $2.1217$  &   $1.8950$ \\[0.2ex]
$D_{1S0}^1$      & $-26.9836$  & $-16.7678$  & $-24.7288$  & $-17.6295$ \\[0.2ex]
$D_{1S0}^2$      &   $3.7402$  & $-2.5565$  &   $0.8214$  &  $-2.0771$ \\[1ex]
\hline
  &  &  &  &  \\[-1.5ex]
$\tilde C_{3S1}$ &  $-0.1498$  &   $0.1782$  &  $-0.1599$  &   $0.0746$ \\[0.2ex]
$C_{3S1}$        &   $0.4144$  &  $-0.9058$  &   $0.6275$  &  $-0.3557$ \\[0.2ex]
$D_{3S1}^1$      & $-26.3516$  & $-13.4902$  & $-23.8555$  & $-12.4078$ \\[0.2ex]
$D_{3S1}^2$      &   $4.8091$  &   $2.6661$  &   $4.3807$  &   $1.8895$  \\[1ex]
\hline \hline
\end{tabular}
\vskip 0.3 true cm
\parbox{16.5cm}{
\caption{The S---wave LECs $\tilde C_i$, $C_i$ and $D_i$ at N$^3$LO 
for the different cut--off combinations 
$\big\{ \Lambda \, [\mbox{MeV}], \; \tilde \Lambda \, [\mbox{MeV}] \big\}$.
The values of the $\tilde C_i$ are in $10^4$ GeV$^{-2}$, of the $C_i$ in $10^4$ GeV$^{-4}$ 
and of the $D_i$ in $10^4$ GeV$^{-6}$. \label{tab:LEC1}}
}
\end{center}
\end{table*}

\begin{table}[H] 
\begin{center}
\begin{tabular}{||c||r|r|r|r||}
\hline \hline
  &  &  &  &  \\[-1.5ex]
LEC   &  $\{ 450, \; 500\}$   &  $\{ 600, \; 600\}$  &  $\{ 450, \; 700\}$  &  $\{ 600, \; 700\}$ \\[0.8ex]
\hline \hline 
  &  &  &  &  \\[-1.5ex]
$C_{1P1}$      &   $0.1862$  &  $0.3374$  &   $0.2072$  &   $0.3444$ \\[0.2ex]
$D_{1P1}$      &   $2.3257$  &  $1.9180$  & $2.3968$  & $1.9213$ \\[1ex]
\hline
  &  &  &  &  \\[-1.5ex]
$C_{3P0}$      &   $1.1729$  &  $1.2034$  &   $1.1913$  &   $1.2031$ \\[0.2ex]
$D_{3P0}$      &   $1.0892$  &  $1.2500$  &   $1.2190$  &   $1.4116$ \\[1ex]
\hline
  &  &  &  &  \\[-1.5ex]
$C_{3P1}$      &   $-0.6334$  &  $-0.6602$  &   $-0.7576$  &   $-0.7193$ \\[0.2ex]
$D_{3P1}$      &   $4.2369$   &  $3.8465$   &   $4.2099$   &   $3.8756$ \\[1ex]
\hline
  &  &  &  &  \\[-1.5ex]
$C_{3P2}$      &   $-0.5542$  &  $-0.5812$  &   $-0.6217$  &   $-0.6114$ \\[0.2ex]
$D_{3P2}$      &   $4.1956$   &  $4.2270$   &   $4.0340$   &   $4.1723$ \\[1ex]
\hline
  &  &  &  &  \\[-1.5ex]
$C_{\epsilon 1}$        &   $-0.4516$  &  $-0.2726$  &   $-0.5045$  &   $-0.3352$ \\[0.2ex]
$D_{\epsilon 1}$        &   $2.6303$   &  $1.7686$   &   $2.0296$   &   $1.5516$ \\[1ex]
\hline \hline
\end{tabular}
\vskip 0.3 true cm
\parbox{16.5cm}{
\caption{The LECs $C_i$ and $D_i$ in the P--waves and $\epsilon_1$ 
at N$^3$LO for the different cut--off combinations 
$\big\{ \Lambda \, [\mbox{MeV}], \; \tilde \Lambda \, [\mbox{MeV}] \big\}$.
The values of the $C_i$ ($D_i$) are in $10^4$ GeV$^{-4}$ 
($10^4$ GeV$^{-6}$). \label{tab:LEC2}}
}
\end{center}
\end{table}

\begin{table}[H] 
\begin{center}
\begin{tabular}{||c||r|r|r|r||}
\hline \hline
  &  &  &  &  \\[-1.5ex]
LEC   &  $\{ 450, \; 500\}$   &  $\{ 600, \; 600\}$  &  $\{ 450, \; 700\}$  &  $\{ 600, \; 700\}$ \\[0.8ex]
\hline \hline 
  &  &  &  &  \\[-1.5ex]
$D_{1D2}$      &   $-2.2450$  &  $-2.1874$  & $-2.3398$  & $-2.2203$ \\[1ex]
\hline
  &  &  &  &  \\[-1.5ex]
$D_{3D1}$      &   $-1.3988$  &  $-1.7483$  &   $-1.2250$  &   $-1.6620$ \\[1ex]
\hline
  &  &  &  &  \\[-1.5ex]
$D_{3D2}$      &   $-1.4180$   &  $-0.9023$   &   $-1.3578$   &   $-0.8580$ \\[1ex]
\hline
  &  &  &  &  \\[-1.5ex]
$D_{3D3}$      &   $-2.0792$   &  $-1.5493$   &   $-1.7522$   &   $-1.4841$ \\[1ex]
\hline
  &  &  &  &  \\[-1.5ex]
$D_{\epsilon 2}$        &   $0.2333$   &  $0.2901$   &   $0.2274$   &   $0.2892$ \\[1ex]
\hline \hline
\end{tabular}
\vskip 0.3 true cm
\parbox{16.5cm}{
\caption{The D--wave LECs $D_i$ at N$^3$LO for the different cut--off combinations 
$\big\{ \Lambda \, [\mbox{MeV}], \; \tilde \Lambda \, [\mbox{MeV}] \big\}$.
The values of the $D_i$ are in $10^4$ GeV$^{-6}$. \label{tab:LEC3}}
}
\end{center}
\end{table}

\begin{table*}[htb] 
\begin{center}
\begin{tabular}{||c||c|c|c||r|r|r||}
\hline \hline
 {} &   \multicolumn{3}{|c||}{}  &   \multicolumn{3}{c||}{}  \\[-1.5ex]
    &   \multicolumn{3}{|c||}{Chiral N$^3$LO}  &   \multicolumn{3}{c||}{Nijmegen PWA} \\[1ex]
\hline  
{} & {} &  {} & {}    & {} &  {} & {}            \\[-1.8ex]
 $i$   & $\Delta^\gamma_i$ & $\Delta^\pi_i$ & $\Delta_i$ 
    & $\Delta^\gamma_i$ & $\Delta^\pi_i$ & $\Delta_i$   \\[0.8ex]
\hline \hline
{} & {} &  {} & {}    & {} &  {} & {}            \\[-1.5ex]
 $^3P_0$ & $0.359 \ldots  0.360$ & $-0.456 \ldots -0.454$ & $-0.096 \ldots -0.094$ & $0.371$  & $-0.447$ & $-0.076$   \\  
 $^3P_1$ & $-0.187 \ldots -0.184$ & $ 0.191 \ldots  0.192$ & $ 0.005 \ldots  0.008$ & $-0.186$ & $0.183$  & $-0.003$   \\  
 $^3P_2$ & $0.092 \ldots  0.093$ & $-0.031 \ldots -0.030$ & $ 0.061 \ldots  0.062$ & $0.092$  & $-0.035$ & $0.057$   \\  
 $^1D_2$ & $0.014 \ldots  0.015$ & $-0.024 \ldots -0.024$ & $-0.010 \ldots -0.009$ & $0.014$  & $-0.023$ & $-0.009$   \\[1ex]
\hline \hline
  \end{tabular}
\vspace{0.1cm}
\caption{Effects (in degrees) on the phase shifts $\delta_i$, $i = \{ ^3P_0, \; ^3P_1, \; ^3P_2, \;  ^1D_2 \}$,  due to 
removal of the Coulomb interactions ($\Delta^\gamma_i$)
and subsequently including the pion mass differences in the 1PE potential
($\Delta^\pi_i$) at $E_{\rm lab} = 10$ MeV. 
The shifts $\Delta^\gamma_i$, $\Delta^\pi_i$ and $\Delta_i$ are defined as follows: $\Delta^\gamma_i = (\bar \delta_{\rm pp})_i
- (\delta_{\rm pp})_i$, $\Delta^\pi_i = (\delta_{\rm np})_i - (\bar \delta_{\rm pp})_i$,
$\Delta_i \equiv \Delta^\gamma_i + \Delta^\pi_i = (\delta_{\rm np})_i - (\delta_{\rm pp})_i$ and $(\bar \delta_{\rm pp})_i$
denotes the {\it pp} phase shifts calculated in the absence of the Coulomb interaction. 
The Nijmegen PWA results are from \cite{nijpwa}. The cut--offs $\Lambda$ and $\tilde \Lambda$ are varied 
as specified in eq.~(\ref{cutoffs}).
\label{tab1a}}
\end{center}
\end{table*}

\vskip 1 true cm
\begin{table*}[htb] 
\begin{center}
\begin{tabular}{||c||c|c|c||r|r|r||}
\hline \hline
 {} &   \multicolumn{3}{|c||}{}  &   \multicolumn{3}{c||}{}  \\[-1.5ex]
    &   \multicolumn{3}{|c||}{Chiral N$^3$LO}  &   \multicolumn{3}{c||}{Nijmegen PWA} \\[1ex]
\hline  \hline
{} & {} &  {} & {}    & {} &  {} & {}            \\[-1.8ex]
 $i$   & $\Delta^\gamma_i$ & $\Delta^\pi_i$ & $\Delta_i$ 
    & $\Delta^\gamma_i$ & $\Delta^\pi_i$ & $\Delta_i$   \\[0.8ex]
\hline
{} & {} &  {} & {}    & {} &  {} & {}            \\[-1.5ex]
 $^3P_0$ & $0.320 \ldots  0.325$ & $-0.789 \ldots -0.784$ & $-0.465 \ldots -0.464$ & $0.342$  & $-0.785$ & $-0.443$   \\  
 $^3P_1$ & $-0.222 \ldots -0.218$ & $ 0.293 \ldots  0.293$ & $ 0.071 \ldots  0.075$ & $-0.221$ & $0.275$  & $0.054$   \\  
 $^3P_2$ & $0.185 \ldots  0.190$ & $-0.099 \ldots -0.093$ & $ 0.088 \ldots  0.096$ & $0.184$  & $-0.115$ & $0.069$   \\  
 $^1D_2$ & $0.029 \ldots  0.031$ & $-0.048 \ldots -0.048$ & $-0.020 \ldots -0.017$ & $0.031$  & $-0.046$ & $-0.015$   \\[1ex]
\hline \hline
  \end{tabular}
\vspace{0.1cm}
\caption{Effects on the phase shifts $\delta_i$, $i = \{ ^3P_0, \; ^3P_1, \; ^3P_2, \;  ^1D_2 \}$,  due to 
removal of the Coulomb interactions ($\Delta^\gamma_i$)
and subsequently including the pion mass differences in the 1PE potential
($\Delta^\pi_i$) at $E_{\rm lab} = 25$ MeV. For notations see Table \ref{tab1a}.
\label{tab1b}}
\end{center}
\end{table*}

\begin{table*}[htb] 
\vspace{0.6cm}
\begin{center}
\begin{tabular}{||c||c|c|c||r|r|r||}
\hline \hline
 {} &   \multicolumn{3}{|c||}{}  &   \multicolumn{3}{c||}{}  \\[-1.5ex]
    &   \multicolumn{3}{|c||}{Chiral N$^3$LO}  &   \multicolumn{3}{c||}{Nijmegen PWA} \\[1ex]
\hline  \hline
{} & {} &  {} & {}    & {} &  {} & {}            \\[-1.8ex]
 $i$   & $\Delta^\gamma_i$ & $\Delta^\pi_i$ & $\Delta_i$ 
    & $\Delta^\gamma_i$ & $\Delta^\pi_i$ & $\Delta_i$   \\[0.8ex]
\hline
{} & {} &  {} & {}    & {} &  {} & {}            \\[-1.5ex]
 $^3P_0$ & $0.091 \ldots  0.109$ & $-0.879 \ldots -0.866$ & $-0.775 \ldots -0.770$ & $0.119$  & $-0.896$ & $-0.777$   \\  
 $^3P_1$ & $-0.236 \ldots -0.227$ & $ 0.320 \ldots  0.323$ & $ 0.084 \ldots  0.095$ & $-0.233$ & $0.297$  & $0.064$   \\  
 $^3P_2$ & $0.251 \ldots  0.260$ & $-0.175 \ldots -0.161$ & $ 0.078 \ldots  0.098$ & $0.253$  & $-0.221$ & $0.032$   \\  
 $^1D_2$ & $0.045 \ldots  0.050$ & $-0.041 \ldots -0.041$ & $ 0.003 \ldots  0.008$ & $0.049$  & $-0.034$ & $0.015$   \\[1ex]
\hline  \hline
  \end{tabular}
\vspace{0.3cm}
\parbox{16cm}{
\caption{Effects on the phase shifts $\delta_i$, $i = \{ ^3P_0, \; ^3P_1, \; ^3P_2, \;  ^1D_2 \}$,  due to 
removal of the Coulomb interactions ($\Delta^\gamma_i$)
and subsequently including the pion mass differences in the 1PE potential
($\Delta^\pi_i$) at $E_{\rm lab} = 50$ MeV. For notations see Table \ref{tab1a}.
}\label{tab1c}}
\end{center}
\end{table*}

%\vspace{1cm}

\begin{table*}[htb] 
\vspace{0.6cm}
\begin{center}
\begin{tabular}{||l||c|c|c||c||}
\hline \hline
{} & {} &  {} & {} & {}\\[-1.5ex]
                & {NLO}                 & {NNLO}                & {N$^3$LO} &{Nijmegen PWA} \\[1ex]
\hline  \hline
{} & {} &  {} & {} & {}\\[-1.5ex]
$a$ [fm]        & $-23.447 \ldots -23.522$  & $-23.497 \ldots -23.689$  & $-23.585 \ldots -23.736$  & $-$23.739  \\[1ex] 
$r$ [fm]        & $2.60 \ldots  2.62$       & $ 2.62  \ldots     2.67$  & $2.64 \ldots 2.68$  & 2.68       \\[1ex]
$v_2$ [fm$^3$]  & $-0.46 \ldots  - 0.47$    & $- 0.48 \ldots  - 0.52 $  & $-0.49 \ldots -0.51$  & $-$0.48    \\[1ex]
$v_3$ [fm$^5$]  & $4.3  \ldots   4.4$       & $ 4.0   \ldots     4.2 $  & $4.0 \ldots 4.1$  & 4.0        \\[1ex]
$v_4$ [fm$^7$]  & $- 20.7 \ldots  -21.0$    & $-19.9  \ldots   - 20.5$  & $-19.8 \ldots -20.2$  & $-$20.0    \\[1ex]
\hline  \hline
  \end{tabular}
\vspace{0.3cm}
\parbox{16cm}{
\caption{Scattering length and range parameters for the $^1S_0$ partial wave using the
NLO and NNLO potential \cite{EGMs2} compared to the N$^3$LO results and to the 
Nijmegen phase shift analysis  (PWA). The values 
$v_{2,3,4}$ are based on the {\it np} Nijm II potential and the values of the scattering length 
and the effective range are from ref.~\cite{Rentm99}.
}\label{tab1}}
\end{center}
\end{table*}

\begin{table*}[htb] 
\vspace{1.cm}
\begin{center}
\begin{tabular}{||l||c|c|c||c||}
\hline \hline
{} & {} &  {} & {} & {}\\[-1.5ex]
                & {NLO}                 & {NNLO}                & {N$^3$LO} &{Nijmegen PWA} \\[1ex]
\hline  \hline
{} & {} &  {} & {} & {}\\[-1.5ex]
$a$ [fm]        & $5.429 \ldots 5.433$  & $5.424 \ldots 5.427$    & $5.414 \ldots 5.420$  & $5.420$  \\[1ex] 
$r$ [fm]        & $1.710 \ldots 1.722$  & $1.727   \ldots 1.735$  & $1.743 \ldots 1.746$  & $1.753$  \\[1ex]
$v_2$ [fm$^3$]  & $0.06 \ldots  0.07 $  & $0.04 \ldots  0.05 $    & $0.04 \ldots 0.05$  & $0.04$   \\[1ex]
$v_3$ [fm$^5$]  & $0.77  \ldots 0.81 $  & $ 0.71   \ldots  0.76 $  &$0.69 \ldots 0.70$  & $0.67$   \\[1ex]
$v_4$ [fm$^7$]  & $-4.3 \ldots  -4.4 $  & $-4.1  \ldots   - 4.2$  & $-4.0 \ldots -4.1$  & $-4.0$   \\[1ex]
\hline  \hline
  \end{tabular}
\vspace{0.3cm}
\parbox{13cm}{
\caption{Scattering length and range parameters for the $^3S_1$ partial wave
  using the 
CR NLO and NNLO potential \cite{EGMs2} compared to the N$^3$LO results and to the Nijmegen PWA \cite{Swart95}.
\label{tab2}}
}
\end{center}
\end{table*}

\pagebreak
\begin{table*}[htb] 
%\vspace{1.cm}
\begin{center}
\begin{tabular}{||l||c|c|c||c||}
    \hline \hline
{} & {} &  {} &  {} & {}\\[-1.5ex]
    & {NLO}  & {NNLO}  & {N$^3$LO} &  {Exp} \\[1ex]
\hline  \hline
{} & {} &  {} &  {} & {}\\[-1.5ex]
$E_{\rm d}$ [MeV] &   $-2.171 \ldots  -2.186$   & $-2.189 \ldots -2.202$   & $-2.216 \ldots -2.223$  & $-$2.224575(9) \\[1ex]
%    \hline
$Q_{\rm d}$ [fm$^2$] &  $0.273 \ldots 0.275$    & $0.271  \ldots  0.275$   & $0.264 \ldots 0.268$     & 0.2859(3) \\[1ex]
%    \hline
$\eta_{\rm d}$ &     $0.0256 \ldots   0.0257$     & $0.0255 \ldots 0.0256$   & $0.0254 \ldots 0.0255$    & 0.0256(4)\\[1ex]
%    \hline
$\sqrt{\langle r^2 \rangle^{\rm d}_m}$ [fm] 
                     &   $1.973  \ldots  1.974$     & $1.970  \ldots  1.972$   &  $1.973 \ldots 1.985$ 
%(XXXcheck)
& 1.9753(11) \\[1ex]
%    \hline
$A_S$ [fm$^{-1/2}$] & $0.868\ldots 0.873$ & $0.874  \ldots  0.879$   &  $0.882 \ldots 0.883$    & 0.8846(9)\\[1ex]
%    \hline
$P_{\rm d}\; [\%]$ &     $3.46 \ldots  4.29$      & $3.53   \ldots  4.93 $   &  $2.73 \ldots 3.63$   &  -- \\[1ex]
    \hline \hline
  \end{tabular}
\vspace{0.3cm}
\parbox{15.8cm}{\caption{Deuteron properties derived from the chiral potential
    at N$^3$LO 
    compared to the NLO and NNLO results from \cite{EGMs2}
    and the data. Here, $E_{\rm d}$ is the
    binding energy, $Q_{\rm d}$ the quadrupole moment, $\eta_{\rm d}$ the asymptotic
    $D/S$ ratio, $\sqrt{\langle r^2 \rangle^{\rm d}_m}$  the root--mean--square matter radius, $A_S$ the 
    strength of the asymptotic S--wave normalization and $P_{\rm d}$ the D-state 
    probability. The data for $E_{\rm d}$ are from \cite{le82}, for $Q_{\rm d}$ from \cite{bi79,er83},
    for $\eta_{\rm d}$ from \cite{rod90} and for $A_S$ from \cite{er83}. For the rms--radius we 
    actually show the experimental value for the deuteron ``point--nucleon'' rms--radius from \cite{friar97}.  
    In the N$^3$LO calculation, the cut--offs are varied as specified in eq.~(\ref{cutoffs}).
\label{tab3}}
}
\end{center}
\end{table*}

\vfill
\eject

\newpage

\vspace{1cm}

\begin{figure*}[htb]
\centerline{{\large \bf FIGURES}}%
\vskip 1 true cm
%\vspace{0.5cm}
\centerline{
\psfig{file=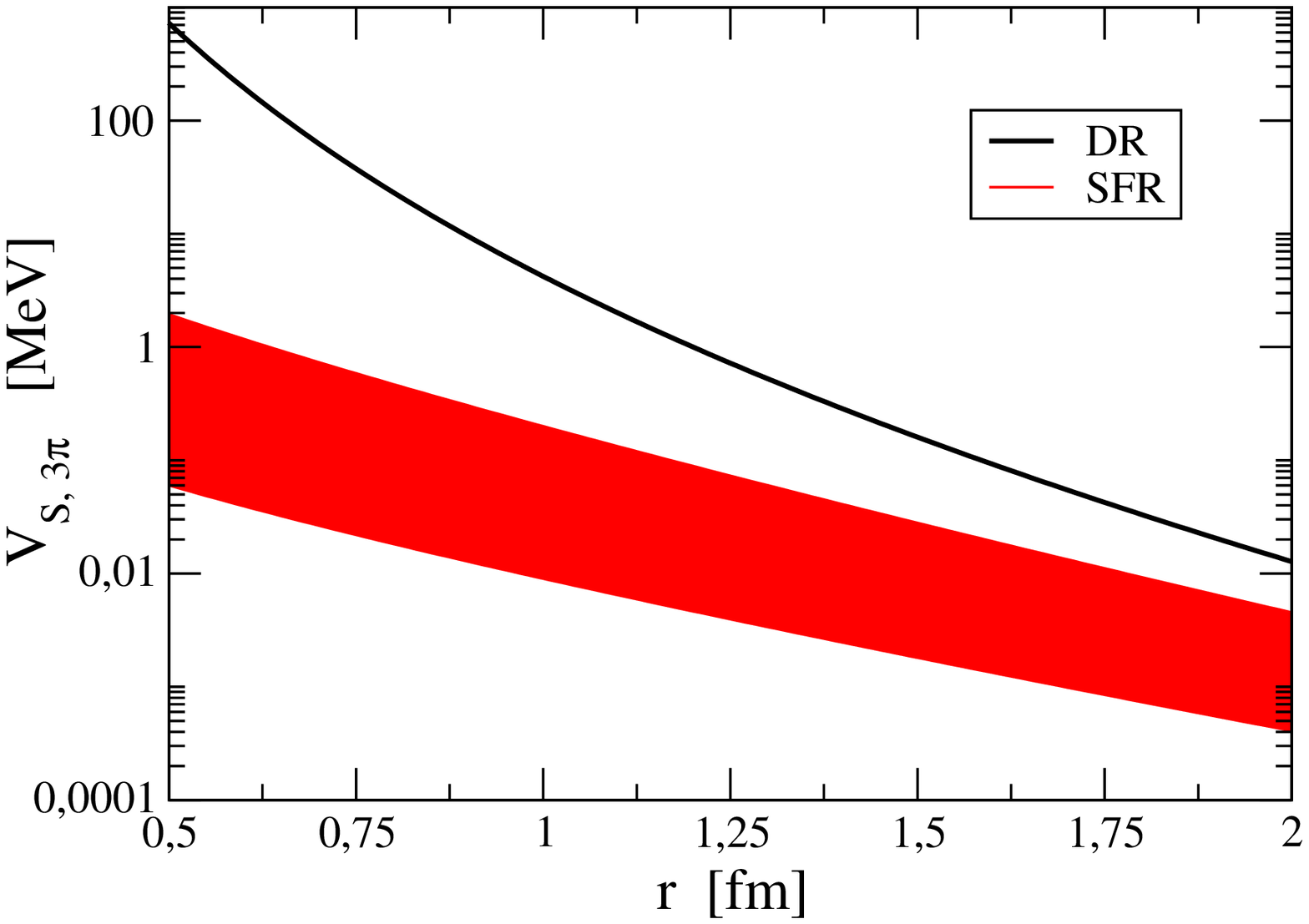,width=11cm}}
\vspace{0.3cm}
\centerline{\parbox{14cm}{
\caption[fig4]{\label{fig:pic1} Isoscalar spin--spin 3PE potential using dimensional (DR) and 
spectral function regularization (SFR). The cut--off in the spectral function varies in the 
range $\tilde \Lambda = 500 \ldots 700$ MeV. 
}}}
\vspace{0.5cm}
%\end{figure*}%
%
%\begin{figure*}[htb]
%\vspace{0.5cm}
\centerline{
\psfig{file=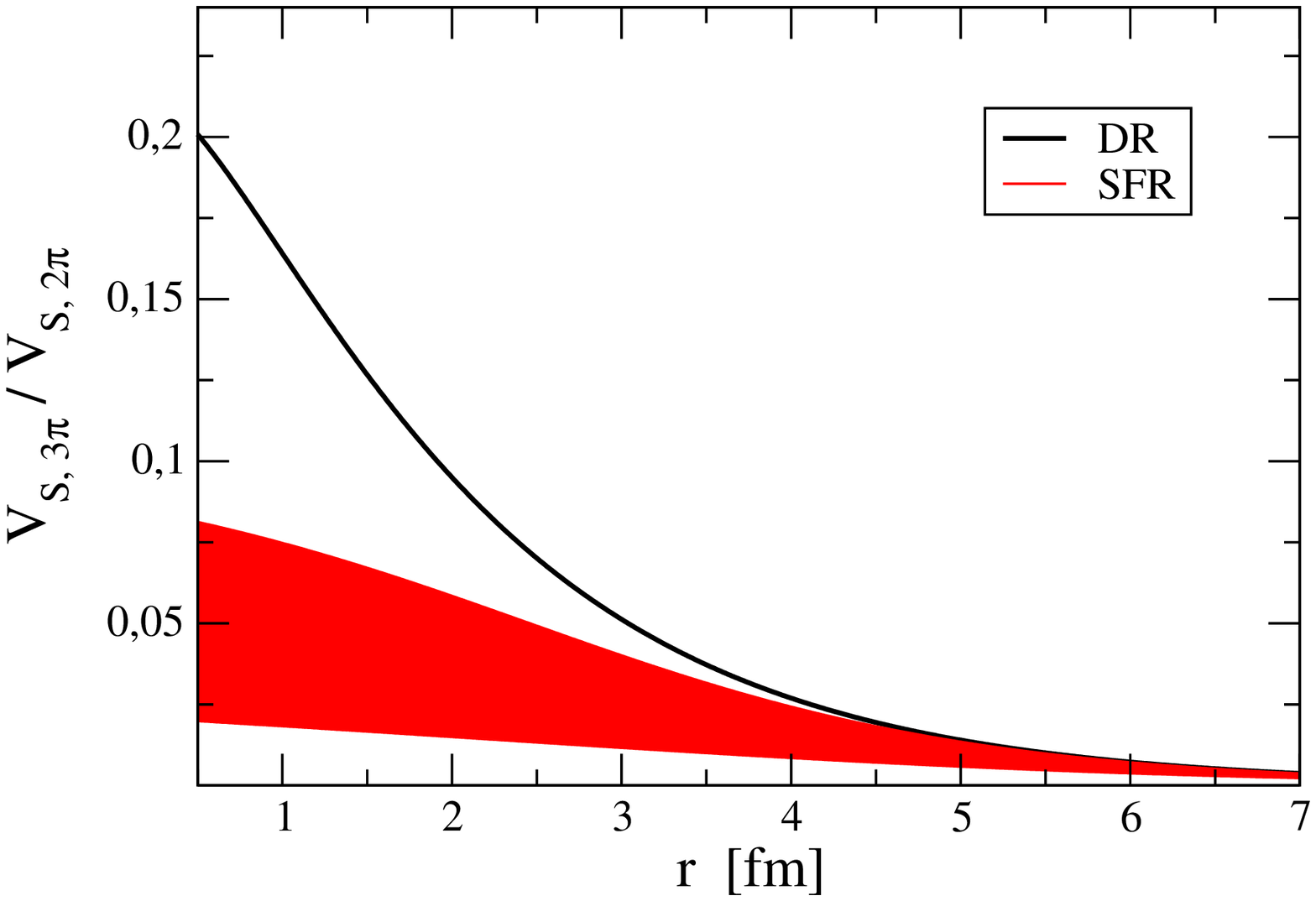,width=11cm}}
\vspace{0.3cm}
\centerline{\parbox{14cm}{
\caption[fig4]{\label{fig:pic2} The ratio of the isoscalar spin--spin 3PE and 
2PE N$^3$LO contributions using dimensional (DR) and spectral function 
regularization (SFR). The cut--off in the spectral function varies in the 
range $\tilde \Lambda = 500 \ldots 700$ MeV. 
}}}
\vspace{0.5cm}
\end{figure*}

\begin{figure*}[htb]
\vspace{0.5cm}
\centerline{
\psfig{file=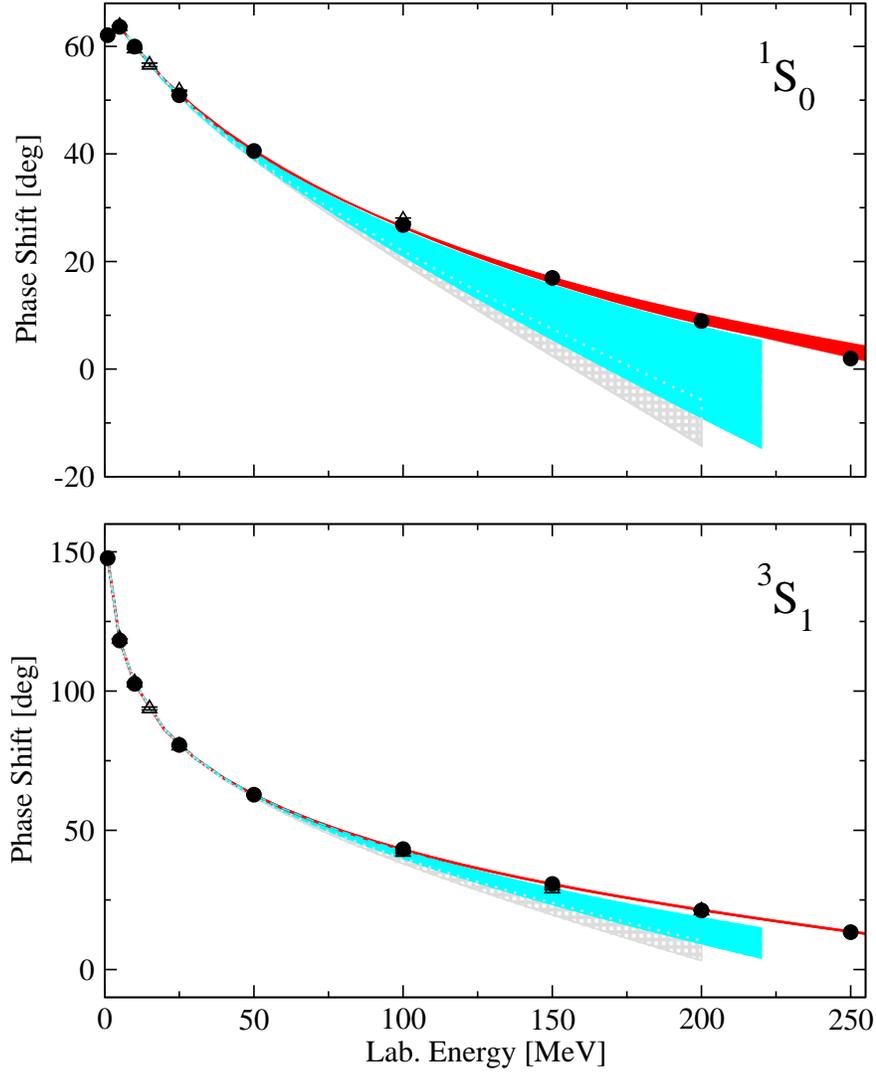,width=12cm}}
\vspace{0.3cm}
\centerline{\parbox{14cm}{
\caption[fig4]{\label{fig1} S--wave {\it np} phase shifts versus the nucleon laboratory energy. The   
grid, light shaded and dark shaded bands show the NLO, NNLO \cite{EGMs2} and N$^3$LO results, respectively.
The cut--offs $\Lambda$ and $\tilde \Lambda$ at N$^3$LO are varied as specified in eq.~(\ref{cutoffs}).
The filled circles depict the Nijmegen PWA results \cite{nijpwa}
and the open triangles are the results from the  Virginia Tech PWA \cite{said}.
}}}
\vspace{0.5cm}
\end{figure*}

\begin{figure*}[htb]
\vspace{0.5cm}
\centerline{
\psfig{file=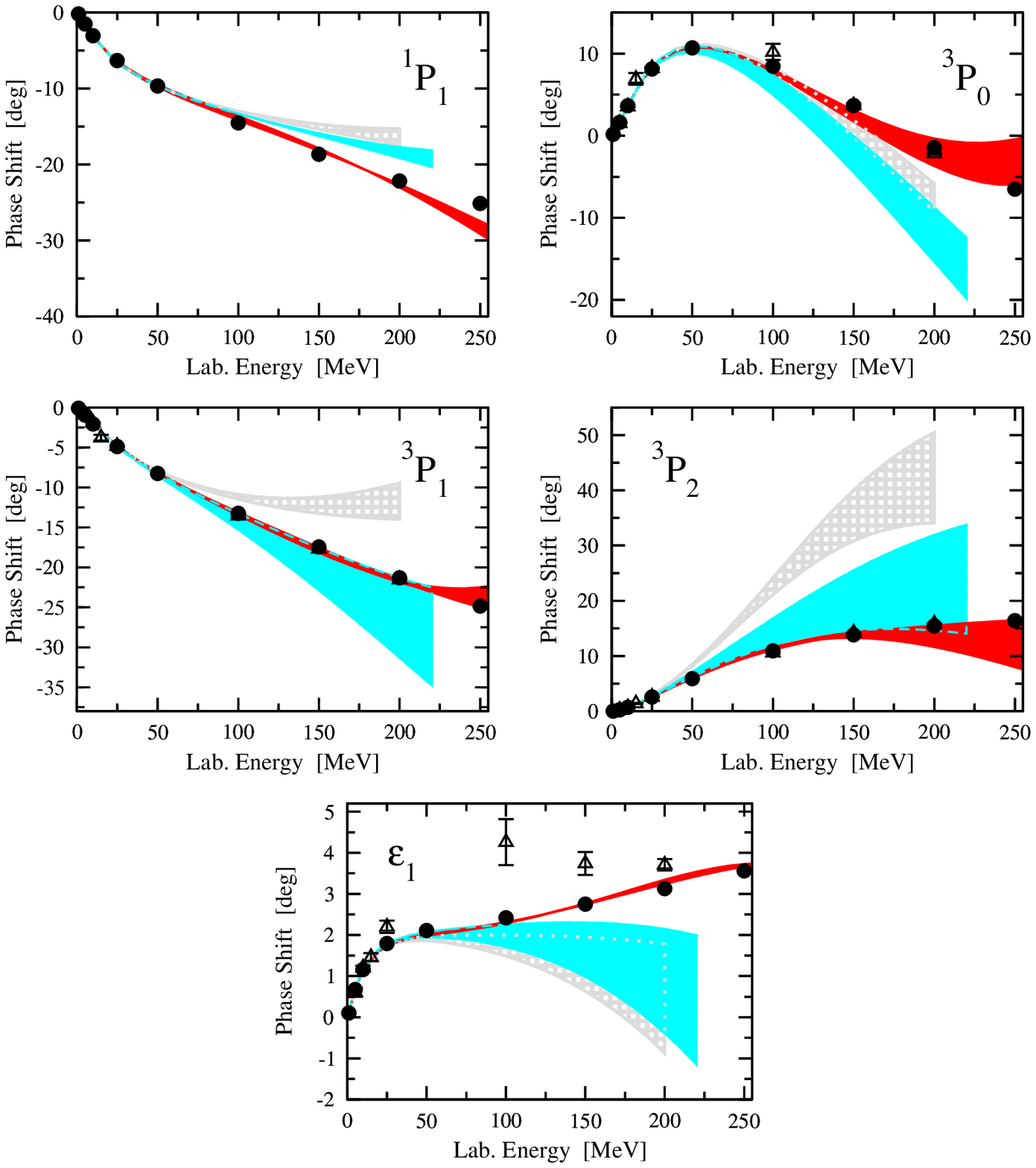,width=15cm}}
\vspace{0.3cm}
\centerline{\parbox{14cm}{
\caption[fig4]{\label{fig2} P--wave {\it np} phase shifts and  mixing angle $\epsilon_1$ versus
the nucleon laboratory energy. For notation see Fig.~\ref{fig1}.
}}}
\vspace{0.5cm}
\end{figure*}

\begin{figure*}[htb]
\vspace{0.5cm}
\centerline{
\psfig{file=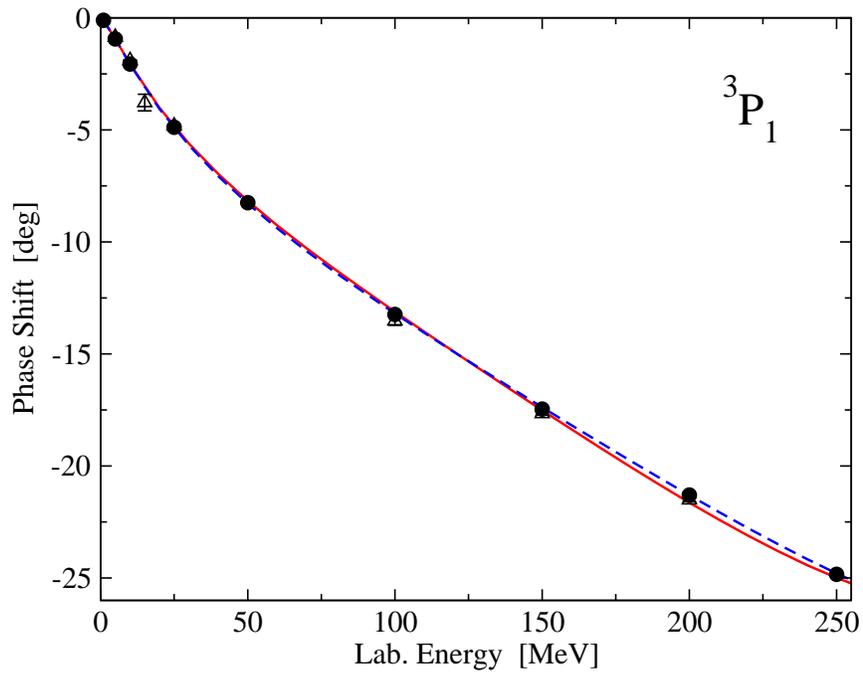,width=12cm}}
\vspace{0.3cm}
\centerline{\parbox{14cm}{
\caption[fig4]{\label{fig:3P1} $^3P_1$--wave {\it np} phase shift versus
the nucleon laboratory energy. The solid and dashed lines correspond 
to the LECs $C_{3P1}$ and $D_{3P1}$ from the first and second lines in eq.~(\ref{LECs_3P1}), respectively.
For remaining notations see Fig.~\ref{fig1}.
}}}
\vspace{0.5cm}
\end{figure*}

\begin{figure*}[htb]
\vspace{0.5cm}
\centerline{
\psfig{file=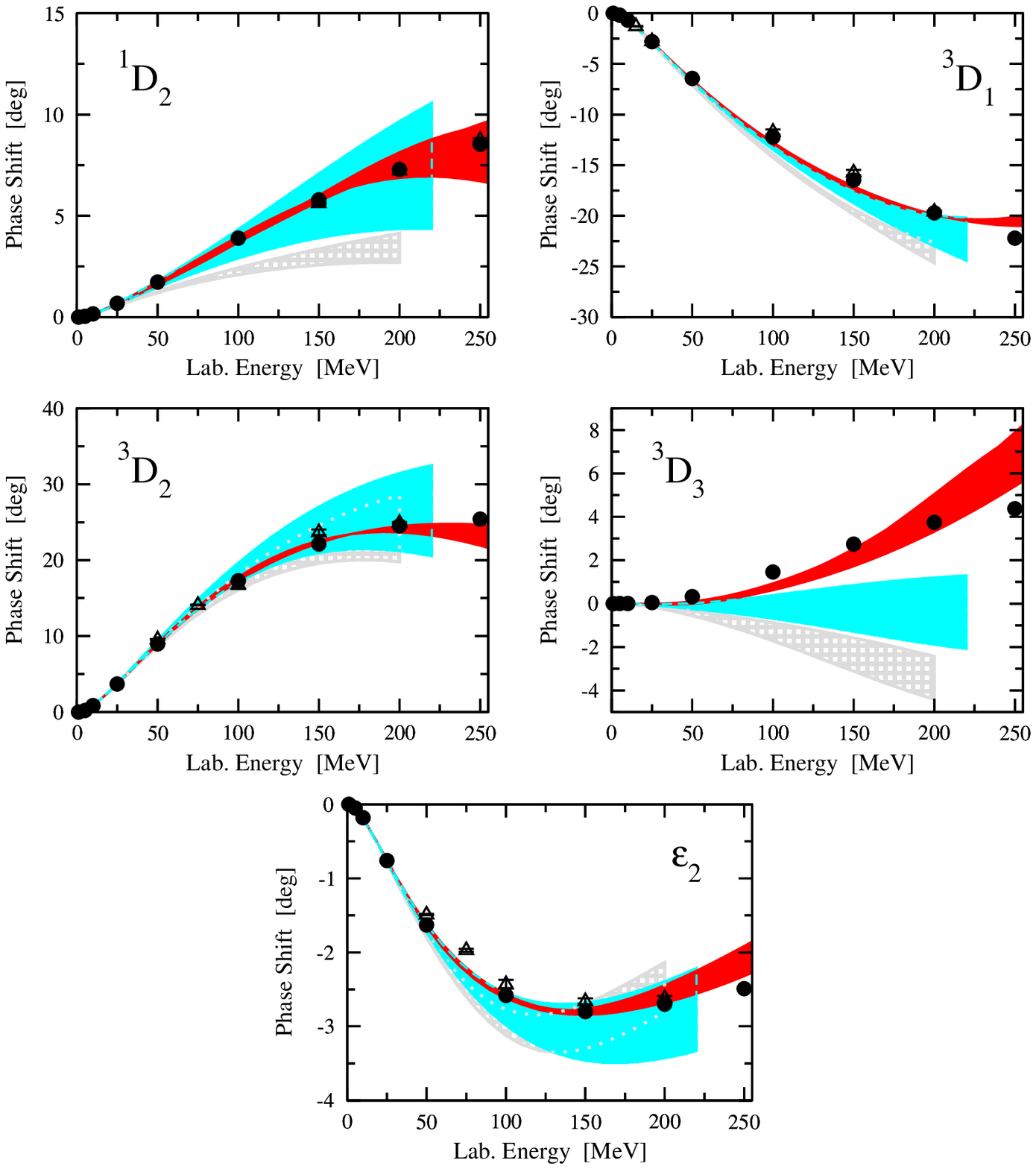,width=15cm}}
\vspace{0.3cm}
\centerline{\parbox{14cm}{
\caption[fig4]{\label{fig3}   D--wave {\it np} phase shifts and  mixing angle $\epsilon_2$ versus
the nucleon laboratory energy. For notation see Fig.~\ref{fig1}. 
}}}
\vspace{0.5cm}
\end{figure*}

\begin{figure*}[htb]
\vspace{0.5cm}
\centerline{
\psfig{file=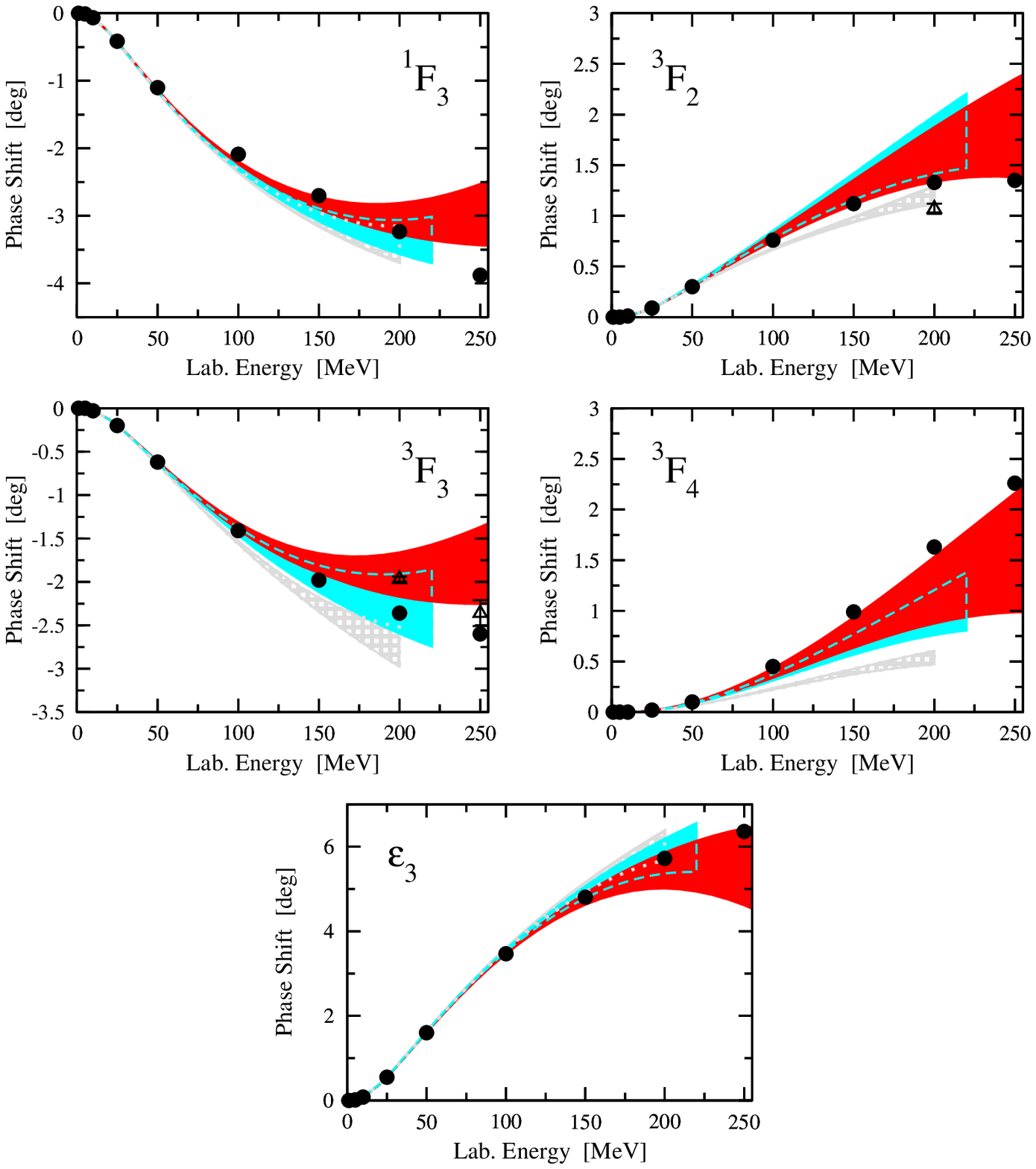,width=15cm}}
\vspace{0.3cm}
\centerline{\parbox{14cm}{
\caption[fig4]{\label{fig4}  F--wave {\it np} phase shifts and  mixing angle $\epsilon_3$ versus
the nucleon laboratory energy. For notation see Fig.~\ref{fig1}.  
}}}
\vspace{0.5cm}
\end{figure*}

\begin{figure*}[htb]
\vspace{0.5cm}
\centerline{
\psfig{file=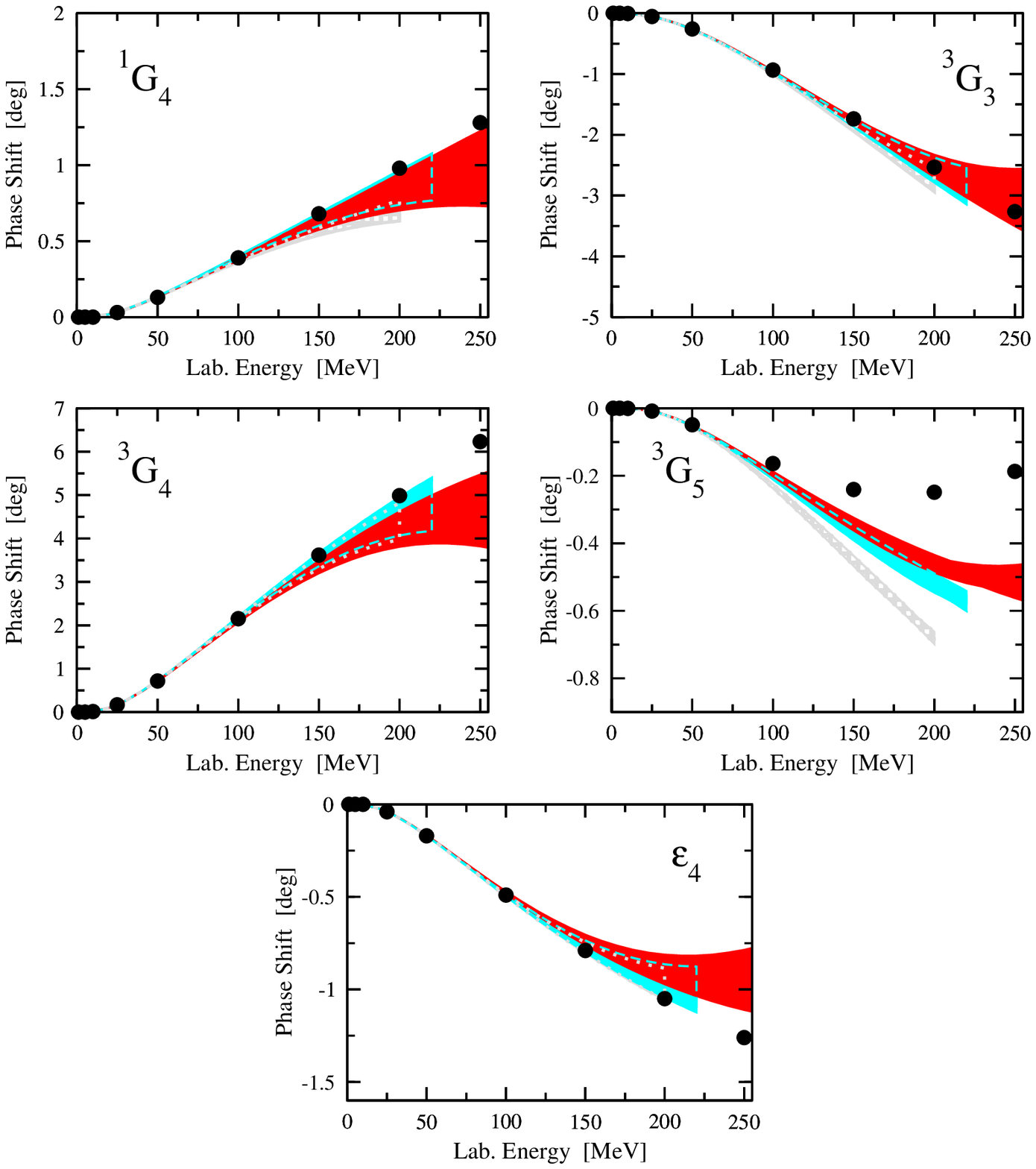,width=15cm}}
\vspace{0.3cm}
\centerline{\parbox{14cm}{
\caption[fig4]{\label{fig5}  G--wave {\it np} phase shifts and  mixing angle $\epsilon_4$ versus
the nucleon laboratory energy. For notation see Fig.~\ref{fig1}.  
}}}
\vspace{0.5cm}
\end{figure*}

\begin{figure*}[htb]
\vspace{0.5cm}
\centerline{
\psfig{file=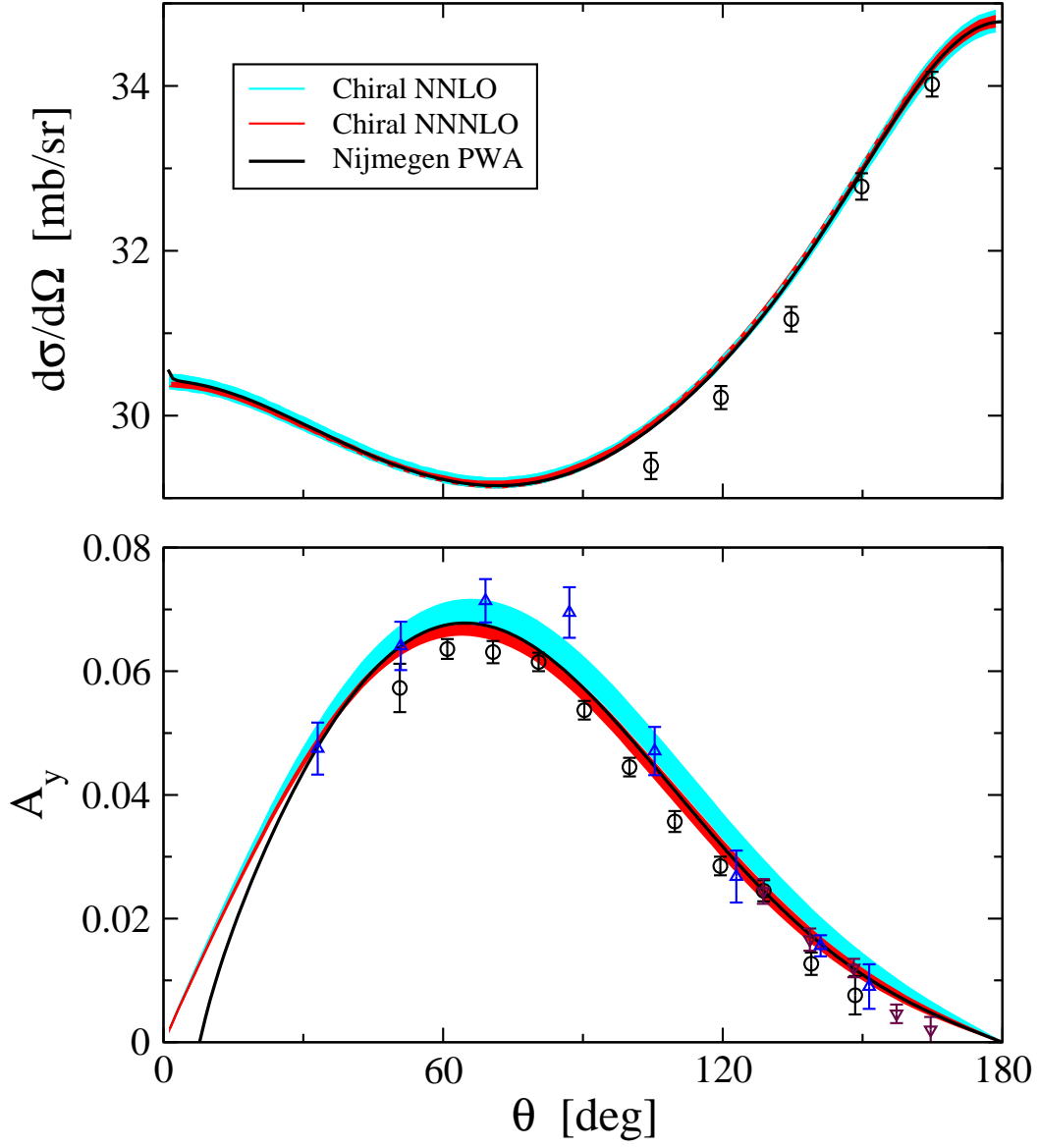,width=15cm}}
\vspace{0.3cm}
\centerline{\parbox{14cm}{
\caption[fig4]{\label{fig6}  
{\it np} differential cross section and vector analyzing power at $E_{\rm lab} = 25$ MeV.
The Nijmegen PWA result is taken from \cite{nnonline}.
Data for the cross section are taken from \cite{fink90} and for the analyzing power
from \cite{sro86,wil84}. The cut--offs $\Lambda$ and $\tilde \Lambda$ are varied 
as specified in eq.~(\ref{cutoffs}).
}}}
\vspace{0.5cm}
\end{figure*}

\begin{figure*}[htb]
\vspace{0.5cm}
\centerline{
\psfig{file=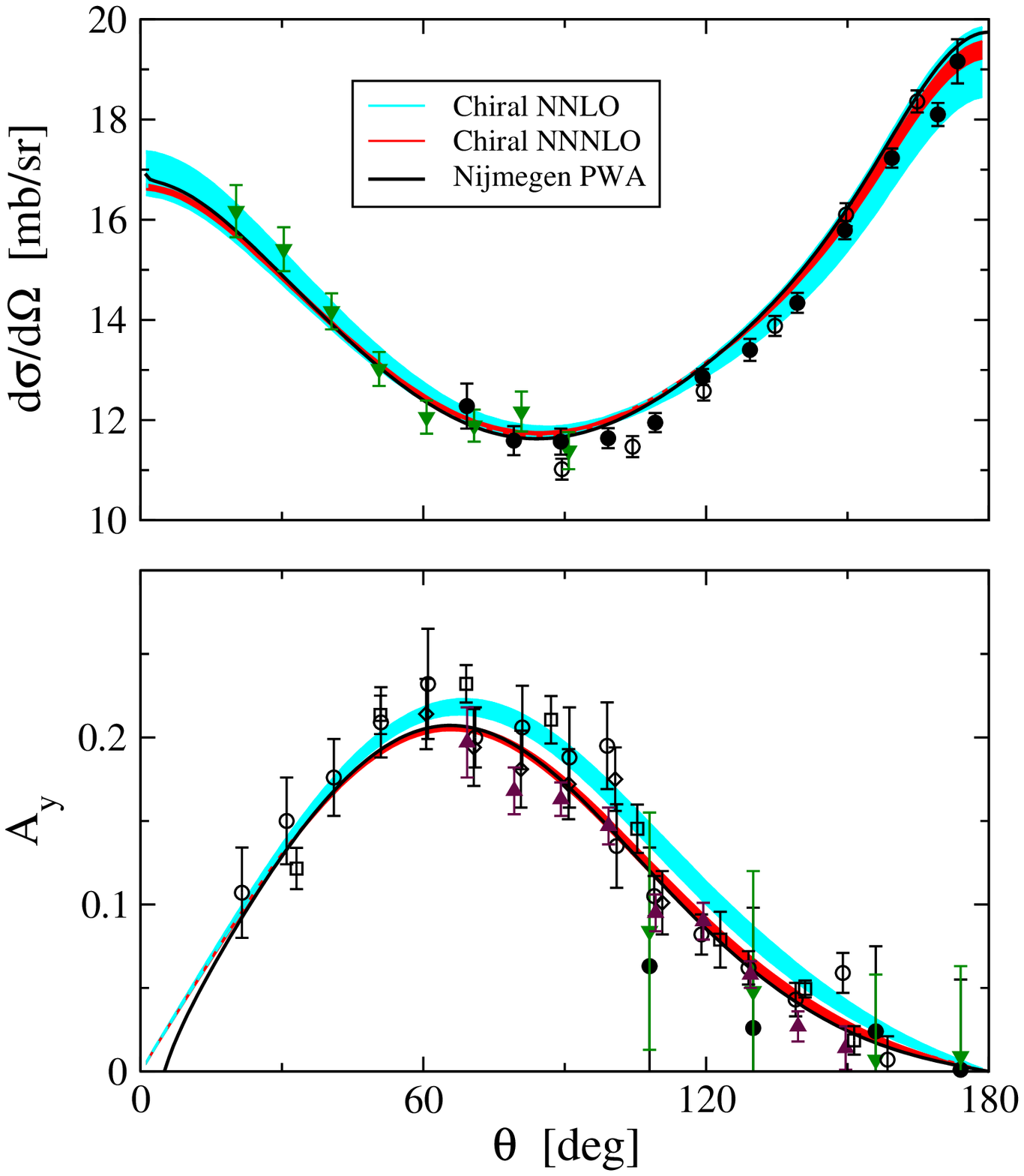,width=15cm}}
\vspace{0.3cm}
\centerline{\parbox{14cm}{
\caption[fig4]{\label{fig7}  
{\it np} differential cross section and vector analyzing power at $E_{\rm lab} = 50$ MeV.
Data for the cross section are taken from \cite{fink90,mont77} and for the analyzing power
from \cite{fitz80,gar80,lang65,rom78,wil84}. For remaining notations see Fig.~\ref{fig6}.
}}}
\vspace{0.5cm}
\end{figure*}

\begin{figure*}[htb]
\vspace{0.5cm}
\centerline{
\psfig{file=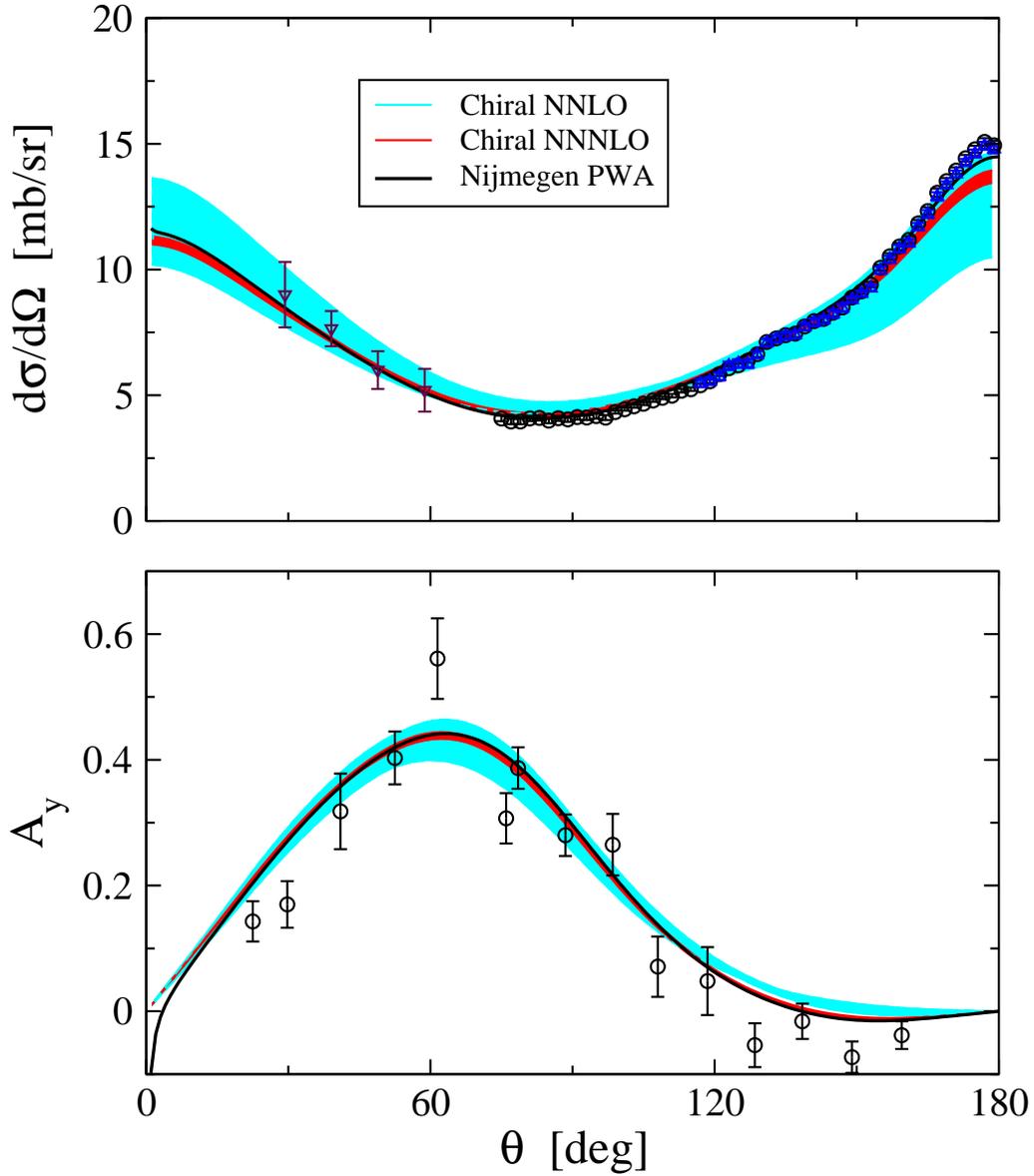,width=15cm}}
\vspace{0.3cm}
\centerline{\parbox{14cm}{
\caption[fig4]{\label{fig8}  
{\it np} differential cross section and vector analyzing power at $E_{\rm lab} = 96$ MeV.
Data for the cross section are taken from \cite{gri58,ra01,roe92}. Data for the analyzing power
are at $E_{\rm lab} = 95$ MeV and taken from \cite{sta57}. For remaining notations see Fig.~\ref{fig6}.
}}}
\vspace{0.5cm}
\end{figure*}

\begin{figure*}[htb]
\vspace{0.5cm}
\centerline{
\psfig{file=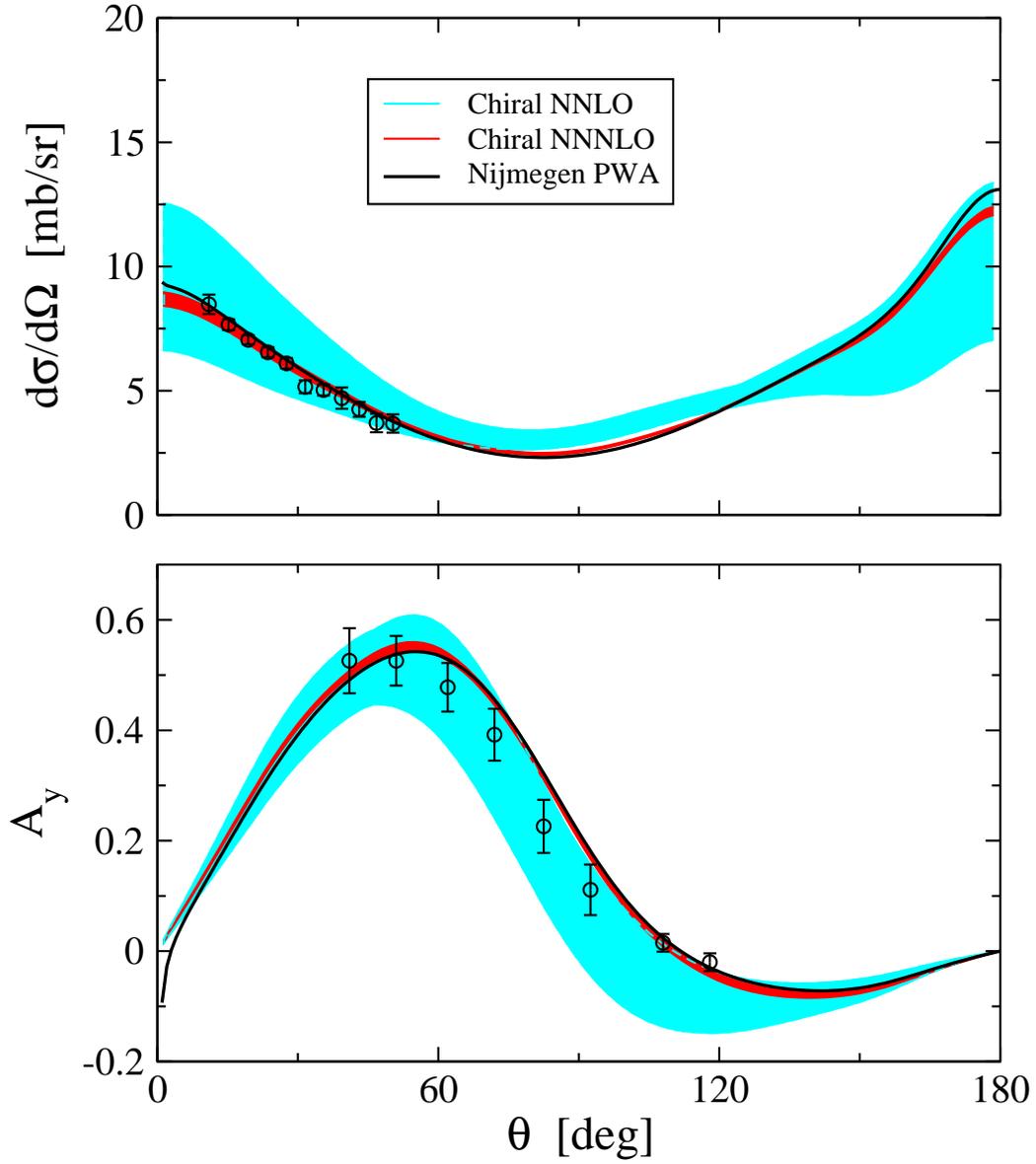,width=15cm}}
\vspace{0.3cm}
\centerline{\parbox{14cm}{
\caption[fig4]{\label{fig9}  
{\it np} differential cross section and vector analyzing power at $E_{\rm lab} = 143$ MeV.
Data for the cross section are at $E_{\rm lab} = 142.8$ MeV and taken from \cite{ber76} and for the analyzing power
from \cite{kuc61}. For remaining notations see Fig.~\ref{fig6}.
}}}
\vspace{0.5cm}
\end{figure*}

\begin{figure*}[htb]
\vspace{0.5cm}
\centerline{
\psfig{file=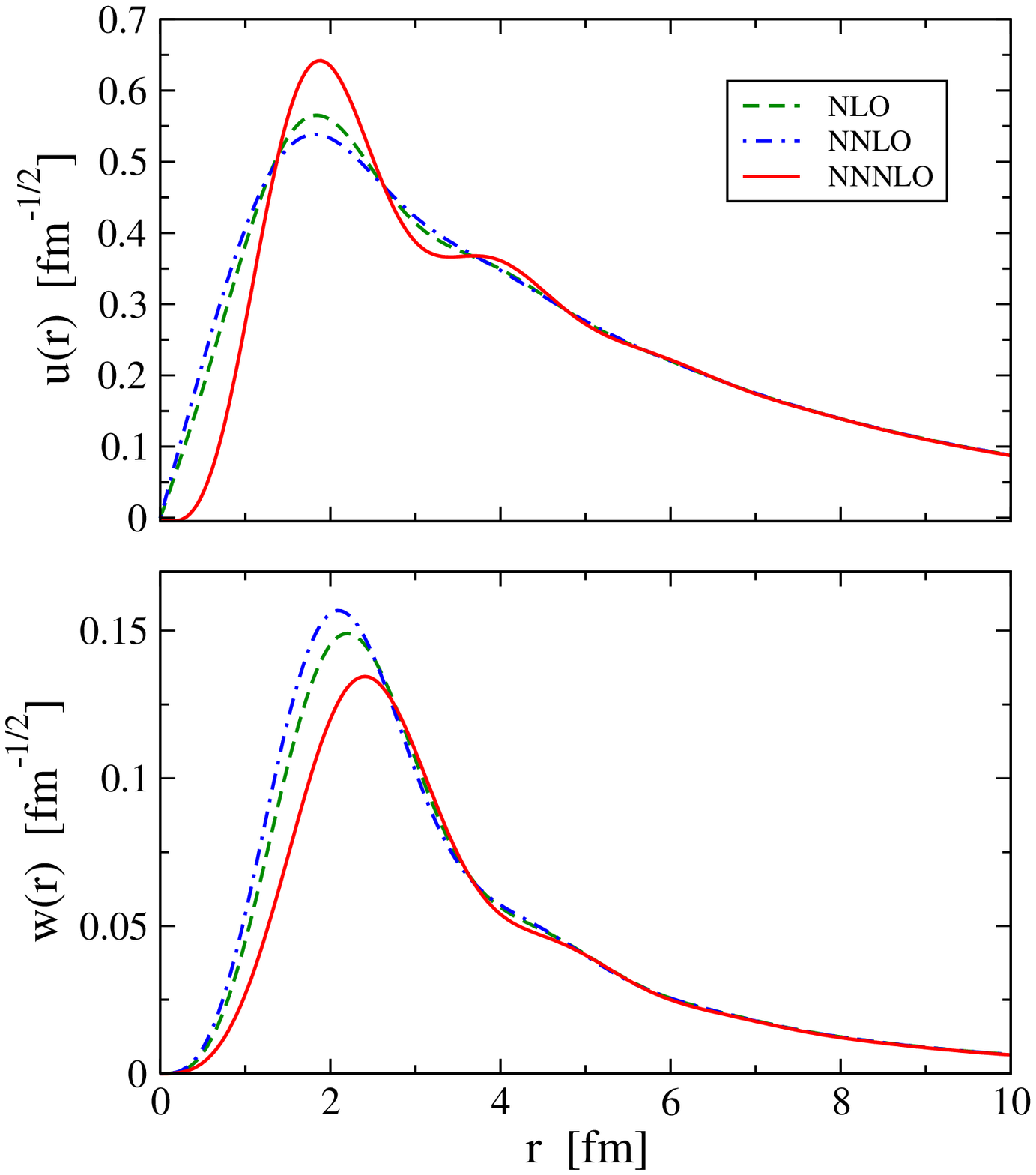,width=15cm}}
\vspace{0.3cm}
\centerline{\parbox{14cm}{
\caption[fig4]{\label{fig10}  
Coordinate space representation of the S-- (upper panel) and D--wave
(lower panel) deuteron wave functions at NLO, NNLO and N$^3$LO for the cut--offs:
$\Lambda = 550$ MeV, $\tilde \Lambda = 600$ MeV.}}}
\vspace{0.5cm}
\end{figure*}


\begin{thebibliography}{99}
\bibitem{wein} S.~Weinberg, Nucl. Phys.  B {\bf 363}  (1991) 3.\vs
\bibitem{border} S.~R.~Beane, P.~F.~Bedaque, W.~C.~Haxton, D.~R.~Phillips and M.~J.~Savage,
  in M.~Shifman: At the frontier of particle physics, vol. 1, p. 133-269 
  (World Scientific, Singapore, 2001).\vs 
\bibitem{BvK} P.~F.~Bedaque and U.~van Kolck,
  Ann.\ Rev.\ Nucl.\ Part.\ Sci.\  {\bf 52} (2002) 339.\vs
\bibitem{E3NF} E.~Epelbaum, A.~Nogga, W.~Gl\"ockle, H.~Kamada, U.-G.~Mei{\ss}ner and H.~Witala,
  Phys.\ Rev.\ C {\bf 66} (2002) 064001.\vs
\bibitem{EM3} D.~R.~Entem and R.~Machleidt, Phys.\ Rev.\ C {\bf 68} (2003) 041001.\vs
\bibitem{Machl_INT} R.~Machleidt, {\it Some Issues Concerning the Nucleon--Nucleon Interaction Based upon 
Chiral Effective Field Theory}, talk given at the INT program Theories of Nuclear Forces 
  and Nuclear Systems, September 29 - December 5 2003, INT, Seattle (USA).\vs
\bibitem{WME} M.~Walzl, U.-G. Mei{\ss}ner and E.~Epelbaum,
  Nucl.\ Phys.\ A {\bf 693} (2001) 663.\vs
\bibitem{Betal} S.~R.~Beane, P.~F.~Bedaque, M.~J.~Savage and U.~van Kolck,
  Nucl.\ Phys.\ A {\bf 700} (2002) 377.\vs
\bibitem{BeSa1} S.~R.~Beane and M.~J.~Savage, Nucl.\ Phys.\ A {\bf 713} (2003) 148.\vs
\bibitem{CHlim} E.~Epelbaum, U.-G.~Mei{\ss}ner and W.~Gl\"ockle, Nucl.\ Phys.\ A {\bf 714} (2003) 535.\vs
\bibitem{BeSa2} S.~R.~Beane and M.~J.~Savage, Nucl.\ Phys.\ A {\bf 717} (2003) 91.\vs
\bibitem{ubi} C.~Ord\'{o}\~{n}ez, L.~Ray and U.~van Kolck, 
              Phys. Rev. C {\bf 53} (1996) 2086.\vs
\bibitem{fr94} J.L.~Friar, S.A.~Coon, Phys. Rev. C {\bf 49} (1994) 1272. \vs
\bibitem{norb} N.~Kaiser, R.~Brockmann and W.~Weise, 
              Nucl. Phys. A {\bf 625} (1997) 758.\vs
\bibitem{EGM1} E.~Epelbaum, W.~Gl\"ockle and U.-G. Mei{\ss}ner,
  Nucl.\ Phys.\ A {\bf 637} (1998) 107.\vs
\bibitem{EGM2} E.~Epelbaum, W.~Gl\"ockle and U.-G. Mei{\ss}ner,
  Nucl.\ Phys.\ A {\bf 671} (2000) 295.\vs
\bibitem{EGMs1} E.~Epelbaum,  W.~Gl\"ockle and U.-G. Mei{\ss}ner,
  Eur.\ Phys.\ J.\ A {\bf 19} (2004) 125.\vs
\bibitem{NK21} N.~Kaiser, Phys.\ Rev.\ C {\bf 64} (2001) 057001.\vs
\bibitem{NK31} N.~Kaiser, Phys.\ Rev.\ C {\bf 61} (2000) 014003.\vs
\bibitem{NK32} N.~Kaiser, Phys.\ Rev.\ C {\bf 62} (2000) 024001.\vs
\bibitem{EGMs2} E.~Epelbaum,  W.~Gl\"ockle and U.-G. Mei{\ss}ner,
  Eur.\ Phys.\ J.\ A {\bf 19} (2004) 401.\vs
\bibitem{Ephd} E.~Epelbaum, doctoral thesis, published in {\it Berichte des Forschungszentrums
  J\"ulich},  No. 3803 (2000).\vs
\bibitem{Fet98} N.~Fettes, U.-G.~Mei{\ss}ner, and S.~Steininger, Nucl. Phys. A {\bf 640} (1998) 199.\vs
\bibitem{Norbert04} N.~Kaiser, private communication.\vs
\bibitem{Friar99} J.L.Friar, Phys. Rev. C {\bf 60} (1999) 034002.\vs
\bibitem{EM2} D.~R.~Entem and R.~Machleidt, Phys.\ Rev.\ C {\bf 66} (2002) 014002.\vs
\bibitem{NK33} N.~Kaiser, Phys.\ Rev.\ C {\bf 63} (2001) 044010.\vs
\bibitem{rob00}  M.R.~Robilotta, Phys. Rev. C {\bf 63} (2001) 044004.\vs
\bibitem{higa03} R.~Higa and M.R.~Robilotta, Phys. Rev. C {\bf 68} (2003) 024004.\vs
\bibitem{bech99} T.~Becher and H.~Leutwyler, Eur. Phys. J. C {\bf 9} (1999) 643.\vs
\bibitem{kolck} U.~van Kolck, Ph.D.~Thesis, University of Texas at Austin, 1993, UMI-94-01021-mc.\vs
\bibitem{vKNij} U.~van Kolck et al., Phys. Rev. Lett. {\bf 80} (1998) 4386.\vs
\bibitem{kolck96} U.~van Kolck, J.L. Friar, T. Goldman, Phys. Lett. B {\bf 371} (1996) 169.\vs
\bibitem{Ep99} E.Epelbaum, U.-G. Mei{\ss}ner, Phys. Lett. B {\bf 461} (1999) 287.\vs
\bibitem{friar03}  J.L. Friar et al.,  Phys. Rev. C {\bf 68} (2003) 024003.\vs
\bibitem{coon96} S.A. Coon, J.A. Niskanen, Phys. Rev. C {\bf 53} (1996) 1154.\vs
\bibitem{nis02} J.A. Niskanen, Phys. Rev. C {\bf 65} (2002) 037001.\vs
\bibitem{nijpwa} V.G.J.~Stoks et al., Phys. Rev. C {\bf 48} (1993) 792.\vs
\bibitem{urech95} R.~Urech, Nucl. Phys. B {\bf 433} (1995) 234.\vs
\bibitem{MS} U.-G.~Mei{\ss}ner and S.~Steininger, Phys.\ Lett.\ B {\bf 419} (1998) 403.\vs
\bibitem{MM} G.~M\"uller and U.-G.~Mei{\ss}ner, Nucl.\ Phys.\ B {\bf 556} (1999) 265.\vs
\bibitem{fet99} N. Fettes, U.-G. Mei{\ss}ner and S. Steininger, Phys. Lett. B {\bf 451} (1999) 233.\vs 
\bibitem{fet00} N. Fettes and U.-G. Mei{\ss}ner, Phys.Rev. C {\bf 63} (2001) 045201.\vs
\bibitem{fet01} N. Fettes and U.-G. Mei{\ss}ner, Nucl. Phys. A {\bf 693} (2001) 693.\vs
\bibitem{Bern}J.~Gasser, M.~A.~Ivanov, E.~Lipartia, M.~Mojzis and A.~Rusetsky,
  Eur.\ Phys.\ J.\ C {\bf 26} (2002) 13.\vs
\bibitem{GLmass}J.~Gasser and H.~Leutwyler, Phys.\ Rept.\  {\bf 87} (1982) 77.\vs
\bibitem{pasc03} V. Pascalutsa, D.R. Phillips, nucl--th/0308065.\vs
\bibitem{FvK}J. Friar and  U. van Kolck, Phys. Rev. C {\bf 60} (1999) 034006.\vs
\bibitem{austin} G.J.M.~Ausin and J.J.~de Swart, Phys. Rev. Lett. {\bf 50} (1983) 2039.\vs
\bibitem{Berg88} J.R.~Bergervoet et al., Phys. Rev. C {\bf 38} (1988) 15.\vs
\bibitem{stoks90} V.G.~Stoks, and J.J.~de Swart, Phys. Rev. C {\bf 42} (1990) 1235.\vs
\bibitem{stoksPhD}  V.G.~Stoks,  Ph.D.~Thesis, Nijmegen, 1990.\vs
\bibitem{ueling} E.A.~Ueling, Phys. Rev. {\bf 48} (1935) 55. \vs
\bibitem{durand} L.~Durand III, Phys. Rev. {\bf 108} (1957) 1597.\vs
\bibitem{ES}D.~Eiras and J.~Soto, Phys.\ Lett.\ B {\bf 491} (2000) 101.\vs
\bibitem{stapp} H.P.~Stapp, T.J.~Ypsilantis and N.~Metropolis,
  Phys. Rev. {\bf 105} (1957) 302.\vs
\bibitem{Bl52} J.M.~Blatt and L.C.~Biedenharn, Phys. Rev. {\bf 86} (1952) 399, 
  Rev. Mod, Phys. {\bf 24} (1952) 258. \vs
\bibitem{KGT} H.~Kamada and W.~Gl\"ockle, Phys. Rev. Lett. {\bf 80} (1998) 2547. \vs
\bibitem{friar79} J.L. Friar, Phys. \ Rev. \ C~{\bf 20} (1979) 325. \vs
\bibitem{kohno83} M. Kohno, J. Phys. G: Nucl. Phys. {\bf 9} (1983) L85.\vs
\bibitem{CP} D.~R.~Phillips and T.~D.~Cohen, Nucl.\ Phys.\ A {\bf 668} (2000) 45.\vs
\bibitem{klar86} S. Klarsfeld et al., Nucl. Phys. A {\bf 456} (1986) 373.\vs
\bibitem{mart95} J. Martorell, D.W.L. Sprung, D.C. Zheng, Phys. Rev. C {\bf 51} (1995) 1127. \vs
\bibitem{friar97} J.L. Friar, J. Martorell, and D.W.L. Sprung, Phys. Rev. A {\bf 56} (1997) 4579.\vs
\bibitem{RR} R.~Rosenfelder, Phys.\ Lett.\ B {\bf 479} (2000) 381.\vs
\bibitem{MR} K.~Melnikov and T.~van Ritbergen, Phys.\ Rev.\ Lett.\  {\bf 84} (2000) 1673.\vs
\bibitem{IS} I.~Sick, Phys.\ Lett.\ B {\bf 576} (2003) 62.\vs
\bibitem{kop95} S. Kopecky et al.,Phys. Rev. Lett. {\bf 74} (1995) 2427.\vs 
\bibitem{jen00} B.K. Jennings, Phys. Rev. C {\bf 62} (2000) 027602.\vs
\bibitem{vanorden} M. Garcon and J.W. Van Orden, Adv. Nucl. Phys. {\bf 26} (2001) 293.\vs
\bibitem{GG} R.~Gilman and F.~Gross, J.\ Phys.\ G {\bf 28} (2002) R37.\vs
\bibitem{rentm03} M.C.M.~Rentmeester, R.G.E.~Timmermans and J.J.~de Swart,  Phys. Rev. C {\bf 67} 
(2003) 044001.\vs
\bibitem{Paul} P. B\"uttiker and U.-G. Mei{\ss}ner, Nucl. Phys. A {\bf 668} (2000) 97.\vs
\bibitem{how98} C.R. Howell et al., Phys. Lett. B {\bf 444} (1998) 252.\vs
\bibitem{gonz99} D.E. Gonz$\acute{a}$lez Trotter et al., Phys. Rev. Lett. {\bf 83} (1999) 3788.\vs
\bibitem{Lepage} P. Lepage, nucl--th/9706029.\vs
\bibitem{Lepage_INT} P.~Lepage, {\it Tutorial: Renormalizing the Schr\"odinger Equation}, 
talk given at the INT program Effective Field Theories and Effective Interactions, 
June 25 - August 5 2000, INT, Seattle (USA).\vs
\bibitem{Geg01} J. Gegelia and G. Japaridze, Phys. Lett. B {\bf 517} (2001) 476.\vs
\bibitem{Geg04} J. Gegelia and S. Scherer, nucl--th/0403052.\vs
\bibitem{beane00} S.R. Beane et al., Phys. Rev. A {\bf 64} (2001) 042103. \vs
\bibitem{ep2002} E. Epelbaum et al., Eur. Phys.J. {\bf A} 15 (2002) 543. \vs
\bibitem{Berg90} J.R.~Bergervoet et al., Phys. Rev. C {\bf 41} (1990) 1435.\vs
\bibitem{nag78} M.M.~Nagels, T.A.~Rijken, and J.J.~de Swart, Phys. Rev. D {\bf 17} (1978) 768.\vs 
\bibitem{Hen79} E.M.~Henley and g.A.~Miller, in {\it Mesons in Nuclei}, edited by M.~Rho and 
D.~Wilkinson, North Holland, 1979, p.406.\vs
\bibitem{V18} R.B. Wiringa et al., Phys. Rev. C {\bf 51} (1995) 38.\vs
\bibitem{Mill90} G.A. Miller, M.K. Nefkens, and I. Slaus, Phys. Rep. {\bf 194} (1990) 1.\vs
\bibitem{Gonz99} D.E. Gonzales Trotter et al., Phys. Rev. Lett. {\bf 83} (1999) 3788.\vs
\bibitem{Hu00} V. Huhn et al., Phys. Rev. C {\bf 63} (2000) 014003.\vs
\bibitem{kong00} X. Kong, and F. Ravndal, Nucl. Phys. A {\bf 665} (2000) 137.\vs
\bibitem{gegelia03} J. Gegelia, Eur. Phys. J. A {\bf 19} (2004) 355.\vs
\bibitem{jackson50} J.D. Jackson, and J.M. Blatt, Rev. Mod. Phys. {\bf 22} (1950) 77\vs
\bibitem{Fachr01} I. Fachruddin, Ch. Elster, and W. Gl\"ockle, Nucl. Phys. A {\bf 689}, (2001) 507c.\vs
\bibitem{Swart95} J.J. de Swart, C.P.F. Terheggen, and V.G.J. Stoks, nucl-th/9509032.\vs
\bibitem{chen99_1} J.-W. Chen, G. Rupak, and M.J. Savage, Nucl. Phys. A {\bf 653} (1999) 386.\vs
\bibitem{chen99_2} J.-W. Chen, G. Rupak, and M.J. Savage, Phys. Lett. B {\bf 464} (1999) 1.\vs
\bibitem{walzl01} M. Walzl and U.-G. Mei{\ss}ner, Phys.Lett. B {\bf 513} (2001) 37.\vs
\bibitem{Er71} K.~Erkelenz, R.~Alzetta and K.~Holinde, Nucl. Phys. A {\bf 176} (1971) 413. \vs
\bibitem{Ja59} M.~Jacob and G.~Wick, Ann. of Phys. {\bf 7} (1959) 404 .\vs
\bibitem{VP}C.~M.~Vincent and S.~C.~Phatak, Phys. Rev. C {\bf 10} (1974) 391.\vs
\bibitem{hink71} O. Hinckelmann and L. Spruch, Phys. Rev. A {\bf 3} (1971) 2. \vs
\bibitem{Rentm99} M.C.M. Rentmeester, private communication, 1999. \vs
\bibitem{le82} C. van der Leun and C. Alderlisten, Nucl. Phys. A {\bf 380} (1982) 261.\vs
\bibitem{bi79} D.M. Bishop and L.M. Cheung, Phys. Rev. A {\bf 20} (1979) 381.\vs
\bibitem{er83} T.E.O. Ericson and M. Rosa--Clot, Nucl. Phys. A {\bf 405} (1983) 497.\vs
\bibitem{rod90} N.L. Rodning and L.D. Knutson, Phys. Rev. C {\bf 41} (1990) 898.\vs
\bibitem{said} SAID on-line program, R.A. Arndt et al., http://gwdac.phys.gwu.edu.\vs
\bibitem{nnonline} Nijmegen NN on-line program, http://nn-online.sci.kun.nl.\vs
\bibitem{fink90} G. Fink et al., Nucl. Phys. A {\bf 518} (1990), 561.\vs
\bibitem{sro86} J. Sromicki et al., Phys. Rev. Lett. {\bf 57} (1986), 2359.\vs
\bibitem{wil84} J. Wilczynski et al., Nucl. Phys. A {\bf 425} (1984), 458.\vs
\bibitem{mont77} T.C. Montgomery et al., Phys. Rev. C {\bf 16} (1977) 499.\vs
\bibitem{fitz80} D.H. Fitzgerald et al., Phys. Rev. C {\bf 21} (1980) 1190.\vs
\bibitem{gar80} R. Garrett et al., Phys. Rev. C {\bf 21} (1980) 1149.\vs
\bibitem{lang65} A. Langsford et al., Nucl. Phys. {\bf 74} (1965), 241.\vs
\bibitem{rom78} J.L. Romero et al., Phys. Rev. C {\bf 17} (1978) 468.\vs
\bibitem{roe92} T. R\"onnqvist et al., Phys. Rev. C {\bf 45} (1992) R496.\vs
\bibitem{sta57} G.H. Stafford et al., Lett. Nuovo Cim. {\bf 5} (1957) 1589.\vs
\bibitem{gri58} T.C. Griffith et al., Proc. Phys. Soc. London, Sect. A {\bf 71} (1958) 305.\vs
\bibitem{ra01} J. Rahm et al., Phys. Rev. C {\bf 63} (2001) 044001.\vs
\bibitem{ber76} A.J. Bersbach et al., Phys. Rev. D {\bf 13} (1976) 535.\vs
\bibitem{kuc61} A.F. Kuckes et al., Phys. Rev. {\bf 121} (1961) 1226.\vs
\end{thebibliography}
\end{document}